\documentclass[a4paper,11pt]{article}

\usepackage[latin1]{inputenc}
\usepackage{amsmath,amssymb,amsthm}
\usepackage{a4wide}
\usepackage{graphicx}
\usepackage{enumerate}
\usepackage{float}
\usepackage{calc}
\usepackage{xcolor,xspace,xargs}
\usepackage{stmaryrd}
\usepackage[pdftitle={Fully dynamic recognition of proper circular-arc graphs},pdfauthor={Francisco Soulignac},pdfcreator={},pdfsubject={Fully dynamic recognition of proper circular-arc graphs},pdfkeywords={dynamic recognition, proper circular-arc graphs, round graphs, co-connectivity},colorlinks=true,linkcolor=blue,citecolor=blue]{hyperref}

\newtheorem{thm}{Theorem}[section]
\newtheorem{lem}[thm]{Lemma}
\newtheorem{cor}[thm]{Corollary}
\newtheorem{obs}[thm]{Observation}

\floatstyle{ruled}
\newfloat{algorithm}{htbp!}{loa}[section]
\floatname{algorithm}{Algorithm}

\newlength{\algindent}
\newenvironment{AlgorithmSteps}{%
  \setlength{\algindent}{0mm}
  \begin{enumerate}[\bf\small 1.]%
  \setlength{\itemsep}{3pt} 
}{%
  \end{enumerate}

}

\newcommand{\Step}[1]{\item\hspace*{\algindent}\parbox[t]{\textwidth-\algindent-\leftmargin}{#1}}

\newcommand{\IncreaseIndent}{\addtolength{\algindent}{7mm}}
\newcommand{\DecreaseIndent}{\addtolength{\algindent}{-7mm}}

\let\Definition=\emph
\newcommand{\B}{\mathcal{B}}
\newcommand{\C}{\mathcal{C}}
\newcommand{\F}{\mathcal{F}}
\newcommand{\G}{\mathcal{G}}
\newcommand{\MH}{\mathcal{H}}
\newcommand{\N}{\mathcal{N}}
\newcommand{\bB}{\mathbb{B}}
\newcommand{\bW}{\mathbb{W}}
\newcommand{\W}{\mathcal{W}}
\newcommand{\X}{\mathcal{X}}
\newcommand{\Y}{\mathcal{Y}}

\newcommand{\eg}{e.g.}
\newcommand{\ie}{i.e.}
\newcommand{\etal}{et al.}
\newcommand{\Etal}{et al}
\newcommand{\Tophi}{\ensuremath{\to_{\Phi}}}
\newcommand{\Topsi}{\ensuremath{\to_{\Psi}}}
\newcommand{\Togamma}{\ensuremath{\to_{\Gamma}}}
\renewcommand{\to}{\ensuremath{\longrightarrow}}
\newcommand{\nto}{\ensuremath{\longarrownot\to}}
\newcommand{\Ntophi}{\ensuremath{\nto_\Phi}}
\newcommand{\Ntopsi}{\ensuremath{\nto_\Psi}}
\newcommand{\Ntogamma}{\ensuremath{\nto_\Gamma}}
\newcommand{\Bip}[2]{\ensuremath{\langle#1,#2\rangle}}
\newcommand{\Cat}{\ensuremath{\bullet}}

\newcommand{\Section}{Section\xspace}
\newcommand{\Sections}{Sections\xspace}
\newcommand{\Figure}{Figure\xspace}

\newcommand{\Table}{Table\xspace}

\newcommand{\Theorem}{Theorem\xspace}
\newcommand{\Theorems}{Theorems\xspace}
\newcommand{\Lemma}{Lemma\xspace}
\newcommand{\Lemmas}{Lemmas\xspace}
\newcommand{\Corollary}{Corollary\xspace}

\title{Fully dynamic recognition of proper circular-arc graphs}
\author{%
  Francisco J.\ Soulignac~\thanks{\small mail: \tt fsoulign@dc.uba.ar}%
}

\date{CONICET and Departamento de Computaci\'on, FCEN, Universidad de Buenos Aires, 
Buenos Aires, Argentina.}

\begin{document}
\maketitle

\begin{abstract}
 We present a fully dynamic algorithm for the recognition of proper circular-arc (PCA) graphs.  The allowed operations on the graph involve the insertion and removal of vertices (together with its incident edges) or edges.  Edge operations cost $O(\log n)$ time, where $n$ is the number of vertices of the graph, while vertex operations cost $O(\log n + d)$ time, where $d$ is the degree of the modified vertex.  We also show incremental and decremental algorithms that work in $O(1)$ time per inserted or removed edge.  As part of our algorithm, fully dynamic connectivity and co-connectivity algorithms that work in $O(\log n)$ time per operation are obtained.  Also, an $O(\Delta)$ time algorithm for determining if a PCA representation corresponds to a co-bipartite graph is provided, where $\Delta$ is the maximum among the degrees of the vertices.  When the graph is co-bipartite, a co-bipartition of each of its co-components is obtained within the same amount of time.  

 \vspace*{3mm} {\bf Keywords: dynamic recognition, proper circular-arc graphs, round graphs, co-connectivity.} 
\end{abstract}

\section{Introduction}
\label{sec:introduction}

The \Definition{dynamic graph recognition and representation problem} for a class of graphs $\C$, or simply the \Definition{dynamic recognition problem} for $\C$, is the problem of maintaining a representation of a dynamically changing graph, while the graph belongs to $\C$.  Its input is a graph $G$ together with the sequence of operations that are to be applied on $G$.  A \Definition{dynamic recognition algorithm} is composed by the algorithm that builds the initial representation of $G$ and the algorithms that apply each update on the representation.  Other kinds of dynamic graph problems have been considered, besides the recognition and representation problems (see \eg~\cite{EppsteinGalilItaliano1999}).

Dynamic recognition problems are classified according to the effects that the operations have on the size of $G$.  A recognition problem that allows no updates is called \Definition{static}.  The input of a static problem is $G$ and the output is a representation of $G$ or an error, according to whether $G \in \C$.  A recognition problem whose updates only increment the size of $G$ is called \Definition{incremental}.  Similarly, a recognition problem that allows only updates that decrement the size of $G$ is called \Definition{decremental}.  Finally, a recognition problem that allows updates of both kinds is called \Definition{fully dynamic}.

Dynamic problems are also classified with respect to the structures that can be inserted or removed. A dynamic problem is \Definition{vertex-only} if only vertices can be inserted or removed, while it is \Definition{edge-only} if only edges can be inserted or removed (sometimes the insertion and/or removal of isolated vertices 
is also allowed in an edge-only problem).  There are problems in which other structures, such as cliques, are included or removed (\eg~\cite{Yrysgul2006}), but we do not deal with such problems in this article.

In the last decade, dynamic recognition algorithms for many classes of graphs have been developed, including, among others, chordal graphs, cographs, directed cographs, distance hereditary graphs, interval graphs, $P_4$-sparse graphs, permutation graphs, proper interval graphs, and split graphs~\cite{Crespelle2010,CrespellePaulDAM2006,CrespellePaulA2010,HeggernesManciniDAM2009,HellShamirSharanSJC2001,IbarraATA2008,Ibarra2009,IbarraA2010,NikolopoulosPaliosPapadopoulos2006,ShamirSharanDAM2004,TedderCorneil2007}.

In this paper we deal with the dynamic recognition problem for proper circular-arc graphs.  A \Definition{circular-arc model} is a family of arcs of some circle.  A graph is said to \Definition{admit} a circular-arc model when its vertices are in a one-to-one correspondence with the arcs of the model in such a way that two vertices of the graph are adjacent if and only if their corresponding arcs have nonempty intersection.  Those graphs that admit a circular-arc model are called \Definition{circular-arc graphs}.  Proper circular-arc graphs and proper interval graphs form two of the most studied subclasses of circular-arc graphs.  A circular-arc model is \Definition{proper} when none of its arcs is properly contained in some other arc of the model, while it is \Definition{interval} when its arcs do not cover the entire circle.  A graph is a \Definition{proper circular-arc (PCA) graph} when it admits a proper circular-arc model, while it is a \Definition{proper interval (PIG) graph} when it admits an interval proper circular-arc model.  

Circular-arc graphs and their subclasses have applications in disciplines as diverse as allocation problems, archeology, artificial intelligence, biology, computer networks, databases, economy, genetics, and traffic light scheduling, among others.  In particular, Hell \etal~\cite{HellShamirSharanSJC2001} describe an application of dynamic PIG graphs in physical mapping of DNA.  Lin and Szwarcfiter~\cite{LinSzwarcfiterDM2009} survey the static recognition problem for several subclasses of circular-arc graphs.  

The static recognition problems for both PIG and PCA graphs require $O(n+m)$ time~\cite{CorneilDAM2004,CorneilKimNatarajanOlariuSpragueIPL1995,DengHellHuangSJC1996,HellHuangSJDM2005,HerreraMeidanisPicininIPL1995}. Here and in the remainder of this section, $n$ and $m$ refer to the number of vertices and edges of the graph, respectively.  The first static recognition and representation algorithm for PCA graphs was given by Deng \etal~\cite{DengHellHuangSJC1996}.  As part of their algorithm, Deng \etal\ developed a vertex-only incremental algorithm for the recognition of connected PIG graphs that runs in $O(d)$ time per vertex insertion, where $d$ is the degree of the inserted vertex.  Later, Hell \etal~\cite{HellShamirSharanSJC2001} extended this algorithm into a fully dynamic algorithm for the recognition of PIG graphs that runs in $O(d + \log n)$ time per vertex update and in $O(\log n)$ time per edge update.  The algorithm by Hell \etal\ can be restricted to solve only the incremental and decremental problems in $O(d)$ time per vertex operation and $O(1)$ time per edge operation.  Even later, Ibarra~\cite{Ibarra2009} developed an edge-only fully dynamic algorithm for the recognition of PIG graphs that also runs in $O(\log n)$ time per edge modification.

In this article we develop the first fully dynamic, incremental, and decremental recognition algorithms for PCA graphs.  Our algorithms build upon the recognition algorithms of PIG graphs given by Hell \Etal.  The time complexity of our algorithms equals the time complexity required by the algorithms by Hell \Etal.  That is, we present: 
\begin{itemize}
 \item a fully dynamic algorithm that runs in $O(d + \log n)$ per vertex update and in $O(\log n)$ time per edge update,
 \item an incremental algorithm that runs in $O(d)$ time per vertex insertion and $O(1)$ time per edge insertion, and
 \item a decremental algorithm that runs in $O(d)$ time per vertex removal and $O(1)$ time per edge removal.
\end{itemize}
The representation maintained by the algorithm is, strictly speaking, not a proper circular-arc model of the graph. Neither the representation maintained by Hell \etal\ for the recognition of PIG graphs is a proper interval model.  Instead, combinatorial structures called straight representations ---for PIG graphs--- and round representation ---for PCA graphs--- are maintained (see \Section~\ref{subsec:round-graphs}).  It is worth to mention that straight and round representations are in a one-to-one correspondence with proper interval and proper circular-arc models, respectively.  Moreover, if required, straight and round representations can be transformed into proper interval and proper circular-arc models in $O(n)$ time.

The organization of the article is as follows.  In \Section~\ref{sec:preliminaries} we introduce the basic terminology and some required tools.  In particular, we define straight and round representations.  \Section~\ref{sec:data structure} briefly overviews the algorithms by Deng \etal\ and by Hell \etal\ for the recognition of PIG graphs.  These algorithms are important for us because of two reasons.  First, they are invoked by our algorithms when PIG graphs need to be recognized.  Second, these algorithms and ours share some fundamental ideas.  In particular, the basic implementation of round representations, that is common to all our algorithms, is a simple generalization of the implementation of straight representations that the algorithms by Hell \etal\ use.  Our implementation of round representations is given in \Section~\ref{sec:data structure}.  The basic algorithms that manipulate round representations are presented in \Section~\ref{sec:basic manipulation}.  These algorithms try to modify as little as possible the input round representation.  In that sense, they can be considered as generalizations of the algorithms by Hell \etal, even though some of algorithms in \Section~\ref{sec:basic manipulation} share no similarities with those in~\cite{HellShamirSharanSJC2001}.  \Section~\ref{sec:co-bipartite} is devoted to co-bipartite PCA graphs.  First we introduce an efficient algorithm for computing all the co-components of the input graph, and then we develop two methods that can be used to traverse all the round representations of a PCA graph.  As a corollary, we obtain a new proof for a theorem by Huang~\cite{Huang1992} that characterizes the structure of round representations.  \Sections \ref{sec:incremental}~and~\ref{sec:decremental} combine all the previous tools into the incremental and decremental algorithms, respectively, while \Section~\ref{sec:dyn-pca connectivity} integrates the incremental and decremental algorithms into a fully dynamic algorithm.  As part of the fully dynamic algorithm, simple connectivity and co-connectivity algorithms for fully dynamic PCA graphs are derived from the work in~\cite{HellShamirSharanSJC2001}.  Finally, some further remarks are given in \Section~\ref{sec:conclusions}.

\section{Preliminaries}
\label{sec:preliminaries}

For a graph $G$, we use $V(G)$ and $E(G)$ to denote the sets of vertices and edges of $G$, respectively, while we use $n$ and $m$ to denote $|V(G)|$ and $|E(G)|$, respectively.  We write $uv$ to represent the edge of $G$ between the pair of adjacent vertices $u$ and $v$. The \Definition{neighborhood} of $v$ is the set $N_G(v)$ of all the neighbors of $v$, and the \Definition{complement neighborhood} of $v$ is the set $\overline{N_G}(v)$ of all the non-neighbors of $v$.  For $V \subseteq V(G)$, we write \Definition{$\N_G(V) = \bigcup_{v \in V}N_G(v)$} and \Definition{$\overline{\N_G}(V) = \bigcup_{v \in V}\overline{N_G}(v)$}.  The cardinality of $N_G(v)$ is the \Definition{degree} of $v$ and is denoted by $d_G(v)$.  The maximum among the degrees of all the vertices is represented by $\Delta(G)$.  For $k \in \mathbb{N}$,  $G$ is said to be a \Definition{$k$-degree} graph when $\Delta(G) \leq k$.  The \Definition{closed neighborhood} of $v$ is the set $N_G[v] = N_G(v) \cup \{v\}$; if $N_G[v] = V(G)$, then $v$ is a \Definition{universal vertex}.  Two vertices $v$ and $w$ are \Definition{twins} when $N_G[v] = N_G[w]$.  We omit the subscripts from $N$, $\N$, and $d$ when there is no ambiguity about $G$.

The subgraph of $G$ \Definition{induced} by $V \subseteq V(G)$, denoted by $G[V]$, is the graph that has $V$ as vertex set and two vertices of $G[V]$ are adjacent if and only if they are adjacent in $G$.  A \Definition{clique} is a subset of pairwise adjacent vertices. We also use the term \Definition{clique} to refer to the corresponding subgraph.  An \Definition{independent set} is a set of pairwise non-adjacent vertices.  A \Definition{semiblock} of $G$ is a nonempty set of twin vertices, and a \Definition{block} of $G$ is a maximal semiblock. A \Definition{hole} is a chordless cycle with at least four vertices.

The \Definition{complement} of $G$, denoted by $\overline{G}$, is the graph that has the same vertices as $G$ and such that two vertices are adjacent in $\overline{G}$ if and only if they are not adjacent in $G$.  Graph $G$ is \Definition{co-connected} when $\overline{G}$ is connected, and each component of $\overline{G}$ is called a \Definition{co-component} of $G$.   The \Definition{union} of two vertex-disjoint graphs $G$ and $H$ is the graph $G \cup H$ with vertex set $V(G) \cup V(H)$ and edge set $E(G) \cup E(H)$.  The \Definition{join} of $G$ and $H$ is the graph $G + H = \overline{\overline{G} \cup \overline{H}}$, \ie, $G + H$ is obtained from $G \cup H$ by inserting all the edges $vw$, for $v \in V(G)$ and $w \in V(H)$. 

A graph $G$ is \Definition{bipartite} when there is a partition $V_1, V_2$ of $V(G)$ such that both $V_1$ and $V_2$ are independent sets.  Contrary to the usual definition of a partition, we allow one of the sets $V_1$ and $V_2$ to be empty.  So, the graph with one vertex is bipartite for us.  The partition of $V(G)$ into $V_1, V_2$, denoted by $\Bip{V_1}{V_2}$, is called a \Definition{bipartition} of $G$.  When $\overline{G}$ is bipartite, $G$ is a \Definition{co-bipartite} graph and each bipartition of $\overline{G}$ is a \Definition{co-bipartition} of $G$.  

A \Definition{semiblock family} is a family formed by pairwise disjoint nonempty sets of vertices.  A \Definition{semiblock graph} $\G$ is a graph whose vertex set is a semiblock family.  To avoid confusions, we refer to $V(\G)$ as the \Definition{semiblock family} of $\G$, and to its elements as \Definition{semiblocks}, instead of calling them the vertex set and vertices of $\G$, respectively.  We note, however, that the notation and terminology of graphs holds for semiblock graphs as well.  For instance, we call $N(B)$ to the family of semiblocks adjacent of $B$, we say that a semiblock is universal, we refer to a family of sets $\B$ as a clique, etc. 

A semiblock graph $\G$ with no twins is called a \Definition{block graph}.  For block graphs we also call $V(\G)$ a \Definition{block family} and refer to its elements as \Definition{blocks}.  The \Definition{extension} of a semiblock graph $\G$ is the graph $G$ with vertex set $\bigcup V(\G)$ such that $v \in B$ is adjacent to $w \in W$ if and only if $B \in N[W]$, for each $B,W \in V(\G)$.  In other words, each set $B$ is transformed into a semiblock, and the edges between the semiblocks are preserved to its vertices.  Observe that each semiblock of $\G$ is also a semiblock of $G$.  Furthermore, $\G$ is a block graph if and only if each block of $\G$ is a block of $G$.  Each semiblock graph $\G$ whose extension is isomorphic to $G$ is called a \Definition{reduction} of $G$.  If $\G$ is a block graph, then $\G$ is the \Definition{block reduction} of $G$.

\subsection{Orderings and ranges}

An \Definition{ordering} is a finite set $S$ that is associated with an enumeration $x_1, \ldots, x_n$ of its elements. Elements $x_1$ and $x_n$ are the \Definition{leftmost} and \Definition{rightmost} elements of $S$, respectively.  The \Definition{reverse} of $S$, denoted by $S^{-1}$, is the ordering $x_n, \ldots, x_1$.  If $T = y_1, \ldots, y_m$ is an ordering, we denote by $S \Cat T$ the ordering $x_1, \ldots, x_n, y_1, \ldots, y_m$.  We also consider each single element $y$ an ordering, thus $S \Cat y$ is the ordering $x_1, \ldots, x_n, y$, and $y \Cat S$ is the ordering $y, x_1, \ldots, x_n$.

In this article we deal with many collections that are of a circular (cyclic) nature, such as circular lists, circular families, etc.  Generally, the objects in a collection are labeled with some kind of index that identifies the position of the object inside the collection.  Unless otherwise stated, we assume that all the operations on these indices are taken modulo the length of the collection.  Furthermore, we may refer to negative indices and to indices greater than the length of the collection.  In these cases, indices should also be understood modulo the length of the collection.  For instance, the element $x_{kn+i}$ of the ordering $x_1, \ldots, x_n$ is $x_i$, for any $1 \leq i \leq n$ and $k \in \mathbb{Z}$.

The above assumption allows us to work with orderings as if they were circular orderings.  We use the standard interval notation applied to orderings, though we call them ranges to avoid confusions with interval graphs.  Let $S = x_1, \ldots, x_n$ be an ordering.  For $x_i, x_j \in X$, the \Definition{range} $[x_i, x_j]$ is defined as the ordering $x_i, x_{i+1}, \ldots, x_{j-1}, x_j$ where, as said before, all the operations are calculated modulo $n$.  Notice that $x_i$ and $x_j$ are the leftmost and rightmost of $[x_i, x_j]$, respectively.  Similarly, the range $[x_i, x_j)$ is obtained by removing the last element from $[x_i, x_j]$, the range $(x_i, x_j]$ is obtained by removing the first element from $[x_i, x_j]$, and $(x_i, x_j)$ is obtained by removing both the first and last elements from $[x_i, x_j]$.  

The range notation that we use clashes with the usual notation for ordered pairs.  Thus, we write $\Bip{x}{y}$ to denote the ordered pair $(x, y)$.  The unordered pair formed by $x$ and $y$ is, as usual, denoted by $\{x,y\}$.  Also, for the sake of notation, we sometimes write $\#S$ to denote the cardinality of a range $S$.  

\subsection{Round graphs}
\label{subsec:round-graphs}

\begin{figure}
  \centering
  $\Phi = \Bip{\{B_1, \ldots, B_5\}}{\{B_1 \stackrel{F_r}{\to} B_2, B_2 \stackrel{F_r}{\to} B_4, B_3 \stackrel{F_r}{\to} B_4, B_4 \stackrel{F_r}{\to} B_5, B_5 \stackrel{F_r}{\to} B_5\}}$
  
  ~

  \includegraphics{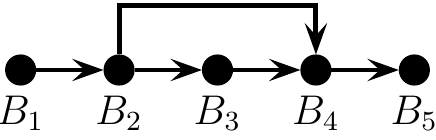}
  \caption{A round representation $\Phi$ and its associated $\Tophi$ relation.}\label{fig:round representation}
\end{figure}

A \Definition{round representation} is a pair $\Phi = \Bip{\B(\Phi)}{F_r^\Phi}$ where $\B(\Phi) = B_1, \ldots, B_k$ is an ordered semiblock family, and $F_r^\Phi$ is a mapping from $\B(\Phi)$ to $\B(\Phi)$ such that $F_r^\Phi(B_i) \in [B_i, F_r^\Phi(B_{i+1})]$, for every $B_i \in \B(\Phi)$.  For each $B \in \B(\Phi)$, the semiblock $F_r^\Phi(B)$ is called the \Definition{right far neighbor} of $B$.  We use a convenient notation for dealing with the range $(B, F_r^\Phi(B)]$.  For $B, W \in \B(\Phi)$, we write $B \Tophi W$ to mean that $W \in (B, F_r^\Phi(B)]$.  Similarly, write $B \Ntophi W$ to indicate that $W \not \in (B, F_r^\Phi(B)]$.  As usual, we do not write the subscript and superscript $\Phi$ when $\Phi$ is clear by context.  \Figure~\ref{fig:round representation} depicts a round representation and its corresponding $\to$ relation.

Every round representation $\Phi$ is associated with several mappings that are useful for the dynamic algorithms.  Let $\B(\Phi) = B_1, \ldots, B_n$.  For $B_i \in \B(\Phi)$, define:
\begin{itemize}
  \item the \Definition{right semiblock} of $B_i$, denoted by $R^\Phi(B_i)$, as $B_{i+1}$,
  \item the \Definition{left semiblock} of $B_i$, denoted by $L^\Phi(B_i)$, as $B_{i-1}$,
  \item the \Definition{left far neighbor} of $B_i$, denoted by $F_l^\Phi(B_i)$, as the unique $B_j \in \B(\Phi)$ such that (a) $B_j \to B_i$ or $B_j = B_i$ and (b) $B_{j-1} = B_i$ or $B_{j-1} \nto B$.
  \item the \Definition{right near neighbor} of $B_i$, denoted by $N_r^\Phi(B_i)$, as $B_{i+1}$ if $B \to B_{i+1}$, and as $B_i$ otherwise.
  \item the \Definition{left near neighbor} of $B_i$, denoted by $N_l^\Phi(B_i)$, as $B_{i-1}$ if $B_{i-1} \to B_i$, and as $B_i$ otherwise.
  \item the \Definition{right unreached semiblock} of $B_i$, denoted by $U_r^\Phi(B_i)$, as $R^\Phi(F_r^\Phi(B_i))$, and
  \item the \Definition{left unreached semiblock} of $B_i$, denoted by $U_l^\Phi(B_i)$, as $L^\Phi(F_l^\Phi(B_i))$.
\end{itemize}
As usual, we omit the superscript $\Phi$ when $\Phi$ is clear from the context.

The following observation shows equivalent definitions of round representations.

\begin{observation}
  The following statements are equivalent for $\Phi = \Bip{\B(\Phi)}{F_r}$.
  \begin{itemize}
    \item $\Phi$ is a round representation.
    \item For every $B_l, B_m, B_r \in \B(\Phi)$, if $B_m \in (B_l, B_r)$ and $B_l  \to B_r$, then $B_m \to B_r$.
    \item For every $B \in \B(\Phi)$, $B = F_r(N_l(B))$ or $B \to F_r(N_l(B))$.
  \end{itemize}
\end{observation}

Through this article, we deal with two types of round representations of interest.  A \Definition{normal round representation} is a round representation $\Phi$ such that $B \in [F_l(B), F_r(B)]$, for every $B \in \B(\Phi)$.  In other words, $\Phi$ is normal if either $B \nto W$ or $W \nto B$, for every pair $B, W \in \B(\Phi)$.  For the sake of simplicity, from now on, whenever we write that $\Phi$ is a round representation, we mean that $\Phi$ is a normal round representation.  A \Definition{straight representation} is a round representation $\Phi$ such that $F_r(B) = B$, for some $B \in \B(\Phi)$.  

\begin{figure}
 
  \hspace*{\stretch{1}}
  \includegraphics{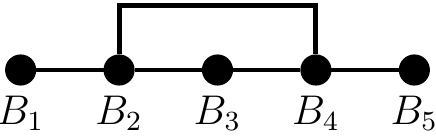} 
  \hspace*{\stretch{1}} 
  \includegraphics{fig-straight-graph-directed} 
  \hspace*{\stretch{1}}

  \caption{A round graph $\G$ and the relation $\Tophi$ for some round representation $\Phi$ of $\G$. }\label{fig:round graph}
\end{figure}

A semiblock graph $\G$ is a \Definition{round graph} if there exists a round representation $\Phi$ such that $\B(\Phi)$ is an ordering of $V(\G)$ and $N[B] = [F_l(B), F_r(B)]$, for every $B \in V(\G)$.  For the round graph $\G$, we say that $\G$ \Definition{admits} the round representation $\Phi$, and that $\Phi$ \Definition{represents} $\G$.  A round graph that admits a straight representation is also called a \Definition{straight graph}.  \Figure~\ref{fig:round graph} shows a round graph together with the $\to$ relation associated to some of its round representations.  

A round graph may admit several round representations.  On the other hand, each round representation represents exactly one round graph.  Indeed, the round graph $\G$ represented by $\Phi$ has $\B(\Phi)$ as its semiblock family, while $B$ and $W$ are adjacent if and only if $B \Tophi W$ or $W \Tophi B$.  We write $\G(\Phi)$ to denote the unique round graph represented by $\Phi$.  

The concept of induced representation plays a central role in the dynamic algorithms, so it is better to define it in a constructive manner. Let $\B = [B_l, B_r]$ be a range of $\B(\Phi)$.  The \Definition{restriction} of $F_r$ to $\B$, denoted by $F_r|\B$, is the mapping $F$ from $\B$ to $\B$ such that $F(B) = F_r(B)$ if $F_r(B) \in \B$, while $F(B) = B_r$ otherwise.  The representation of $\Phi$ \Definition{induced by} $\B$, denoted by $\Phi|\B$, is the pair $(\B, F_r|\B)$.  In other words, $\Phi|\B$ is obtained from $\Phi$ by keeping only those blocks inside $\B$, and then adapting $F_r^\Phi$.  Observe that any $\B \subseteq \B(\Phi)$ can be described with a sequence of ranges $\B_1, \ldots, \B_k$ of $\B(\Phi)$ such that $\B_{i+1} \subseteq B_{i}$ and $\B_k = \B$.  Thus, the concept of an induced representation is generalized to $\B$ as $\Phi|\B = (\ldots(\Phi|\B_1)|\ldots)|\B_k$.  Also, we write $F_r \setminus \B = F_r|(\B(\Phi) \setminus \B)$ and $\Phi \setminus \B = \Phi|(\B(\Phi) \setminus \B)$.  That is, $F_r \setminus \B$ and $\Phi \setminus \B$ are obtained from $F_r$ and $\Phi$ by removing $\B$, respectively.  

\begin{observation}
 For each $\B \subseteq \B(\Phi)$, $\G(\Phi|\B) = \G(\Phi)[\B]$ and $\G(\Phi \setminus \B) = \G(\Phi) \setminus \B$.
\end{observation}

Hell \etal~\cite{HellShamirSharanSJC2001} introduce the concept of a contig (round representations are related to DNA sequences) to deal with the straight representations of each component.  We slightly change the meaning of a contig to fit better for our purposes.  Let $\Phi$ be a round representation of a round graph $\G$, and $\B$ be a range of $\B(\Phi)$.  Say that $\B$ is a \Definition{contig range} when $\G[\B]$ is a component of $\G$.  In such case, $\Phi|\B$ is a \Definition{contig} of $\Phi$ \Definition{representing} $\G[\B]$.  We also refer to $\Phi$ as a \Definition{contig} to indicate that $\G$ is connected, and as a \Definition{block contig} to indicate that $\G$ is also a  block graph.  The following is a well know property of round representations.

\begin{observation}
  Every component of $\G$ is represented by a contig.
\end{observation}

We classify contigs into \Definition{linear contigs} and \Definition{circular contigs} according to whether the contigs are straight or not, respectively.  Each linear contig has two special semiblocks: the \Definition{left end semiblock} is the semiblock $B$ such that $F_l(B) = B$, and the \Definition{right end semiblock} is the semiblock $B$ such that $F_r(B) = B$. 

Two semiblocks $B, W$ of a round representation $\Phi$ are \Definition{indistinguishable} when $F_l(B) = F_l(W)$ and $F_r(B) = F_r(W)$.  Clearly, if $B \to W$, then all the semiblocks in $[B, W]$ are pairwise indistinguishable in $\Phi$.  We say that $\Phi$ is \Definition{compressed} when it contains no pair of indistinguishable semiblocks.  The \Definition{compression} of $\Phi$ is the round enumeration that is obtained by iteratively moving the elements of $W$ to $B$, and then removing $W$, for some pair of indistinguishable semiblocks $B$ and $W$, until $\Phi$ is compressed.  It is not hard to see that $B$ and $W$ are twins in $\G(\Phi)$ when they are indistinguishable in $\Phi$.  The converse is not true, but almost.  The following lemmas resume the situation.

\begin{lemma}[\eg~\cite{HuangJCTSB1995}]\label{lem:indistinguishable pig}
  Two semiblocks of a straight representation\/ $\Phi$ are twins in $\G(\Phi)$ if and only if they are indistinguishable in\/ $\Phi$.
\end{lemma}

\begin{lemma}[\eg~\cite{LinSoulignacSzwarcfiter2011}]\label{lem:indistinguishable}
  Two semiblocks of a round representation\/ $\Phi$ are twins in $\G(\Phi)$ if and only if they are both universal in $\G(\Phi)$ or indistinguishable in\/ $\Phi$.
\end{lemma}

These lemmas show an important property of round graphs.  If at most one universal semiblock is admitted, then twin semiblocks can be identified as indistinguishable semiblocks.  For any $u \in \mathbb{N}_0$, say that a semiblock graph is \Definition{$u$-universal} when it contains at most $u$ universal semiblocks.  Similarly, say that $\Phi$ is a \Definition{$u$-universal} round representation when $\G(\Phi)$ is $u$-universal. The following is a simple corollary of \Lemma~\ref{lem:indistinguishable}.

\begin{corollary}\label{cor:indistinguishable}
  Let\/ $\Phi$ be a round representation.  Then, $\G(\Phi)$ is a block graph if and only if\/ $\Phi$ is compressed and\/ $1$-universal.
\end{corollary}

Say that two round representations are \Definition{equal} when one can be obtained from the other by permuting indistinguishable semiblocks.  In other words, two round representations are equal when their compressions are equal.  By definition, if $\Phi$ and $\Psi$ are equal round representations, then $\G(\Phi)$ and $\G(\Psi)$ are isomorphic.   

Notice that if $\Phi$ is a round representation, then $\Psi = \Bip{\B(\Phi)^{-1}}{F_l^\Phi}$ is also a round representation of $\G(\Phi)$.  Furthermore, $F_l^\Psi = F_r^\Phi$, $L^\Psi = R^\Phi$, $R^\Psi = L^\Phi$, $N_l^\Psi = N_l^\Psi$, $N_r^\Psi = N_l^\Phi$, $U_r^\Psi = U_l^\Phi$ and $U_l^\Psi = U_r^\Phi$.  The representation $\Psi$ is the \Definition{reverse} of $\Phi$, and we denote it by $\Phi^{-1}$.  The following theorems show that many round graphs admit only two non-equal round representations.  

\begin{theorem}[\cite{Roberts1969}]\label{thm:unique-PIG-models}
  Connected straight graphs admit at most two straight representations, one the reverse of the other.
\end{theorem}
 
\begin{theorem}[\cite{HuangJCTSB1995}]\label{thm:unique-PCA-models}
  Connected and co-connected round graph admit at most two round representations, one the reverse of the other.
\end{theorem}

\subsection{Proper circular-arc graphs}

For the sake of simplicity, in this paper we use an alternative definition of proper circular-arc and proper interval graphs. These definitions follow from~\cite{DengHellHuangSJC1996,HuangJCTSB1995}.  

For each round representation $\Phi$, write \Definition{$G(\Phi)$} to denote the extension of $\G(\Phi)$.  A graph is a \Definition{proper circular-arc (PCA)} graph if it is isomorphic to $G(\Phi)$, for some round representation $\Phi$.  Clearly, all the reductions of a PCA graph are round graphs.  As for round graphs, $G(\Phi)$ is said to \Definition{admit} $\Phi$, while $\Phi$ \Definition{represents} $G(\Phi)$.  When $\G(\Phi)$ is a block graph, we also refer to $\Phi$ as a \Definition{round block representation} of $G(\Phi)$.  

PCA graphs are characterized by a family of minimal forbidden induced subgraphs, as in \Theorem~\ref{thm:forbiddens PCA}.  There, $H^*$ denotes the graph that is obtained from $H$ by inserting an isolated vertex.  Graph $\overline{C_3^*}$ is also denoted by \Definition{$K_{1,3}$}.

\begin{figure}
\centering
 \begin{tabular}{c@{\hspace{8mm}}c@{\hspace{8mm}}c@{\hspace{8mm}}c@{\hspace{8mm}}c}
  \includegraphics{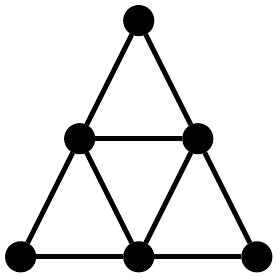} & \includegraphics[angle=90]{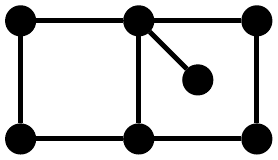} & \includegraphics[angle=90]{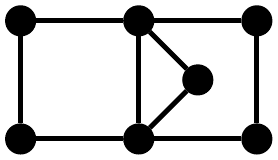} &  \includegraphics[angle=90]{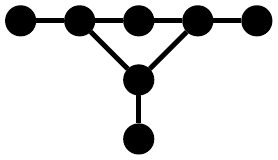} & \includegraphics[angle=90]{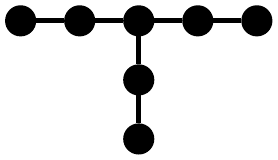} \\
  $S_3$ & $H_2$ & $H_3$ & $H_4$ & $H_5$
 \end{tabular}
 \caption{Complements of the forbidden induced subgraphs for PCA graphs}\label{fig:forbiddens PCA}
\end{figure}

\begin{theorem}[\cite{TuckerDM1974}]\label{thm:forbiddens PCA}
  A graph is a PCA graph if and only if it does not contain as induced subgraphs any of the following graphs: $C_n^*$ for $n \geq 4$, $\overline{C_{2n}}$ for $n \geq 3$, $\overline{C^*_{2n+1}}$ for $n \geq 1$, and the graphs $\overline{S_3}$, $\overline{H_2}$, $\overline{H_3}$, $\overline{H_4}$, $\overline{H_5}$ and $S_3^*$ (see \Figure~\ref{fig:forbiddens PCA}).
\end{theorem}

Proper interval graphs are defined as PCA graphs, by replacing round representations with straight representations.  That is, a graph is a \Definition{proper interval graph (PIG)} graph when it is isomorphic to $G(\Phi)$ for some straight representation $\Phi$.  PIG graphs are also characterized by minimal forbidden induced subgraphs.

\begin{theorem}[\cite{LekkerkerkerBolandFM1962/1963}]\label{thm:forbiddens PIG}
  A PCA graph is a PIG graph if and only if it does not contain $C_k$ for $k \geq 4$, and $S_3$ as induced subgraphs.
\end{theorem}

\section{The data structure}
\label{sec:data structure}

In this section we describe the base data structure used by the dynamic algorithms for the recognition of PCA graphs.  Before presenting the data structure for PCA graphs, we give a brief overview of the data structures used by Deng \etal\ and Hell \etal\ for the recognition of PIG graphs.  This overview is important because of two reasons.  First, it describes some of the design issues of these algorithms and how are they solved.  Second, our dynamic data structures are based on those by Hell \etal, which are in turn based on the data structure by Deng \Etal.

\subsection{The DHH and HSS algorithms: an overview}
\label{sec:DHH and HSS overview}

In~\cite{DengHellHuangSJC1996}, Deng \etal\ developed an incremental algorithm, from now on called the \Definition{DHH algorithm}, for the recognition of connected PIG graphs.  The dynamic representation maintained by the algorithm is a linear block contig $\Phi$ representing the input graph $G$.  When a new vertex $v$ is inserted into $G$, there are two possibilities.  If $v$ has some twin in some block of $\B(\Phi)$, then $v$ is inserted into this block and the algorithm halts.  Otherwise, a new block has to be created for $v$ and a new linear block contig $\Psi$ representing $G \cup \{v\}$ has to be generated.  Recall that $\Gamma = \Psi \setminus \{v\}$ is a linear contig representing $G$.  Observe that, since $\B(\Psi)$ contains only blocks of $G \cup \{v\}$, every semiblock of $\Gamma$ is equal to either $B \cap N(v)$ or $B \setminus N(v)$, for some $B \in \B(\Phi)$.  So, each block of $\Phi$ is either a block of $\Gamma$, or the union of two semiblocks of $\Gamma$.  By \Lemma~\ref{lem:indistinguishable pig} and \Theorem~\ref{thm:unique-PIG-models}, $\Gamma$ is rather similar to $\Phi$ in the sense that $\Gamma$ is obtained from $\Phi$ just by splitting some blocks into consecutive indistinguishable semiblocks.  Then, knowing that $\Psi$ is a block contig representing $G \cup \{v\}$, we obtain that $v$ simultaneously has neighbors and non-neighbors in at most two blocks of $\Phi$, and that these blocks are of the form $B \cup L^\Psi(B)$ and $W \cup R^\Psi(W)$.  Even more, $v$ has to be adjacent to all the vertices in the blocks inside $(B, W)$.  So, \[\B(\Phi) = (R^\Psi(W), L^\Psi(B)) \Cat B \cup L^\Psi(B) \Cat (B, W) \Cat W \cup R^\Psi(W).\]  Of course, there are other cases in which $v$ has no neighbors in $L^\Psi(B)$ or $R^\Psi(W)$.  The DHH algorithm finds the blocks $B \cup L^\Psi(B)$ and $W \cup R^\Psi(W)$ of $\Phi$ and the position where $\{v\}$ is to be inserted in $\Phi$, and it inserts $\{v\}$ by updating $F_r^\Phi$ into $F_r^\Psi$.  

\begin{figure}[htb!]
 \centering
  \includegraphics{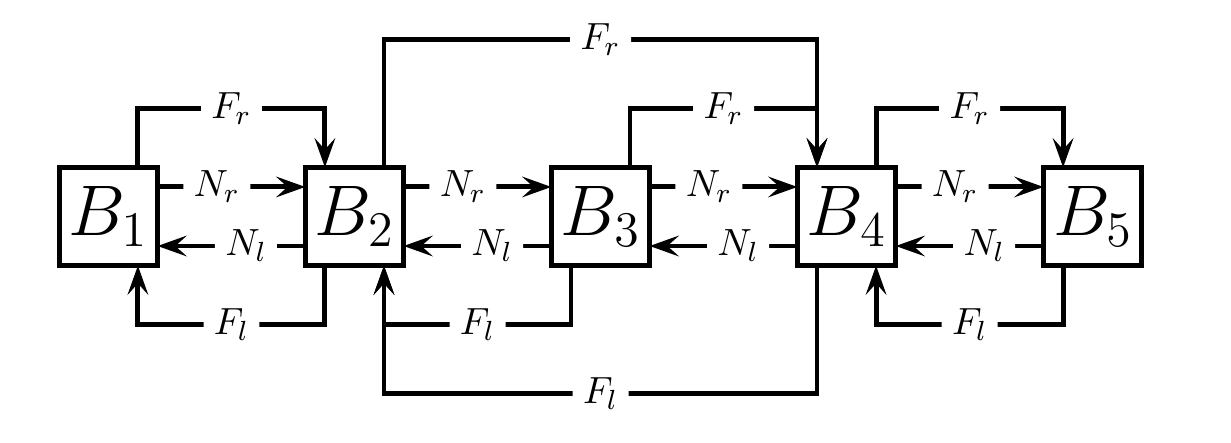}
 
 \caption{Data structure implementing the block contig depicted in \Figure~\ref{fig:round representation}.}\label{fig:straight-data-structure-dhh}
\end{figure}

The implementation of the linear contigs $\Phi$ used in this algorithm is simple (see \Figure~\ref{fig:straight-data-structure-dhh}). There is doubly-linked list of blocks representing $\B(\Phi)$, where each $B \in \B(\Phi)$ has two \Definition{near pointers} $N_l(B)$ and $N_r(B)$ and two \Definition{far pointers} $F_l(B)$ and $F_r(B)$.  These pointers encode the mappings $N_l, N_r, F_l$ and $F_r$, respectively; the overloaded notation is intentional.  Also, every vertex has a pointer to its block.  When $\{v\}$ is inserted as a new block into $\Phi$, the blocks $B \cup L^\Psi(B)$ and $W \cup R^\Psi(W)$ in the above paragraph have to be updated, as well as the far pointers of all the resulting blocks inside $[B, W]$.  All these operations are done in $O(d(v))$ time, \ie, $O(1)$ time per edge insertion, which is optimal.

The DHH algorithm was extended by Hell \etal~\cite{HellShamirSharanSJC2001} to handle the case in which the input graph is not connected.  In this case, $G$ admits an exponential number of straight block representations which can be constructed by permuting and reversing the block contigs of its components.  To handle this situation, the \Definition{vertex-only incremental HSS} algorithm keeps both linear block contigs representing each component, as implied by \Theorem~\ref{thm:unique-PIG-models}; recall these contigs are one the reverse of the other.  When a new vertex $v$ is inserted, there are two possibilities.  Either $N(v)$ is included in one component $G_1$ of $G$, or $N(v)$ intersects exactly two components $G_1$ and $G_2$ of $G$.  In the former case, $v$ is inserted into the contigs representing $G_1$ as in the DHH algorithm.  In the latter case, $G_1$ and $G_2$ have to be combined into a new component, and the block contigs representing $G_1$ and $G_2$ have to be replaced with the two linear block contigs representing $G_1 \cup G_2 \cup \{v\}$.  Let $\Psi$ be a linear block contig representing $G_1 \cup G_2 \cup \{v\}$, and $B$ and $W$ be the left and right end blocks in $\Psi$, respectively.  Again, we know that $\Gamma = \Psi \setminus \{v\}$ is a linear contig representing of $G_1 \cup G_2$.  Even more, $F_r^\Gamma$ maps semiblocks in $[B,\{v\})$ to semiblocks in $[B, \{v\})$, and semiblocks in $(\{v\}, W]$ to semiblocks in $(\{v\}, W]$.  Thus, $\Gamma|[B, \{v\})$ and $\Gamma|(\{v\}, W]$ represent one of the components each.  Also, $v$ has neighbors and non-neighbors in at most one block $B_l$ of $G_1$, and in at most one block $W_r$ of $G_2$.  

\begin{figure}[htb!]
 \centering
 \includegraphics{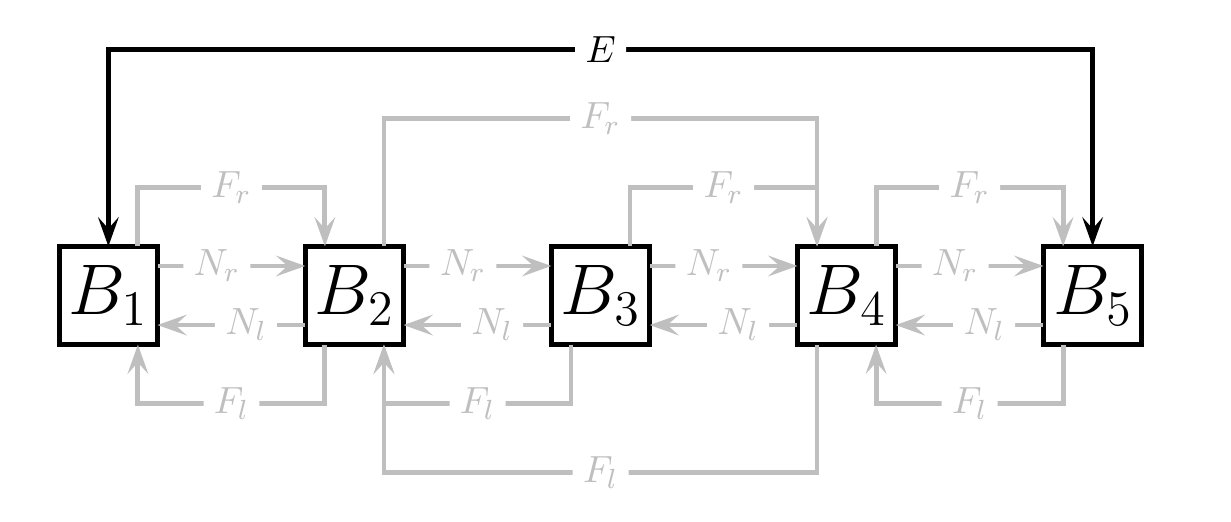}  
 \caption{Data structure with end pointers for the block contig of \Figure~\ref{fig:round representation}.}\label{fig:straight-data-structure-ep}
\end{figure}

A method similar to the DHH algorithm is enough to insert the new block for $v$ once $G_1$, $G_2$, $B_l$, and $W_r$ are known.  However, it is not easy to find $G_1$ and $G_2$ if $\Phi$ is implemented as in the DHH algorithm.  To find $G_1$ and $G_2$, the simplest way is to first locate the ranges of blocks with neighbors of $v$.  For this purpose, $N(v)$ is first traversed and the blocks with neighbors of $v$ are marked.  Then, the contigs are traversed to the right and to the left, starting from a marked block $B$.  The traversal stops either when a block not marked is found or when all the blocks in the contig have been traversed.  The family of traversed blocks form a range of blocks, all of which have neighbors of $v$.  In case that two maximal ranges are found, then $G \cup \{v\}$ is a PIG graph only if these ranges fall in different contigs, and each of these ranges contains at least one of the end blocks.  To test if two ranges, both containing at least one end block, belong to the same contig, an \Definition{end pointer} $E^{\Phi}(B)$ is stored for each block $B \in \B(\Phi)$.  If $B$ is not an end block, then $E^{\Phi}$ points to NULL; otherwise it points to the other end block of its contig (see \Figure~\ref{fig:straight-data-structure-ep}).  With this new data structure, the HSS algorithm handles the insertion of a vertex in $O(d(v))$ time.

The vertex-only incremental HSS algorithm can be adapted to allow the insertion of edges as well.  Suppose some edge $vw$ is to be inserted into $G$.  We consider here only the case in which $G$ is connected.  Let $\Phi$ be a linear block contig representing $G$ and suppose $v \in B$ and $w \in W$, for $B, W \in \B(\Phi)$.  In $\G(\Phi)$, the block $B$ is adjacent to all the blocks in $(B, F_r(B)]$, while the block $W$ is adjacent to all the blocks in $[F_l(W), W)$.  For $G \cup \{vw\}$ to be a PIG graph, $F_r(B)$ must be equal to $L(W)$ and $F_l(W)$ must be equal to $R(B)$, or vice versa.  We have at least two possibilities for the insertion of the edge.  Either $v$ becomes a member of $R(B)$ or $v$ gets separated from $B$ to form a new block $\{v\}$ that lies between $B$ and $R(B)$.  In the latter case, the far pointers of all those blocks referencing $B$ have to be updated so as to reference $\{v\}$.

\begin{figure}[htb!]
 \centering
 \includegraphics{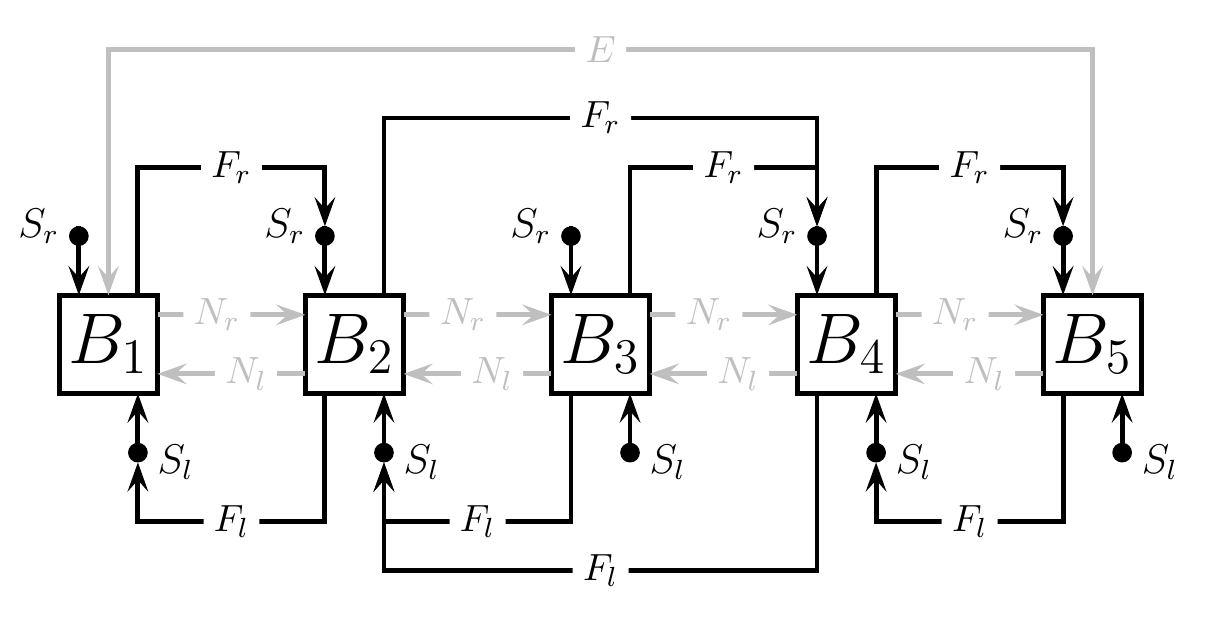}  
 \caption{Data structure with self pointers for the block contig of \Figure~\ref{fig:round representation}.  Now, every far pointer references a self pointer.}\label{fig:straight-data-structure-sp}
\end{figure}

To update these far pointers to reference the new block $\{v\}$ in $O(1)$ time, the HSS algorithm uses the technique of \Definition{nested pointers}.  For each block $B$, two \Definition{self pointers} $S^{\Phi}_l(B)$ and $S^{\Phi}_r(B)$ that point to $B$ are stored.  Every far pointer that was previously referencing $B$ now references $S^{\Phi}_l(B)$.  Similarly, every far pointer previously referencing $B$ now references $S^{\Phi}_r(B)$ (see \Figure~\ref{fig:straight-data-structure-sp}).  To move all the right far pointers referencing $B$ so as to reference $\{v\}$, we only need to exchange the value of $S^{\Phi}_r(B)$ so as to point to $S^\Psi_r(\{v\})$.

Up to this point we have discussed the incremental algorithms for the recognition of PIG graphs.  The decremental algorithms for the removal of vertices and edges are similar to the incremental ones.  However, end pointers have to be removed from the data structures that implement contigs.  This is because when two components result from the removal of a vertex or an edge, the new end pointers cannot be computed efficiently.  On the other hand, without the end pointers, a vertex $v$ can be removed in $O(d(v))$ time, while an edge $vw$ can be removed in $O(1)$ time.

Finally, Hell \etal\ developed a fully dynamic recognition algorithm in where insertions and removals of vertices and edges are unrestricted.  The algorithm is simply the combination of the incremental and decremental algorithms that we described above.  However, there is an incompatibility with respect to the use of the end pointers.  They are needed by the incremental algorithm to test whether two blocks belong to the same contig, while they are harmful for the decremental algorithm.  To solve this problem, Hell \etal\ propose a \Definition{dynamic connectivity structure}, supporting an operation to test if two blocks belong to the same contig, that can be queried and updated in $O(\log n)$ time per operation on the PIG graph.

\Table~\ref{table:HSS} summarizes the time complexities of the HSS algorithms.  Each column of the table indicates the data structure that is implemented by the dynamic algorithm.  No connectivity means that there is no way to test if two blocks belong to the same contig.  End pointers indicates that there is one end pointer for each block of the contig.  Finally, connectivity structure means that there is a dynamic data structure to test if any two blocks belong to the same contig or not.

\begin{table}\label{table:HSS}
\centering
\begin{tabular}{|l|c|c|c|}
 \hline
 Operation        & No connectivity & End pointers  & Connectivity structure \\ \hline
 Vertex insertion & not allowed     & $O(d(v))$     & $O(d(v) + \log n)$   \\ 
 Edge insertion   & not allowed     & $O(1)$        & $O(\log n)$          \\
 Vertex removal   & $O(d(v))$       & not allowed   & $O(d(v) + \log n)$   \\ 
 Edge removal     & $O(1)$          & not allowed   & $O(\log n)$          \\ \hline
\end{tabular}
\caption{Time complexities of the HSS algorithms.}
\end{table}

\subsection{The base data structure}
\label{sec:dyn-pca data structure}

In the previous section we saw that three different data structures are used by the HSS algorithms.  There is one with end pointers for the incremental algorithm, one with no support for connectivity queries for the decremental algorithm, and one with a connectivity structure for the fully dynamic algorithm.  We will extend these data structures for our algorithms, so as to implement general contigs instead of linear contigs.  In this section, however, we describe only the \Definition{base round representation}, which is common to all the algorithms in this article.

The implementation of each contig $\Phi$ is almost the same as the one used by the HSS algorithm.  The main difference is that near pointers now may represent a circular list instead of a linear list.  That is, the following data is stored to implement $\Phi$ for each semiblock $B \in \Phi$:
\begin{enumerate}
 \item The vertices that compose $B$.
 \item Left and right \Definition{near pointers}, $N_l^{\Phi}(B)$ and $N_r^{\Phi}(B)$, referencing the left and right near neighbors of $B$, respectively.
 \item Left and right \Definition{self pointers}, $S_l^{\Phi}(B)$ and $S_r^{\Phi}(B)$, pointing to $B$.
 \item Left and right \Definition{far pointers}, $F_l^{\Phi}(B)$ and $F_r^{\Phi}(B)$, referencing the left and right far neighbors of $B$, respectively.
\end{enumerate}
As usual, we omit the superscript $\Phi$ when no confusions arise.  The overloaded notation for $N_l$, $N_r$, $F_l$ and $F_r$ as both pointers and mappings is intentional.  So, depending on the context, we may write, for instance, $F_r(B)$ to mean both a block or a self pointer.  Recall that $N_l(B) = F_l(B) = B$ whenever $B$ is the left end semiblock and $N_r(B) = F_r(B) = B$ whenever $B$ is the right end semiblock.  Notice that $\Phi$ is linear if and only if the linked list described by its near pointers is actually a linear list.  Thus, it is trivial to query whether $\Phi$ is linear or not, and such a query takes $O(1)$ time.  We refer to $\Phi$ as a \Definition{base contig} to emphasize that $\Phi$ is a contig implemented with the above data.

Every round representation $\Phi$ is implemented as a family of base contigs.  The order between the contigs is not important for the recognition algorithm.   Thus, $\Phi^{-1}$ is just implemented as the family $\{\Gamma^{-1} \mid \Gamma \text{ is a contig of } \Phi\}$.  We refer to $\Phi$ as a \Definition{base round representation} to emphasize that $\Phi$ is implemented in this way.  Say that $\Phi$ satisfies the \emph{straightness property} when either $\Phi$ is straight or $\G(\Phi)$ is not straight.  Clearly, $\Phi$ satisfies the straightness property if and only if all its contigs satisfy the straightness property as well.

Following the ideas by Deng \etal\ and Hell \etal, two round block representations $\Phi, \Phi^{-1}$ satisfying the straightness property are stored to implement a dynamic PCA graph $G$.  The reason behind the straightness property is that the HSS algorithms can be applied on $\Phi$ and $\Phi^{-1}$ whenever $G$ is a PIG graph.  Furthermore, as in the HSS algorithms, the implementation of base contigs is specialized differently for the incremental, decremental, and fully dynamic problems.  For the incremental problem, each base contig is augmented with end pointers and other data (see \Section~\ref{sec:incremental}).  Similarly, for the decremental algorithms each base contig is extended with some useful information about co-contigs (see \Section~\ref{sec:decremental}).  Finally, for the fully dynamic algorithm the implementation of $G$ is extended with a data structure that solves some connectivity problems (see \Section~\ref{sec:dyn-pca connectivity}).  \Table~\ref{table:non-straight} is a preview of the time complexities of the algorithms, according to which implementation is used.

\begin{table}\label{table:non-straight}
\centering
 \begin{tabular}{|l|c|c|c|}
  \hline
  Operation        & Decremental DS  & Incremental DS & Fully-dynamic DS        \\ \hline
  Vertex insertion & not allowed     & $O(d(v))$      & $O(d(v) + \log n)$      \\
  Edge insertion   & not allowed     & $O(1)$         & $O(\log n)$             \\ 
  Vertex removal   & $O(d(v))$       & not allowed    & $O(d(v) + \log n)$      \\
  Edge removal     & $O(1)$          & not allowed    & $O(\log n)$             \\ \hline
 \end{tabular}
\caption{Time complexities of the dynamic recognition algorithms for PCA graphs.}
\end{table}

\section{Basic manipulation of contigs}
\label{sec:basic manipulation}

In this section we design several algorithms that will be used later for implementing the dynamic operations on the graph.  Most of these algorithms are generalizations of those by Deng \etal\ and Hell \etal\ from linear contigs to general (or circular) contigs.  Their goal is to allow the insertion and removal of semiblocks, as well as the insertion and removal of connections between semiblocks, without changing much of the input contig. 

For the removal of a semiblock we are given a compressed contig $\Psi$ and a semiblock $W$, and the goal is to build the compression of $\Phi = \Psi \setminus W$.  The insertion of a semiblock follows the inverse path.  We are given a compressed contig $\Phi$ and a semiblock $W$ (together with its family of neighbors) and the goal is to find a compressed contig $\Psi$ that contains $W$ such that $\Phi = \Psi \setminus W$, whenever possible.  It is worth noting that the proposed algorithms do not require $W$ to belong to $\B(\Psi)$; $W$ could be properly included in some semiblock of $\B(\Psi)$.  In such case, $\Phi$ is a compressed round representation of $G(\Psi) \setminus W$.

The connection and disconnection of semiblocks have similar definitions.  For the disconnection, we are given two semiblocks $B_l$ and $B_r$ of a compressed contig $\Psi$ that are adjacent in $\G(\Psi)$, and the goal is to compute a compressed round representation $\Phi$ of $G(\Psi) \setminus \{vw \mid v \in B_l\mbox{, } w \in B_r\}$, if possible, in such a way that $\B(\Phi)$ and $\B(\Psi)$ are almost the same orderings. The connection operation is just the inverse of the disconnection; we are given $B_l$ and $B_r$ as semiblocks of $\Phi$, and $\Psi$ is expected as the output.

In \Section~\ref{subsec:semiblock operations}, we present an algorithm for computing $\Phi = \Psi \setminus \{W\}$, with $W$ as input, without caring about the compression of $\Psi$ or $\Phi$.  Next, we deal with the inverse operation: given $W$ and $N(W)$, compute $\Psi$.  For these insertion and removal operations, is enough to solve the case in which $\Psi$ is circular.  Nevertheless, the described algorithms can be used to solve other cases as well. In \Section~\ref{subsec:compaction and separation} we show an algorithm that can be used to transform any contig into its compression, by \Definition{compacting} consecutive indistinguishable semiblocks.  The inverse operation is also provided, \ie, given one semiblock, separate it into two consecutive indistinguishable semiblocks.  Following, \Section~\ref{subsec:compressed operations} combines the previous algorithms so as to remove and insert semiblocks to compressed contigs.  For the sake of simplicity, in this part we restrict ourselves to contigs with few universal semiblocks.  Finally, in \Section~\ref{subsec:connection and disconnection} we define the pairs of semiblocks that can be disconnected from $\Psi$ and show how to actually disconnect these  semiblocks.  Its inverse operation, namely the connection of semiblocks, is also discussed.

We remark that the algorithms in this section do not require nor assure the straightness property.  So, for instance, the compressed removal algorithm could generate a circular contig representing a PIG graph.  This ignorance about the straightness property is desired because it allows the generation of all the possible contigs that represent a graph.

\subsection{Removal and insertion of semiblocks}
\label{subsec:semiblock operations}

We begin describing the simplest operation on contigs: the removal of a semiblock.  Given a semiblock $W$ of a contig $\Psi$, the goal is to compute the round representation $\Phi = \Psi \setminus \{W\}$.  Algorithm~\ref{alg:semiblock removal} is invoked to fulfill this goal.

\begin{algorithm}
  \caption{Removal of a semiblock.}\label{alg:semiblock removal}

  \textbf{Input:} a semiblock $W$ of a base contig $\Psi$.

  \textbf{Output:} $\Psi$ is transformed into the base $\Psi \setminus \{W\}$.

  \mbox{}

  \begin{AlgorithmSteps}
    \Step{Set $F_r(B) = N_l(W)$ for every $B \in [F_l(W), W)$ such that $F_r(B) = W$.}\label{alg:semiblock removal:right}
    \Step{Set $F_l(B) = N_r(W)$ for every $B \in (W, F_r(W)]$ such that $F_l(B) = W$.}\label{alg:semiblock removal:left}
    \Step{Remove $W$ from $\B(\Psi)$.}
  \end{AlgorithmSteps}
\end{algorithm}

For the correctness of Algorithm~\ref{alg:semiblock removal}, recall how $F_r^\Phi$ and $F_l^\Phi$ are defined.  For every $B \in \B(\Phi)$, $F_r^\Phi(B) = F_r^\Psi(B)$ if $F_r^\Psi(B) \neq W$, while $F_r^\Phi(B) = L^\Psi(W)$ otherwise.  Notice that if $F_r^\Psi(B) = W$, then (i) $W$ is not the left end semiblock of $\Psi$ and (ii) $B \in [F_l(W), W)$. By (i), $L^\Psi(W) = N_l^\Psi(W)$, hence Step~\ref{alg:semiblock removal:right} correctly updates all the right far pointers.  An analogous reasoning on the reverse of $\Psi$ is enough to conclude that Step~\ref{alg:semiblock removal:left} correctly updates the left far pointers.  Therefore, Algorithm~\ref{alg:semiblock removal} is correct.  With respect to the time complexity, only the semiblocks in $[F_l^\Psi(W), F_r^\Psi(W)] = N_{\G(\Psi)}[W]$ are traversed.

Thought Algorithm~\ref{alg:semiblock removal} is simple, it is not much efficient when the removed semiblock has large degree in $\G(\Psi)$.  Another way to remove a semiblock is by taking advantage of the self pointers.  Observe that by moving $S_r(W)$ so as to point to $N_l(W)$ we are actually moving all the right far pointers referencing $W$ so as to reference $N_l(W)$.  Hence, the first step of Algorithm~\ref{alg:semiblock removal} takes $O(1)$ time with this approach.  The inconvenient is that all those semiblocks that were previously pointing to $S_r(N_l(W))$ need to be updated so as to point to the new self pointer of $N_l(W)$.  Algorithm~\ref{alg:universal removal} implements this new idea.  Steps \ref{alg:universal removal:right}~and~\ref{alg:universal removal:left} preemptively restore the far pointers, and then Step~\ref{alg:universal removal:move} emulates the moving of the far pointers done by Algorithm~\ref{alg:semiblock removal}.

\begin{algorithm}[!htb]
  \caption{Removal of a semiblock of large degree.}\label{alg:universal removal}

  \textbf{Input:} a semiblock $W$ of a base contig $\Psi$.

  \textbf{Output:} $\Psi$ is transformed into the base $\Psi \setminus W$.

  \mbox{}

  \begin{AlgorithmSteps}
    \Step{Set $F_r(B) := S_r(W)$ for every $B \in [F_l(N_l(W)), F_l(W))$.}\label{alg:universal removal:right}
    \Step{Set $F_l(B) := S_r(W)$ for every $B \in (F_r(W), F_r(N_r(W))]$.}\label{alg:universal removal:left}
    \Step{Set $S_r(N_l(W)) := S_r(W)$ and $S_l(N_r(W)) := S_l(W)$.}\label{alg:universal removal:move}
    \Step{Remove $W$ from $\B(\Psi)$.}
  \end{AlgorithmSteps}
\end{algorithm}

With respect to the time complexity of Algorithm~\ref{alg:universal removal}, Steps \ref{alg:universal removal:right}~and~\ref{alg:universal removal:left} both take $O(n+u-d_{\G(\Psi)}(W))$ time when $\Psi$ is $u$-universal, as follows from the next lemma applied on both $\Psi$ and $\Psi^{-1}$.  

\begin{lemma}\label{lem:universal removal 2}
 If\/ $\Psi$ is a $u$-universal contig and\/ $W \in \B(\Psi)$, then \[\#[F_l(N_l(W)), F_l(W)) = O(n+u-d_{\G(\Psi)}(W)).\]
\end{lemma}

\begin{figure}[htb!]
  \centering
  \begin{tabular}{c@{\hspace{1cm}}c}
    \includegraphics{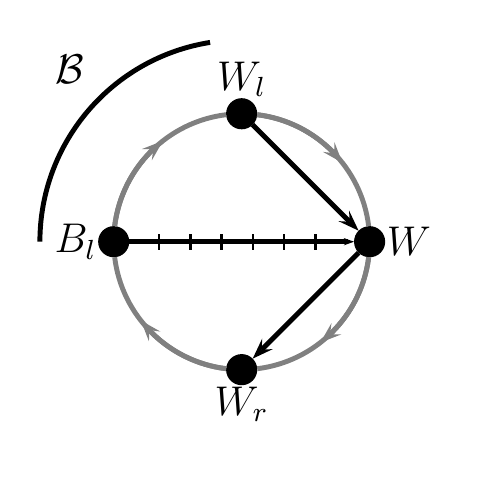} & \includegraphics{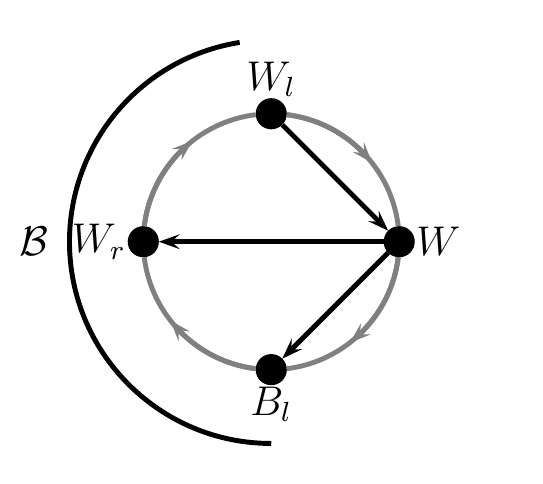} \\
    (a) & (b)
  \end{tabular}  
  \caption{Configurations of \Lemma~\ref{lem:universal removal 2}: (a) $B_l \in (W_r, W_l)$ and (b) $B_l \not\in (W_r,W_l)$.  Crossed lines are used to indicate missing edges.}\label{fig:universal removal}
\end{figure}

\begin{proof}
  Let $B_l = F_l(N_l(W))$, $W_r = F_r(W)$, $W_l = F_l(W)$, $\B = [B_l, W_l)$ (see \Figure~\ref{fig:universal removal}), and $q$ be the number of semiblocks in $\B \setminus (W_r, W_l)$ .  Clearly, $|\B| \leq \#(W_r, W_l)+q = n-d_{\G(\Psi)}(W)+q$.  If $q > 0$, then $\B \setminus (W_r, W_l) = (B_l, W_r]$. Let $B \in (B_l, W_r]$.  Since $W \to W_r$, it follows that $W \to B$, while since $B_l \to N_l(W)$, it follows that $B \to N_l(W)$.  Therefore, $B$ is a universal semiblock, which implies $q \leq u$.
\end{proof}

Combining Algorithms \ref{alg:semiblock removal}~and~\ref{alg:universal removal} with a simple check of the degrees, the following lemma is obtained.

\begin{lemma}\label{lem:semiblock removal}
  If\/ $\Psi$ is a $u$-universal base contig and\/ $W \in \B(\Psi)$, then the base\/ $\Psi \setminus \{W\}$ can be computed in $O(\min\{d_{\G(\Psi)}(W), n+u-d_{\G(\Psi)}(W))$ time, when\/ $W$ is given as input.  
\end{lemma}

The insertion of a semiblock is not as straightforward as the removal is.  For the sake of simplicity, we only discuss those insertions on contigs in which the inserted semiblock does not terminate as an end semiblock.  The other types of insertions are quite similar, and were already discussed in~\cite{DengHellHuangSJC1996,HellShamirSharanSJC2001}.  Let $\Psi$ be a contig in which $W$ is not an end semiblock, and suppose $\Phi = \Psi \setminus W$ is also a contig.   Also, let $B_l = F_l^\Psi(W)$ and $B_r = F_r^\Psi(W)$.  Since $W$ is not an end semiblock, $B_l \neq B_r$, and both $B_l$ and $B_r$ belong to $\B(\Phi)$.  We refer to $\Bip{B_l}{B_r}$ as \Definition{receptive} in $\Phi$, and to $\Psi$ as a \Definition{$W$-reception of $\Bip{B_l}{B_r}$} in $\Phi$.  Notice that the order between $B_l$ and $B_r$ is important; $\Bip{B_l}{B_r}$ could be receptive, even when $\Bip{B_r}{B_l}$ is not.  Observe also that all the $W$-receptions of $\Bip{B_l}{B_r}$ represent the same round graph.  Indeed, the neighborhood of $W$ in such round graph is $[B_l, B_r]$.  Also, it matters not which are the elements of $W$ (as long as $\B(\Phi) \cup \{W\}$ is a semiblock family).  Therefore, the property of being receptive depends exclusively on the election of $B_l$ and $B_r$ and not on $\Psi$ and $W$.  

\begin{figure}[htb!]
  \centering
  \begin{tabular}{c@{\hspace{1cm}}c}
    \includegraphics{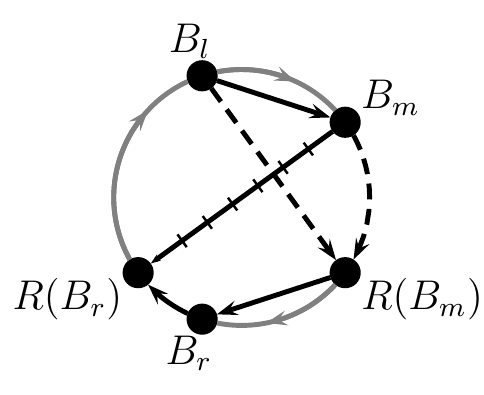} & \includegraphics{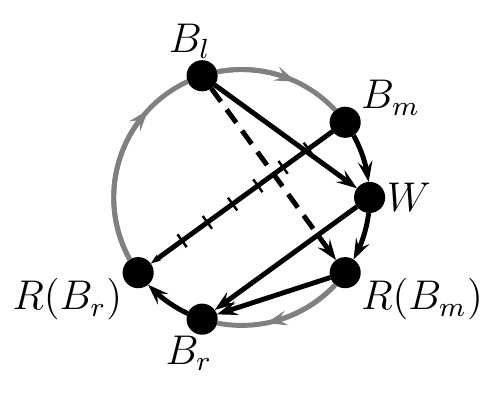} \\
    (a) & (b)
  \end{tabular}
  \caption{Example of a receptive contig (a) and its $W$-reception (b).  A dashed arrow between $B$ and $B'$ indicates that either $B \to B'$ or $B \nto B'$.}\label{fig:receptive}
\end{figure}

The contig $\Psi$ is an evidence that $\Bip{B_l}{B_r}$ is receptive in $\Phi$.  The goal of the \Definition{reception problem} is to determine whether a pair $\Bip{B_l}{B_r}$ is receptive in the absence of such a certificate.  That is, given $\Phi$ and $B_l, B_r \in \B(\Phi)$, determine whether $\Bip{B_l}{B_r}$ is receptive in $\Phi$. If so, a $W$-reception of $\Bip{B_l}{B_r}$ is desired.  The following lemma exhibits a solution for this problem (see \Figure~\ref{fig:receptive}).

\begin{lemma}\label{lem:receptive representation}
  Let\/ $\Phi$ be a contig, $B_l \neq  B_r$ be semiblocks of $\B(\Phi)$.  Then, $\Bip{B_l}{B_r}$ is receptive in\/ $\Phi$ if and only if there exists $B_m \in \{F_r(B_l), U_l(R(B_r))\}$ such that
  \begin{enumerate}[(i)]
    \item $B_m \in [B_l, B_r)$ and $F_r(B_m) \in [B_l, B_r]$, and
    \item if $F_l(B_l) \neq R(B_m)$, then $F_r(R(B_m)) \not\in [B_l, B_r)$.
  \end{enumerate}
  Furthermore, if\/ $\Bip{B_l}{B_r}$ is receptive in\/ $\Phi$, then\/ $\Psi = \Bip{(B_m, B_m] \Cat W}{ F_r^\Psi}$ is a\/ $W$-reception of\/ $\Bip{B_l}{B_r}$ in\/ $\Phi$, for any semiblock\/ $W$ such that $\B(\Phi) \cup \{W\}$ is a semiblock family, where $F_r^\Psi(W) = B_r$ and, for $B \in \B(\Phi)$,
  \[
    F_r^\Psi(B) = 
    \begin{cases}
      W & \text{if } B \in [B_l, B_m] \text{ and } F_r^\Phi(B) = B_m \\
      F_r^\Phi(B) & \text{otherwise}.
    \end{cases}
  \]
\end{lemma}

\begin{proof}
  First suppose $\Bip{B_l}{B_r}$ is receptive in $\Phi$ and let $\Psi$ be a $W$-reception of $\Phi$.  Note that if $W$ and $R^\Psi(W)$ are indistinguishable, then the contig obtained by changing $W$ and $R^\Psi(W)$ in $\Psi$ is also a $W$-reception of $\Phi$.  Hence, we can assume that $W$ and $R^\Psi(W)$ are not indistinguishable.  By the definition of receptive, $\Phi = \Psi \setminus W$, $B_l = F_l^\Psi(W)$, $B_r = F_r^\Psi(W)$, and $W \in [B_l, B_r]$ in $\Psi$.  Let $B_m = L^\Psi(W)$.  If $W = F_r^\Psi(B_l)$, then $B_m = F_r^\Phi(B_l)$.  Otherwise, $B_l \Topsi R^\Psi(W)$ and, since $W$ and $R^\Psi(W)$ are not indistinguishable, it follows that $R^\Psi(W) \Topsi R^\Psi(B_r)$.  Consequently, since $W \Ntopsi R^\Psi(B_r)$, we obtain that $W = U_l^\Psi(R^\Psi(B_r))$, which implies that $B_m = U_l^\Phi(R^\Phi(B_r))$.  

  Consider conditions (i)~and~(ii).  By definition, $B_m \in [B_l, B_r)$, while, since $B_m \Topsi W$ and $W \Ntopsi R^\Psi(B_r) = R^\Phi(B_r)$, we obtain that $F_r^\Phi(B_m) \in [B_m, B_r]$.  Hence, (i) follows.  Furthermore, since $W \Topsi B_r$, then $R^\Psi(W) = R^\Phi(B_m) \Tophi B_r$, while if $R^\Phi(B_m) \Tophi B_l$, then $B_l$ is universal in $\G(\Phi)$ and $F_l^\Phi(B_l) = R^\Phi(B_m)$.  Therefore, (ii) holds as well.

  For the converse, we claim that $\Psi$, as defined in the furthermore part, is a contig.  Clearly, $\G(\Psi)$ is connected because $\G(\Phi)$ is connected.  Then, we only need to prove that, for every $B \in \B(\Phi)$, either $B = F_r^\Psi(N_l^\Psi(B))$ or $B \Topsi F_r^\Psi(N_l^\Psi(B))$.  For this, let $B \in \B(\Psi)$, $N = N_l^\Psi(B)$, and $F = F_r^\Psi(N)$ be such that $B \neq F$, and consider the following cases.

  \begin{description}
    \item [Case 1:] $N = B_m$, thus $B = W$.  In this case, since $F \neq W$ and (i) holds, it follows that $F \in (B_m, B_r]$, thus $W \Topsi F$.

    \item [Case 2:] $N = W$, thus $B = N_r^\Phi(B_m)$ and $F = B_r$.  In this case, since $B \neq F$, we obtain, by (ii), that $B \Topsi F$.

    \item [Case 3:] $N \in (W, B_l)$.  In this case, $F = F_r^\Phi(N)$ while either $B \in (W, B_l)$ or $B = B_l$.  In the former case $F_r^\Psi(B) = F_r^\Phi(B)$ and $B \Tophi F$, thus $B \Topsi F$.  In the latter case, $F_r^\Psi(B) = W$ thus $B \Topsi F$.

    \item [Case 4:] $N \in [B_l, F_l^\Psi(R^\Psi(W)))$.  In this case, $F = W$, while either $B \in [B_l, F_l^\Psi(R^\Psi(W)))$ or $B = F_l^\Psi(R^\Psi(W))$.  Then, either $F_r^\Psi(B) = F$ (in the former case) or $F_r^\Psi(B) = R^\Psi(W)$ (in the latter case), thus $B \Topsi F$.

    \item [Case 5:] $N \in [F_l^\Psi(R^\Psi(W)), B_m)$.  In this case, $F = F_r^\Phi(N)$ and $F_r^\Psi(B) = F_r^\Phi(B)$, thus the claim follows.
  \end{description}
  
 Now, since $\Psi$ is a contig, we obtain that $\Phi = \Psi \setminus W$ and, by definition, $F_l^\Psi(W) = B_l$ and $F_r^\Psi(W) = B_r$.  In other words, $\Psi$ is a $W$-reception of $\Bip{B_l}{B_r}$ in $\Phi$, as desired.  
\end{proof}

Algorithm~\ref{alg:semiblock insertion} solves the reception problem.  Its inputs are two different semiblocks $B_l, B_r$ of a contig $\Phi$, and a semiblock $W$ such that $\B(\Phi) \cup \{W\}$ is a semiblock family.  If $\Bip{B_l}{B_r}$ is receptive in $\Phi$, then the output is the $W$-reception of $\Bip{B_l}{B_r}$ defined in the furthermore part of \Lemma~\ref{lem:receptive representation}.  Otherwise, an error message is obtained.  Step~\ref{alg:semiblock insertion:chk} looks for the semiblock $B_m$ that satisfies conditions (i)~and~(ii) of \Lemma~\ref{lem:receptive representation}, while Steps \ref{alg:semiblock insertion:ext init}--\ref{alg:semiblock insertion:ext end} build the $W$-reception of $\Bip{B_l}{B_r}$ when $\Phi$ is receptive.

\begin{algorithm}[!ht]
  \caption{Insertion of a new semiblock.}\label{alg:semiblock insertion}

  \textbf{Input:} Two different semiblocks $B_l, B_r$ of a base contig $\Phi$, and a semiblock $W$ such that $\B(\Phi) \cup \{W\}$ is a semiblock family.

  \textbf{Output:} if $\Bip{B_l}{B_r}$ is receptive in $\Phi$, then $\Phi$ is transformed into the base $W$-reception of $\Bip{B_l}{B_r}$ defined in \Lemma~\ref{lem:receptive representation}.  Otherwise, an error message is obtained.  

  \mbox{}

  \begin{AlgorithmSteps}
    \Step{Set a $1$ mark in all the semiblocks in $[B_l, B_r)$ and a $2$ mark in $B_r$.}\label{alg:semiblock insertion:chk init}
    \Step{Determine whether $\{F_r(B_l), U_l(R(B_r))\}$ has a semiblock $B_m$ marked with $1$ such that: (i) $F_r(B_m)$ is marked and (ii) $R(B_m) = F_l(B_l)$ or $F_r(R(B_m))$ is not marked with $1$.  If  false, then output an error message and halt.}\label{alg:semiblock insertion:chk}
    \Step{Insert $W$ between $B_m$ and $R(B_m)$, updating the near pointers.}\label{alg:semiblock insertion:ext init}
    \Step{Set $F_r(W) := B_r$ and $F_l(W) := B_l$.}
    \Step{Set $F_r(B) := W$ for every $B \in [B_l, B_m]$ such that $F_r^\Phi(B) = B_m$.}
    \Step{Set $F_l(B) := W$ for every $B \in [R(B_m), B_r]$ such that $F_l^\Phi(B) = R(B_m)$.}\label{alg:semiblock insertion:ext end}
  \end{AlgorithmSteps}
\end{algorithm}

Discuss the time complexity of Algorithm~\ref{alg:semiblock insertion}.  First note that, by \Lemma~\ref{lem:receptive representation}, either (a) $F_r(B_l)$ and $F_l(B_r)$ are the right and left end semiblocks of $\Phi$, respectively, or (b) $[B_l, B_r]$ has no end semiblocks, or (c) $\Bip{B_l}{B_r}$ is not receptive.  As a preprocessing, $[B_l, B_r]$ is traversed, in $O(\#[B_l, B_r])$ time, to evaluate if $\Phi$ satisfies either condition (a) or (b).  If $\Phi$ satisfies neither condition, the algorithm is halted.  Thus, suppose either (a) or (b) holds for $\Phi$ when Algorithm~\ref{alg:semiblock insertion} is invoked.  If $B_r$ is the right end semiblock, then $U_l(R(B_r)) = B_r$ is not marked with $1$ at Step~\ref{alg:semiblock insertion:chk init}.  Thus $U_l(R(B_r))$ needs not be considered in this case.  For the other case, there are two possibilities according to whether $F_l(R(B_r))$ is an end block or not.  In the former case, $U_l(R(B_r)) = N_l(F_l(N_r(B_r)))$.  In the latter case, (a) holds, thus $U_l(R(B_r)) = F_r(B_l)$.  Whichever the case, $U_l(R(B_r))$ is obtainable in $O(1)$ time.  Now consider how conditions (i)~and~(ii) are evaluated for $B_m$ in Step~\ref{alg:semiblock insertion:chk} when $B_m$ is marked with $1$.  If $B_m$ is not the right end semiblock, then $R(B_m) = N_r(B_m)$; otherwise (a) holds and $R(B_m) = F_l(B_r)$.  Therefore, Step~\ref{alg:semiblock insertion:chk} takes $O(1)$ time.  The remaining steps can be executed in $O(\#[B_l, B_r])$ time with an standard implementation. 

As it happens with the removal of semiblocks, the insertion problem can be solved more efficiently when the inserted semiblock has large degree in $\Psi$, \ie, when $\#[B_l, B_r] > \#(B_r, B_l)$.  Algorithm~\ref{alg:universal insertion} can be used in this case.  This time, the semiblock $B_m$ satisfying conditions (i) and (ii) of \Lemma~\ref{lem:receptive representation} is looked for at Step \ref{alg:universal insertion:chk}.  Following, if $\Bip{B_l}{B_r}$ is receptive, Steps \ref{alg:universal insertion:ext init}--\ref{alg:universal insertion:ext end} insert $W$ between $B_m$ and $R(B_m)$ and update the far pointers undoing the path taken by Algorithm~\ref{alg:universal removal:left} for the removal.  That is, first $S_r(B_m)$ is updated to refer to $W$ so that all the semiblocks whose right far pointer were referencing $B_m$ now reference $W$.  Analogously, $S_l(R(B_m))$ is updated to refer to $W$.  Finally, the far pointers of the semiblocks inside $(F_l(B_m), B_l)$ and $(B_r, F_r(R(B_m)))$ are corrected so that they do not refer to $W$.  

\begin{algorithm}[!ht]
  \caption{Insertion of a new semiblock of large degree.}\label{alg:universal insertion}

  \textbf{Input:} two different semiblocks $B_l, B_r$ of a contig $\Phi$, and a semiblock $W$ such that $\B(\Phi) \cup \{W\}$ is a semiblock family.

  \textbf{Output:} if $\Bip{B_l}{B_r}$ is receptive in $\Phi$, then $\Phi$ is transformed into the $W$-reception of $\Bip{B_l}{B_r}$ defined in \Lemma~\ref{lem:receptive representation}.  Otherwise, an error message is obtained.  

  \mbox{}

  \begin{AlgorithmSteps}
    \Step{Set a $1$ mark in all the semiblocks in $(B_r, B_l)$ and a $2$ mark in $B_r$.}\label{alg:universal insertion:chk init}
    \Step{Determine whether $\{F_r(B_l), U_l(R(B_r))\}$ has a semiblock $B_m$ not marked such that: (i) $F_r(B_m)$ is not marked with $1$ and (ii) $R(B_m) = F_l(B_l)$ or $F_r(R(B_m))$ is marked.  If  false, then output an error message and halt.}\label{alg:universal insertion:chk}
    \Step{Insert $W$ between $B_m$ and $R(B_m)$, updating the near pointers.}\label{alg:universal insertion:ext init}
    \Step{Set $S_r(W) := S_r(B_m)$ and $S_l(W) := S_l(R(B_m))$.}
    \Step{Set $F_r(B) := B_m$ for every $B \in [F_l(B_m), B_l)$.}
    \Step{Set $F_l(B) := R(B_m)$ for every $B \in (B_r, F_r(R(B_m))]$.}\label{alg:universal insertion:ext end}
  \end{AlgorithmSteps}
\end{algorithm}

For the implementation of Algorithm~\ref{alg:universal insertion}, a preprocessing step is executed to check whether the input satisfies conditions (a) or (b) as in Algorithm~\ref{alg:semiblock insertion}. Note that $[B_l, B_r]$ has no end semiblocks if and only if $\Phi$ is circular or $(B_r, B_l)$ has both end semiblocks.  Thus, for the preprocessing step it is enough to traverse $(B_r, B_l)$ in $O(\#(B_l, B_r))$ time.  Once the preprocessing step is concluded, Algorithm~\ref{alg:universal insertion} is invoked.  Step~\ref{alg:universal insertion:chk} takes $O(1)$ time with an implementation similar to the one discussed for Algorithm~\ref{alg:semiblock insertion}.  Hence, all the steps in Algorithm~\ref{alg:universal insertion} take $O(\#(B_r, B_l))$ time.  

If $\#[B_l, B_r]$ is given together with $B_l$ and $B_r$, then Algorithms \ref{alg:semiblock insertion}~and~\ref{alg:universal insertion} can be combined so as to obtain the following lemma.

\begin{lemma}\label{lem:semiblock insertion}
  Let\/ $\Phi$ be a base contig, and $B_l \neq B_r$ be semiblocks of\/ $\Phi$.  Then, it takes $O(\min\{\#[B_l, B_r], \#(B_r, B_l)\})$ time to determine whether\/ $\Bip{B_l}{B_r}$ is receptive in\/ $\Phi$, when $B_l$, $B_r$, and $\#[B_l, B_r]$ are given as input.  Furthermore, if\/ $\Bip{B_l}{B_r}$ is receptive, then a base\/ $W$-reception\/ $\Psi$ of\/ $\Bip{B_l}{B_r}$ can be obtained in $O(\min\{d_{\G(\Psi)}(W), n-d_{\G(\Psi)}(W)\})$ time, for any semiblock\/ $W$ such that\/ $\B(\Phi) \cup \{W\}$ is a semiblock family.
\end{lemma}

\subsection{Separation and compaction of semiblocks}
\label{subsec:compaction and separation}

\begin{figure}
  \centering
  \begin{tabular}{c@{\hspace{1cm}}c}
    \includegraphics{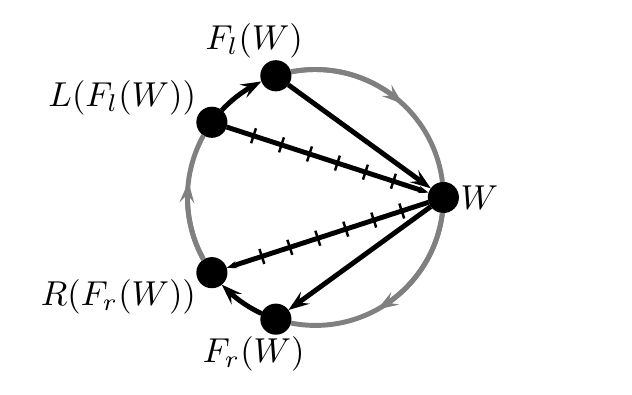} & \includegraphics{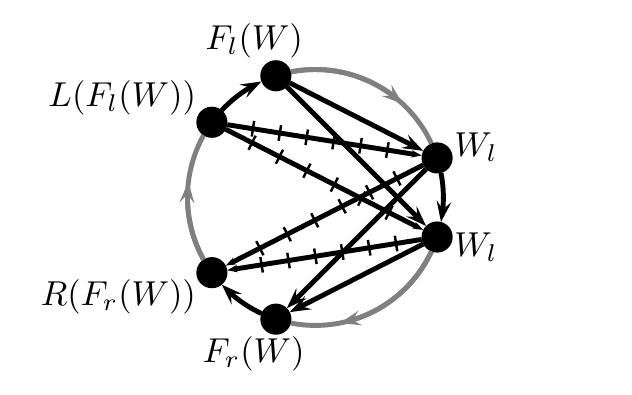} \\
    (a) & (b)
  \end{tabular}
  \caption{(a) A contig $\Phi$ and (b) the separation $\Psi$ of $W$ into $\{W_l, W_r\}$.  Notice that $\Phi$ is the compaction of $\{W_l, W_r\}$ in $\Psi$.}\label{fig:separation}
\end{figure}

In rough words, separating a semiblock $W$ means replacing $B$ with two consecutive semiblocks that partition $W$.  Let $\Phi$ be a contig, and $W_l$ and $W_r$ be two disjoint semiblocks such that $W = W_l \cup W_r$, for some $W \in \B(\Phi)$.  The \Definition{separation of\/ $W$ into\/ $\Bip{W_l}{W_r}$}, see \Figure~\ref{fig:separation}, is the contig $\Psi = \Bip{(W, L(W)] \Cat W_l \Cat W_r}{F_r^\Psi}$ such that, for any $B \in \B(\Psi)$,
\[
  F_r^\Psi(B) = 
  \begin{cases}
    W_r & \text{if } B \in \{W_l, W_r\} \text{ and } F_r^\Phi(W) = W \\
    W_r & \text{if } B \not\in \{W_l, W_r\} \text{ and } F_r^\Phi(B) = W \\
    F_r^\Phi(W) & \text{if } B \in \{W_l, W_r\} \text{ and } F_r^\Phi(W) \neq W \\
    F_r^\Phi(B) & \text{otherwise}
  \end{cases}
\]
Notice that the order of $W_l$ and $W_r$ is important; the separation of $W$ into $\Bip{W_l}{W_r}$ is not the same as the separation of $W$ into $\Bip{W_r}{W_l}$.  We say that $\Psi$ is a \Definition{separation of $W$ in $\Phi$} to mean that there exist $W_l, W_r \in \B(\Psi)$ such that $\Psi$ is the separation of $W$ into $\Bip{W_l}{W_r}$. The next observation follows easily.

\begin{observation}
  $W_l$ and\/ $W_r$ are indistinguishable in the separation of\/ $W_l \cup W_r$ into\/ $\Bip{W_l}{W_r}$.
\end{observation}

For the sake of simplicity, we extend the definition of separation for the case in which either $W_l = \emptyset$ or $W_r = \emptyset$.  Define $\Phi$ to be both the \Definition{separation of $W$ into $\Bip{W}{\emptyset}$} and the \Definition{separation of $W$ into $\Bip{\emptyset}{W}$}.

The separation of $W = W_l \cup W_r$ into $\Bip{W_l}{W_r}$ can be computed as in Algorithm~\ref{alg:separation}. Note that only $W$ and $W_r$ are given as input; $W_l$ is simply $W \setminus W_r$.  Step~\ref{alg:separation:sep} moves the elements of $W_r$ out of $W$, so that $W$ gets transformed into $W_l$.  Step~\ref{alg:separation:self pointers} applies the technique of self pointers for updating the right far pointers.  Observe that any block whose right far pointer was pointing to $W$ has to be updated so as to point to $W_r$.  Clearly, the most time expensive step of Algorithm~\ref{alg:separation} is Step~\ref{alg:separation:sep}, which costs $O(|W_r|)$ time.

\begin{algorithm}[!ht]
 \caption{Separation of a semiblock.}\label{alg:separation}

 \textbf{Input:} A semiblock $W$ of a base contig $\Phi$, and a semiblock $W_r \subseteq B$.

 \textbf{Output:} $\Phi$ is transformed into the base separation of $W$ into $\Bip{W \setminus W_r}{W_r}$.

 \mbox{}

 \begin{AlgorithmSteps}
  \Step{If either $W_r = \emptyset$ or $W_r = W$, then halt.}
  \Step{Move the elements of $W_r$ into a new semiblock lying immediately to the right of $W$.}\label{alg:separation:sep}
  \Step{Set $F_l(W_r) := F_l(W)$ and $F_r(W_r) := F_r(W)$.}
  \Step{Set $S_r(W_r) := S_r(W)$, and $S_r(W) := New$.}\label{alg:separation:self pointers}
 \end{AlgorithmSteps}
\end{algorithm}

Observe that, instead of moving the elements of $W_r$ out of $W$, we could have moved the elements of $W_l$ out of $W$.  This would yield a similar algorithm with temporal cost $O(|W_l|)$ instead of $O(|W_r|)$.  Of course, the input would have been $W_l$ instead of $W_r$.  Combining these algorithms with a simply cardinality check, we obtain the next lemma.

\begin{lemma}\label{lem:semiblock separation}
  Let\/ $\Phi$ be a base contig, $W \in \B(\Phi)$, and\/ $W_m \subseteq W$.  Then, both the separation of\/ $W$ into\/ $\{W \setminus W_m, W_m\}$ and the separation of\/ $W$ into\/ $\{W_m, W \setminus W_m\}$ can be computed in $O(\min\{|W_m|, |W \setminus W_m|\})$ time when\/ $W$ and $W_m$ are given as input.
\end{lemma}

The inverse of the separation is the compaction.  Let $\Phi$ be a contig, and suppose $W_l \in \B(\Phi)$ is indistinguishable with $W_r = N_r(W_l)$.  The \Definition{compaction of $\Bip{W_l}{W_r}$} in $\Phi$, see \Figure~\ref{fig:separation}, is the contig $\Psi = \Bip{(W_r, W_l) \Cat W_l \cup W_r}{F_r^\Psi}$ such that, for any $B \in \B(\Psi)$,
\[
  F_r^\Psi(B) = 
  \begin{cases}
    W_l \cup W_r & \text{if } B = W_l \cup W_r \text{ and } F_r^\Phi(W_r) = W_r \\
    W_l \cup W_r & \text{if } B \neq W_l \cup W_r \text{ and } F_r^\Phi(B) = W_r \\
    F_r^\Phi(W_r) & \text{if } B = W_l \cup W_r \text{ and } F_r^\Phi(W_r) \neq W_r \\
    F_r^\Phi(B) & \text{otherwise.}
  \end{cases}
\]

\begin{observation}
  The compaction and the separation are inverse operations.  That is, $\Psi$ is equal to the compaction of\/ $\Bip{W_l}{W_r}$ in the separation of\/ $\Bip{W_l}{W_r}$ in $\Psi$, for any\/ $W_l \cup W_r \in \B(\Psi)$, while\/ $\Phi$ is equal to the separation of\/ $\Bip{W_l}{W_r}$ of the compaction of\/ $\Bip{W_l}{W_r}$ in\/ $\Phi$, for any\/ $W_l \in \B(\Phi)$ that is indistinguishable with\/ $W_r = N_r^\Phi(W_r)$. 
\end{observation}

As done with the separation, it is convenient to define a robust compaction of $\Bip{W_l}{W_r}$ that works even when $W_l$ and $W_r$ are not indistinguishable.  With this in mind, define $\Phi$ to be the \Definition{compaction of $\Bip{W_l}{W_r}$} in $\Phi$ when $W_l$ and $W_r$ are not indistinguishable.

A method for computing the compaction of $\Bip{W_l}{W_r}$ is depicted in Algorithm~\ref{alg:union}.  Note that there are two possibilities when $W_l$ and $W_r$ are indistinguishable, either move the elements from $W_l$ to $W_r$ or move the elements from $W_r$ to $W_l$.  In Algorithm~\ref{alg:union} we take the latter possibility (see Step~\ref{alg:union:join}).  Note that, since $W_l$ and $W_r$ are indistinguishable, then no semiblock of $\Phi$ has neither $W_l$ as its right far neighbor.  Thus, Step~\ref{alg:union:self pointers} is enough to update all the right far pointers of the contig.  

\begin{algorithm}[!ht]
 \caption{Compaction of two consecutive semiblocks}\label{alg:union}

 \textbf{Input:} A semiblock $W$ of a base contig $\Phi$.

 \textbf{Output:} $\Phi$ is transformed into the base compaction of $\Bip{W}{N_r(W)}$ in $\Phi$.

 \mbox{}

 \begin{AlgorithmSteps}
  \Step{If $W$ and $N_r(W)$ are not indistinguishable, then halt.}
  \Step{Move the elements of $N_r(W)$ to $W$.}\label{alg:union:join}
  \Step{Set $S_r(W) := S_r(N_r(W))$.}\label{alg:union:self pointers}
  \Step{Remove $N_r(W)$ from $\Phi$.}
 \end{AlgorithmSteps}
\end{algorithm}

The time complexity of Algorithm~\ref{alg:union} is clearly $O(|W_r|)$.  The other possibility for computing the compaction, \ie\ moving the elements from $W_l$ to $W_r$, can be implemented similarly, and it takes $O(|W_r|)$ time.  So, we can decide which elements are moved by comparing $|W_r|$ and $|W_l|$.  In such case, the compaction algorithm takes $O(\min\{|W_l|, |W_r|\})$.  We record this fact in the next lemma.

\begin{lemma}\label{lem:semiblock compaction}
  If a semiblock\/ $W$ of a base contig\/ $\Phi$ is given as input, then the compaction of\/ $\Bip{W}{N_r(W)}$ in\/ $\Phi$ can be computed in $O(\min\{|W|, |N_r(W)|\})$ time.
\end{lemma}

\subsection{Compressed insertion and removal of semiblocks}
\label{subsec:compressed operations}

In this part, we consider the compressed removal and compressed insertion of semiblocks. In its basic form, the goal of the compressed removal operation is to find the compression of $\Gamma = \Psi \setminus W$, when a semiblock $W \in \B(\Psi)$, for some compressed contig $\Psi$, is given.  In this section we consider a generalization of this problem in which $W$ is included in some semiblock of $\B(\Psi)$.  Let $B \in \B(\Psi)$ be a semiblock of $\B(\Psi)$ and $W \subseteq B$.  The \Definition{compressed removal of $W$ from $\Psi$} is the contig $\Phi$ obtained by first separating $B$ into $\Bip{W}{B \setminus W}$, then removing $W$ to obtain $\Gamma$, and finally compressing $\Gamma$.  

\begin{observation}
  $G(\Phi) = G(\Psi) \setminus W$.  
\end{observation}

The following lemma shows how do the indistinguishable semiblocks of $\Gamma$ look like.  

\begin{lemma}\label{lem:decremental refinement}
  Let\/ $\Psi$ be a contig, $W, B_l \in \B(\Psi)$, $\Gamma = \Psi \setminus W$, and $B_r = N_r^\Gamma(B_l)$.  If $B_l$ and $B_r$ are not indistinguishable in\/ $\Psi$ and $B_l$ and $B_r$ are indistinguishable in\/ $\Gamma$, then\/ $\{B_l, B_r\}$ is equal to either\/ $\{U_l^\Psi(W), F_l^\Psi(W)\}$, $\{F_r^\Psi(W), U_r^\Psi(W)\}$, or\/ $\{N_l^\Psi(W), N_r^\Psi(W)\}$.  Furthermore, $\{B_l, B_r\} \neq \{N_l^\Psi(W), N_r^\Psi(W)\}$ when $N[W]$ contains at most one universal semiblock of $\G(\Psi)$.
\end{lemma}

\begin{proof}
  Suppose $B_l$ and $B_r$ are indistinguishable in $\Gamma$, and let $W_l = N_l^\Psi(W)$.  Because $B_l$ and $B_r$ are not indistinguishable in $\Psi$, it follows that either $F_r^\Psi(B_l) \neq F_r^\Psi(B_r)$ or $F_l^\Psi(B_l) \neq F_l^\Psi(B_r)$.  Assume the former, since a proof for the latter is obtained by applying the same arguments on $\Psi^{-1}$.  Recall that, for any $B \in \B(\Gamma)$, $F_r^\Gamma(B) \neq F_r^\Psi(B)$ only if $F_r^\Psi(B) = W$ and $F_r^\Gamma(B) = W_l$.  Then, we are left with the following two possibilities.

  \begin{description}
    \item [Case 1:]  $F_r^\Psi(B_l) = W$ and $F_r^\Gamma(B_l) = W_l$.  In this case, since $F_r^\Psi(B_l) \neq F_r^\Psi(B_r)$, we obtain that $F_r^\Psi(B_r) \neq W$, thus $F_r^\Psi(B_r) = F_r^\Gamma(B_r) = F_r^\Gamma(B_l) = W_l$.  Therefore, the only possibility is that $B_l = W_l$ and $B_r = R^\Psi(W)$.  Furthermore, since $B_r = N_r^\Gamma(B_l)$, we obtain that $B_r = N_r^\Psi(W)$, which implies that $B_r$ and $B_l$ are both universal in $\G(\Psi)$ and belong to $N[W]$.

    \item [Case 2:] $F_r^\Psi(B_r) = W$ and $F_r^\Gamma(B_r) = W_l$.  With arguments similar as those used in Case~1, we obtain that $F_r^\Psi(B_l) = F_r^\Gamma(B_l) = W_l$.  Consequently, since $B_r = N_r^\Gamma(B_l)$, it follows that $B_r = F_l^\Psi(W)$ and $B_l = U_l^\Psi(W)$.  Again, if $W \Topsi B_l$, then both $B_r$ and $W$ are universal in $\G(\Psi)$, thus the furthermore part follows.
  \end{description}
\end{proof}

The algorithm for computing the compressed removal of $W$ from $\Psi$ is obtained by simply composing the algorithms in the previous parts of the section.  First separate $B$ into $\Bip{W}{B \setminus W}$, then compute $\Gamma = \Psi \setminus W$, and finally compact the possible pairs of indistinguishable semiblocks of $\Gamma$.  By \Lemma~\ref{lem:decremental refinement}, the possible pairs of indistinguishable semiblocks are $\{U_l^\Gamma(W), F_l^\Gamma(W)\}$, $\{F_r^\Gamma(W), U_r^\Gamma(W)\}$, and $\{N_l^\Gamma(W), N_r^\Gamma(W)\}$.  Observe that we need not compact $F_l^\Gamma(W)$ with $U_l^\Gamma(W)$ (resp.\ $F_r^\Gamma(W)$ with $U_r^\Gamma(W)$) when $F_l^\Gamma(W)$ (resp.\ $F_r^\Gamma(W)$) is the left (resp.\ right) end semiblock.  By \Lemmas~\ref{lem:semiblock removal}, \ref{lem:semiblock separation}~and~\ref{lem:semiblock compaction}, the following corollary is obtained.

\begin{lemma}\label{lem:compressed removal}
  Let\/ $\Psi$ be a compressed and\/ $1$-universal base contig, $B \in \B(\Psi)$, and\/ $W \subseteq B$.  If\/ $W$ is given as input, then the base compressed removal of\/ $W$ from\/ $\Psi$ can be computed in time \[O\left(\begin{array}{l}\min\{d_{\G(\Psi)}(B), n-d_{\G(\Psi)}(B)\} + \min\{|F_r(B)|, |U_r(B)|\} + \min\{|F_l(B)|, |U_l(B)|\} +\\ \min\{|W|, |B \setminus W|\}\end{array}\right).\]
\end{lemma}

The compressed insertion of a semiblock is the inverse operation of the compressed removal.  We only discuss the compressed insertion for those cases in which the resulting representation is a circular contig.  The remaining cases follow from~\cite{DengHellHuangSJC1996,HellShamirSharanSJC2001}.  Furthermore, we are interested only in the case in which the neighborhood of the inserted semiblock contains at most one universal semiblock.  Let $\Psi$ be a compressed circular contig, $B \in \B(\Psi)$ with $N[B]$ containing at most one universal semiblock of $\G(\Psi)$, and $W \subseteq B$.  Suppose $B_l = F_l^\Psi(B)$ and $B_r = F_r^\Psi(B)$, and let $\Phi$ be the compressed removal of $W$ from $\Psi$.  Clearly, $B_l$ and $B_r$ must be included in semiblocks $B_a$ and $B_b$ of $\Phi$, respectively.  Furthermore, $B_a \neq B_b$ since otherwise $B_l$ and $B_r$ would be non-universal twins of $\G(\Psi)$, contradicting \Lemma~\ref{lem:indistinguishable}.  Moreover, since $N[B]$ has at most one universal semiblock of $\G(\Psi)$, at most one semiblock of $\Phi$ in $[B_a, B_b]$ is universal. We refer to $\Bip{B_l}{B_r}$ as \Definition{refinable} in $\Phi$, and to $\Psi$ as a \Definition{$W$-refinement of $\Bip{B_l}{B_r}$} in $\Phi$. (The terms refinable and refinement were introduced in~\cite{HellShamirSharanSJC2001} to refer to similar concepts.)  This definition is similar to the definition of receptive pairs. As with $W$-receptions, all the $W$-refinements of $\Bip{B_l}{B_r}$ represent the same graph.  Indeed, by \Lemma~\ref{lem:decremental refinement}, all the semiblocks of $\Psi \setminus \{B\}$ inside $(B_l, B_r)$ are also semiblocks of $\Phi$.  Consequently, $(B_a, B_b) \cup \{B_l, B_r\}$ is precisely the closed neighborhood of $W$ in the round graph represented by the $W$-refinement of $\Bip{B_l}{B_r}$.  Also, the elements of $W$ are unimportant to determine whether $\Bip{B_l}{B_r}$ is refinable or not.  Therefore, the property of being refinable depends only on the election of $\Bip{B_l}{B_r}$.

Analogous to the reception problem, the \Definition{refinement} problem is to determine whether a pair is refinable.  That is, given $\Phi$ and the semiblocks $B_l \subseteq B_a$ and $B_r \subseteq B_b$, for different semiblocks $B_a, B_b \in \B(\Phi)$, determine whether $\Bip{B_l}{B_r}$ is refinable in $\Phi$.  If so, a $W$-refinement of $\Bip{B_l}{B_r}$ is also desired. Define the \Definition{$\Bip{B_l}{B_r}$-separation} of $\Phi$ as the contig obtained by first separating $B_a$ into $\Bip{B_a \setminus B_l}{B_l}$ and then separating $B_b$ into $\Bip{B_r}{B_b \setminus B_r}$.  The following lemma shows that $\Bip{B_l}{B_r}$ is refinable if and only if $\Bip{B_l}{B_r}$ is receptive in the $\Bip{B_l}{B_r}$-separation of $\Phi$.

\begin{lemma}\label{lem:compressed insertion}
  Let\/ $\Phi$ be a compressed contig, and $B_l \subseteq B_a$ and $B_r \subseteq B_b$ be semiblocks, for different $B_a, B_b \in \B(\Phi)$, such that at most one semiblock in\/ $[B_a, B_b]$ is universal in $\G(\Phi)$. Then, $\Bip{B_l}{B_r}$ is refinable in\/ $\Phi$ if and only if\/ $\Bip{B_l}{B_r}$ is receptive in the\/ $\Bip{B_l}{B_r}$-separation of\/ $\Phi$.  Furthermore, if\/ $\Bip{B_l}{B_r}$ is refinable and\/ $W$ is such that $\B(\Phi) \cup \{W\}$ is a semiblock family, then any\/ $W$-reception of\/ $\Bip{B_l}{B_r}$ in the\/ $\Bip{B_l}{B_r}$-separation of\/ $\Phi$ is equal to a separation of the semiblock containing\/ $W$ in a\/ $W$-refinement of\/ $\Bip{B_l}{B_r}$ in\/ $\Phi$.
\end{lemma}

\begin{proof}
  Suppose $\Bip{B_l}{B_r}$ is refinable in $\Phi$ and let $\Psi$ be a $W$-refinement of $\Bip{B_l}{B_r}$ in $\Phi$ where $W \subseteq B$, for $B \in \B(\Psi)$.  By definition, $N[B]$ has at most one universal semiblock of $\G(\Psi)$, $B_l = F_l^\Psi(B)$, $B_r = F_r^\Psi(B)$ and $\Phi$ is the compressed removal $W$ from $\Psi$.  If $W \neq B$, then $B \setminus W$ is a semiblock of $\Phi$ whose left and right far neighbors are $B_a = B_l$ and $B_b = B_r$.  Therefore, $\Phi$ is exactly the $\Bip{B_l}{B_r}$-separation of $\Phi$, hence $\Bip{B_l}{B_r}$ is receptive in the $\Bip{B_l}{B_r}$-separation of $\Phi$.  On the other hand, if $W = B$, then, by \Lemma~\ref{lem:decremental refinement}, $\{B_l, L^\Psi(B_l)\}$ and $\{B_r, R^\Psi(B_r)\}$, are the only possible pairs of indistinguishable semiblocks of $\Psi \setminus W$, implying that $\Psi \setminus W$ is the $\Bip{B_l}{B_r}$-separation of $\Phi$.  Therefore, $\Bip{B_l}{B_r}$ is receptive in the $\Bip{B_l}{B_r}$-separation of $\Phi$, because $B_l = F_l^\Psi(W)$ and $B_r = F_r^\Psi(W)$.

  For the converse, let $\Gamma$ be the $\Bip{B_l}{B_r}$-separation of $\Phi$, and suppose $\Bip{B_l}{B_r}$ is receptive in $\Gamma$.  Let $\Psi$ be some $W$-reception of $\Bip{B_l}{B_r}$ in $\Gamma$, and call $B$ to the block of $G(\Psi)$ containing $W$.  If $B_l \neq B_a$, then $F_r^\Psi(B_l) \neq F_r^\Psi(B_a \setminus B_l)$, while if $B_r \neq B_b$, then $F_l^\Psi(B_r) \neq F_l^\Psi(B_b \setminus B_r)$.  Hence, since $\{B_l, B_a \setminus B_l\}$ and $\{B_r, B_b \setminus B_r\}$ are the only possible pairs of indistinguishable semiblocks of $\Gamma$, it follows that $\Psi$ has at most one pair of indistinguishable semiblocks, namely $\{N_l^\Psi(W), W\}$ or $\{W, N_r^\Psi(W)\}$, whose union yields $B$.  Even more, $N[B]$ has at most one universal semiblock of $\G(\Psi)$, because at most one semiblock of $\Phi$ in $[B_a, B_b]$ is universal in $\G(\Phi)$, and no semiblock in $(B_b, B_a)$ is adjacent to $W$ in $\Psi$.  Consequently,  as desired, $\Psi$ is a separation of $B$ in a $W$-refinement of $\Bip{B_l}{B_r}$ in $\Phi$.
\end{proof}

The algorithm for testing if $\Bip{B_l}{B_r}$ is refinable in $\Phi$, and obtaining a $W$-refinement if so, is also obtained by combining algorithms of the previous parts.  First, build the $\Bip{B_l}{B_r}$-separation of $\Phi$.  Next, check whether $\Bip{B_l}{B_r}$ is receptive in the $\Bip{B_l}{B_r}$-separation of $\Phi$.  If successful, then a $W$-reception $\Gamma$ of $\Bip{B_l}{B_r}$ is obtained.  The final step, then, is to compact $\Bip{N_l^\Gamma(W)}{W}$ and $\Bip{W}{N_r^\Gamma(W)}$ to obtain the contig $\Psi$.  By \Lemma~\ref{lem:compressed insertion}, $\Psi$ is a $W$-refinement of $\Bip{B_l}{B_r}$ in $\Phi$.  By \Lemmas~\ref{lem:semiblock insertion}, \ref{lem:semiblock separation}~and~\ref{lem:semiblock compaction}, the time required by this algorithm is as in the lemma below.

\begin{lemma}\label{lem:refinable complexity}
  Let\/ $\Phi$ be a compressed base contig, and $B_l \subseteq B_a$ and $B_r \subseteq B_b$ be semiblocks, for different $B_a, B_b \in \B(\Phi)$, such that at most one semiblock in\/ $[B_a, B_b]$ is universal in $\G(\Phi)$. If $B_a$, $B_b$, $B_l$, $B_r$ and $\#[B_a, B_b]$ are given as input, then it takes \[O\left(\min\{\#[B_a, B_b], \#(B_b, B_a)\} + \min\{|B_l|, |B_a \setminus B_l|\} + \min\{|B_r|, |B_b \setminus B_r|\}\right)\] time to determine whether\/ $\Bip{B_l}{B_r}$ is refinable in\/ $\Phi$. Furthermore, if\/ $\Bip{B_l}{B_r}$ is refinable in\/ $\Phi$ and\/ $W$ is a semiblock such that $\B(\Phi) \cup \{W\}$ is a semiblock family, then a base\/ $W$-refinement\/ $\Psi$ of\/ $\Bip{B_l}{B_r}$, in which\/ $W \subseteq B$ for $B \in \B(\Psi)$, can be obtained in time \[O\left(\begin {array}{l}\min\{d_{\G(\Psi)}(B), n-d_{\G(\Psi)}(B)\} + \min\{|B_l|, |B_a \setminus B_l|\} + \min\{|B_r|, |B_b \setminus B_r|\} +\\ \min\{|W|, |B \setminus W|\}\end{array}\right).\]
\end{lemma}

\subsection{Disconnection and connection of semiblocks}
\label{subsec:connection and disconnection}

To end this section, we show two simple algorithms that can be used to insert and remove edges from a graph, when a contig is provided.  

\begin{figure}[htb!]
  \centering
  \begin{tabular}{c@{\hspace{1cm}}c}
    \includegraphics{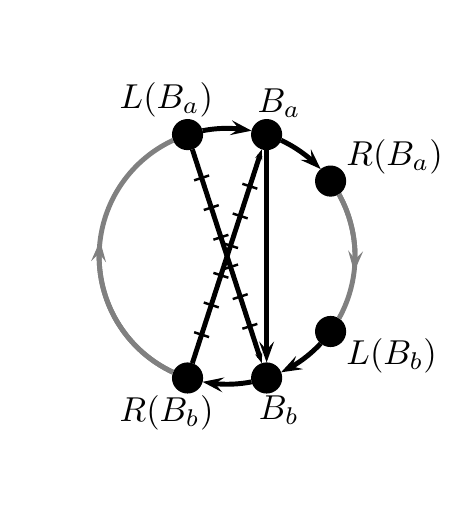} & \includegraphics{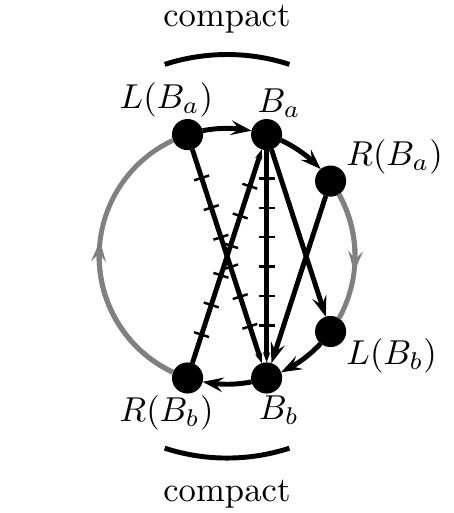} \\
    (a) & (b)
  \end{tabular}
  \caption{(a) A contig $\Psi$ with a disconnectable pair $\Bip{B_a}{B_b}$ and (b) the disconnection $\Phi$ of $\Bip{B_a}{B_b}$ in $\Psi$.  Notice that $\Psi$ is the connection of $\Bip{B_a}{B_b}$ in $\Phi$.}\label{fig:disconnection}
\end{figure}

Let $\Psi$ be a contig and $B_a \neq B_b$ be semiblocks of $\B(\Psi)$.  Say that $\Bip{B_a}{B_b}$ is \Definition{disconnectable} when $B_b = F_r(B_a)$ and $B_a = F_l(B_b)$.  Define $\Gamma$ as the round representation $\Bip{\B(\Psi)}{F_r^\Gamma}$ such that $F_r^\Gamma(B_a) = N_l^\Psi(B_b)$ and $F_r^\Gamma(B) = F_r^\Psi(B)$ for every $B \in \B(\Psi) \setminus \{B_a\}$.  Notice that $\Gamma$ is well defined if and only if $\Bip{B_a}{B_b}$ is disconnectable.  Define the \Definition{disconnection of $\Bip{B_a}{B_b}$ in $\Psi$} to be the compression $\Phi$ of $\Gamma$; see \Figure~\ref{fig:disconnection}.  

\begin{observation}
  $G(\Phi) = G(\Psi) \setminus \{vw \mid v \in B_a\mbox{, }w \in B_b\}$.
\end{observation}

The disconnection of semiblocks can be generalized to subsets of semiblocks.  Recall that, for semiblocks $B_l \subset B_a$ and $B_r \subset B_b$, the $\Bip{B_r}{B_l}$-separation of $\Psi$ is the contig $\Lambda$ obtained by first separating $B_b$ into $\Bip{B_b \setminus B_r}{B_r}$ and then separating $B_a$ into $\Bip{B_l}{B_a \setminus B_l}$.  By the separation definition, $\Bip{B_l}{B_r}$ is disconnectable in $\Lambda$ if and only if $\Bip{B_a}{B_b}$ is disconnectable in $\Psi$.  Define the \Definition{disconnection of $\Bip{B_l}{B_r}$ in $\Psi$} to be the disconnection $\Phi$ of $\Bip{B_l}{B_r}$ in $\Lambda$.  The following lemma generalizes the relation between $G(\Psi)$ and $G(\Phi)$.

\begin{lemma}\label{lem:disconnection}
  Let\/ $\Psi$ be a contig, and $B_l \subseteq B_a$ and $B_r \subseteq B_b$ be semiblocks, for different $B_a, B_b \in \B(\Psi)$.  If\/ $\Bip{B_a}{B_b}$ is disconnectable and\/ $\Phi$ is the disconnection of\/ $\Bip{B_l}{B_r}$ in\/ $\Psi$, then $G(\Phi) = G(\Psi) \setminus \{vw \mid v \in B_l\mbox{, }w \in B_r\}$.
\end{lemma}

The algorithm to obtain the disconnection $\Phi$ of $\Bip{B_l}{B_r}$ in a compressed contig $\Psi$, whenever $\Bip{B_a}{B_b}$ is disconnectable, is obtained by composing algorithms of the previous parts.  For the first step, compute the $\Bip{B_r}{B_l}$-separation $\Gamma$ of $\Psi$.  Next, set $F_r^\Gamma(B_l) = N_l^\Gamma(B_r)$ and $F_l^\Gamma(B_r) = N_r^\Gamma(B_l)$.  Finally, just compact the possible indistinguishable semiblocks of $\Gamma$.  To determine which are the possible indistinguishable semiblocks of $\Gamma$, observe that if $B_l$ is not a right end semiblock, then $B_l$ and $N_r^\Gamma(B_l)$ are not indistinguishable.  Indeed, since $B_a \Topsi B_b$, it follows that $N_r^\Gamma(B_l) \Togamma B_r$ while $B_l \Ntogamma B_r$.  Similarly, if $B_r$ is not a left end semiblock, then $B_r$ and $N_l^\Gamma(B_r)$ are not indistinguishable.  Consequently, the only possible pairs of indistinguishable semiblocks of $\Gamma$ are $\{L^\Gamma(B_l), B_l\}$ and $\{B_r, R^\Gamma(B_r)\}$.  Therefore, by \Lemmas \ref{lem:semiblock separation}~and~\ref{lem:semiblock compaction}, we obtain the following bound on the time required by the algorithm.

\begin{lemma}\label{lem:disconnectable complexity}
  Let\/ $\Psi$ be a compressed base contig, and $B_l \subseteq B_a$ and $B_r \subseteq B_b$ be semiblocks, for different $B_a, B_b \in \B(\Psi)$.  If\/ $\Bip{B_a}{B_b}$ is disconnectable, then the base disconnection of\/ $\Bip{B_l}{B_r}$ in\/ $\Psi$, when $B_a$, $B_b$, $B_l$, and $B_r$ are given as input, can be computed in time \[O\left(\min\{|B_a|, |B_l|\} + \min\{|B_b|, |B_r|\} + \min\{|B_l|, |N_l(B_a)|\} + \min\{|B_r|, |N_r(B_b)|\}\right).\]
\end{lemma}

The connection operation is the inverse of the disconnection operation.  We describe the connection operation only for semiblocks that belong to the same contig; for the other case, see~\cite{HellShamirSharanSJC2001}.  Let $\Phi$ be a contig and $B_a \neq B_b$ be semiblocks of $\B(\Psi)$.  Say that $\Bip{B_a}{B_b}$ is \Definition{connectable} when $B_b = U_r(B_a)$ and $B_a = U_l(B_b)$.  Let $B_l, B_r$ be semiblocks such that $B_l \subseteq B_a$ and $B_r\subseteq B_b$, and let $\Lambda$ be the $\Bip{B_l}{B_r}$-separation of $\Phi$.  Its not hard to see that $\Bip{B_l} {B_r}$ is connectable in $\Lambda$.  Define $\Gamma$ as the contig $\Bip{\B(\Lambda)}{F_r^\Gamma}$ such that $F_r^\Gamma(B_l) = B_r$ and $F_r^\Gamma(B) = F_r^\Lambda(B)$ for every $B \in \B(\Gamma) \setminus \{B_l\}$.  Notice that $\Gamma$ is well defined if and only if $\Bip{B_l}{B_r}$ is connectable.  Define the \Definition{connection of $\Bip{B_l}{B_r}$ in $\Phi$} to be the compression $\Psi$ of $\Gamma$ (see \Figure~\ref{fig:disconnection}).  Opposite to the disconnection, the connection represents the insertion of edges to $G(\Phi)$.

\begin{lemma}\label{lem:connection}
  Let\/ $\Phi$ be a contig, and $B_l \subseteq B_a$ and $B_r \subseteq B_b$ be semiblocks, for different $B_a, B_b \in \B(\Psi)$.  If\/ $\Bip{B_a}{B_b}$ is connectable and\/ $\Psi$ is the connection of\/ $\Bip{B_l}{B_r}$ in\/ $\Phi$, then $G(\Psi) = G(\Phi) \cup \{vw \mid v \in B_l\mbox{, }w \in B_r\}$.
\end{lemma}

The algorithm to obtain the connection $\Psi$ of $\Bip{B_l}{B_r}$ in $\Phi$, whenever $\Bip{B_a}{B_b}$ is connectable, is rather similar to the disconnection algorithm.  For the first step, compute the $\Bip{B_l}{B_r}$-separation $\Gamma$ of $\Phi$.  Next, set $F_r^\Gamma(B_l) = B_r$ and $F_l^\Gamma(B_r) = B_l$.  Finally, just compact the indistinguishable semiblocks of $\Gamma$. In this case, the possible pairs of indistinguishable semiblocks of $\Gamma$ are $\{B_l, N_r^\Gamma(B_l)\}$ and $\{N_l^\Gamma(B_r), B_r\}$.  Therefore, by \Lemmas \ref{lem:semiblock separation}~and~\ref{lem:semiblock compaction}, we obtain the following corollary.

\begin{lemma}\label{lem:connection complexity}
  Let\/ $\Phi$ be a compressed base contig, and $B_l \subseteq B_a$ and $B_r \subseteq B_b$ be semiblocks, for different $B_a, B_b \in \B(\Psi)$.  If\/ $\Bip{B_a}{B_b}$ is connectable in\/ $\Phi$, then the base connection of\/ $\Bip{B_l}{B_r}$ in\/ $\Phi$, when $B_a$, $B_b$, $B_l$, and $B_r$ are given as input, can be computed in time \[O\left(\min\{|B_a|, |B_l|\} + \min\{|B_b|, |B_l|\}+\min\{|B_l|,|N_r(B_a)|\}+\min\{|B_r|,|N_l(B_b)|\}\right).\]
\end{lemma}

\section{Co-bipartite round graphs}
\label{sec:co-bipartite}

The incremental algorithm for the recognition of connected PIG graphs by Deng \etal\ takes advantage of the fact that every connected PIG graph admits a unique linear block contig, up to full reversal (\Theorem~\ref{thm:unique-PIG-models}).  For the dynamic recognition of general PIG graphs, Hell \etal\ have to deal with each component in a separate way, since the linear block contigs representing the components can be permuted to form several straight block representations.  For PCA graphs the situation is similar.  Huang~\cite{HuangJCTSB1995} proved that every connected and co-connected round graph admits a unique block contig, up to full reversal (see \Theorem~\ref{thm:unique-PCA-models}).  However, when $G$ is not co-connected, the round block representation of each co-component can be split in two ranges that form a \emph{co-contig}.  As it happens with disconnected PIG graphs, these co-contigs can be permuted so as to form several round block representations of $G$.  

Instead of dealing with the co-components in the data structure, we take a more lazy approach: we compute the co-components only when they are needed.  The advantage of this approach is that we obtain an efficient algorithm for computing the co-components of any round graph.  The disadvantage is that we have to find the co-components fast.  In \Section~\ref{subsec:co-components PCA graphs} we show how to find all the co-contigs in $O(\Delta(\G(\Phi)))$ time, when a round representation $\Phi$ is given.

The algorithm developed on the first part is not efficient enough for our purposes, when semiblocks are inserted to the round graph.  The inconvenient is that we can only spend a time proportional to the degree of the inserted semiblock.  Furthermore, a representation could not exist at all.  \Section~\ref{subsec:co-components incremental} is devoted to this problem.  We show that the inserted semiblock has large degree when it belongs to a round graph that is not co-connected.  Thus, we can adapt the algorithm in \Section~\ref{subsec:co-components PCA graphs} so that, given $\Phi$ and the degree of the inserted semiblock, it either outputs the co-contigs of $\Phi$, or it claims that the modified graph is not round.

Finally, in \Section~\ref{subsec:divide-and-conquer}, we design two algorithms that can be combined so as to traverse all the round representations of a round graph.  The goal of these algorithms is to split a contig $\Phi$ into its co-contigs, and to join these co-contigs to obtain contigs whose represented graphs are isomorphic to $\G(\Phi)$.

Throughout the section, a structure characterization of round representations is obtained.  We remark that the characterization in not new, see \eg~\cite{Huang1992}.  Nevertheless, the algorithmic approach used to obtain such a characterization is new, as far as our knowledge extends.

\subsection{Co-components of round graphs}
\label{subsec:co-components PCA graphs}

Let $B$ be a semiblock of a contig $\Phi$.  The goal of this part is to show how to compute the co-component $\G$ of $\G(\Phi)$ that contains $B$ in $O(d_\G(B))$ time.  The solution to this problem yields an $O(\Delta(\G(\Phi)))$ time algorithm for computing all the co-components of $\G(\Phi)$ (encoded as co-contigs of $\Phi$, cf.~below).   The following proposition, that follows from \Theorem~\ref{thm:forbiddens PCA}, is essential for our purposes.

\begin{lemma}\label{lem:co-connected->co-bipartite}
 If a round graph is not co-connected, then it is co-bipartite.
\end{lemma}

Algorithm~\ref{alg:co-bipartition of CA} outputs the co-component containing $B$, for any co-bipartite semiblock graph $\G$.  Its correctness follows from the following lemma.

\begin{algorithm}
 \caption{Co-bipartition of the co-component containing $B$.}\label{alg:co-bipartition of CA}

 \textbf{Input:} A co-bipartite semiblock graph $\G$, and $B \in V(\G)$.

 \textbf{Output:} The co-bipartition $\Bip{\X}{\Y}$ of the co-component of $\G$ such that $B \in \X$.

 \mbox{}

 \begin{AlgorithmSteps}
  \Step{Set $\X := \{B\}$ and $\Y := \emptyset$. \label{alg:co-bipartition of CA:step:1}}
  \Step{Perform the following operations while $\Y \neq \overline{\N}(\X)$.}  
  \IncreaseIndent
   \Step{Set $\Y := \overline{\N}(\X)$.\label{alg:co-bipartition of CA:step:3}}
   \Step{Set $\X := \overline{\N}(\Y)$.\label{alg:co-bipartition of CA:step:4}}
  \DecreaseIndent
  \Step{Output $\Bip{\X}{\Y}$.}
 \end{AlgorithmSteps}
\end{algorithm}

\begin{lemma}\label{lem:co-bip-alg}
 If $\G$ is a co-bipartite semiblock graph and $\X, \Y$ are the families of Algorithm~\ref{alg:co-bipartition of CA} at some step of its execution, then $\G[\X \cup \Y]$ is co-connected, $\X \cap \Y = \emptyset$, and $B \in \X$.  Moreover, when Algorithm~\ref{alg:co-bipartition of CA} stops, $\Bip{\X}{\overline{\N}(\X)}$ is a co-bipartition of a co-component of $\G$ and $B \in \X$.
\end{lemma}

Let $\Phi$ be a round representation.  A range $[B_l, B_r]$ of $\B(\Phi)$ is a \Definition{co-contig chunk} if $[B_l, B_r] \subseteq \X$ for some co-bipartition $\Bip{\X}{\overline{\N}(\X)}$ of a co-component of $\G(\Phi)$.  The next lemma shows how do $\X$ and $\Y$ look like at each step of Algorithm~\ref{alg:co-bipartition of CA}, when applied to round graphs.

\begin{lemma}\label{lem:overline N}
 If $\X = [B_l, B_r]$ is a co-contig chunk of a round representation\/ $\Phi$, then $\overline{\N}(\X) = (F_r(B_l), F_l(B_r))$.
\end{lemma}

\begin{proof}
 Call $W_l = F_r(B_l)$ and $W_r = F_l(B_r)$.  Suppose, to obtain a contradiction, that $B_l \neq B_r$ and $B_l \nto B_r$.  In this case $B_r \to B_l$ because $\X$ is a clique of $\G(\Phi)$.  Hence, $B_r \to B_j$ for every $B \in (B_r, B_l]$ which implies that $B_r$ is universal.  This is impossible because $B_l$ and $B_r$ belong to the same co-component by definition.  Therefore, either $B_l = B_r$ or $B_l \to B_r$.  Consequently, $B_r \in [B_l, W_l]$ which implies that $B \to B'$ for every $B \in [B_l, B_r]$ and every $B' \in [B_r, W_l]$.  A similar argument can be used to prove that $B' \to B$ for every $B \in [B_l, B_r]$ and every $B' \in [W_r, B_l]$.  Then, all the semiblocks in $[W_r, W_l]$ belong to $N_{\G(\Phi)}[B]$ for every $B \in \X$, thus $\overline{\N}(\X) \subseteq (W_l, W_r)$.

 For the other inclusion, suppose there is some semiblock $W \in (W_l, W_r)$ that is adjacent to all the semiblocks of $\X$ in $\G(\Phi)$.  This implies that $B_l$ and $B_r$ are not universal in $\G(\Phi)$ since otherwise $(W_l, W_r) = \emptyset$.  Hence, $R(W_l)$ is not adjacent to $B_l$ and $L(W_r)$ is not adjacent to $B_r$, so $W \neq R(W_l)$ and $W \neq L(W_r)$.  Consequently, $(W_l, W)$ and $(W, W_r)$ are nonempty ranges.  In particular, both $R(W_l)$ and $L(W_r)$ belong to $\overline{\N}(\X)$, thus there is a path between $R(W_l)$ and $L(W_r)$ in $\overline{\G(\Phi)}$.  Such path must contain three blocks $W_1, B_2, W_3$ such that $W_1$ is not adjacent to $B_2$, $B_2$ is not adjacent to $W_3$, $B_2 \in \X$, $W_1 \in (W_l, W)$, and $W_2 \in (W, W_r)$.  By hypothesis, either $W \to B_2$ or $B_2 \to W$.  The former is impossible because $W_3 \nto B_2$, while the latter is impossible because $B_2 \nto W_1$.  
\end{proof}

\begin{corollary}\label{cor:X and Y co-bipartite}
 Let\/ $\Phi$ be a round representation.  If $\G(\Phi)$ is co-bipartite, then, at each step of Algorithm~\ref{alg:co-bipartition of CA} when applied to $\G(\Phi)$, $\X$ is a co-contig chunk of\/ $\Phi$ and $\Y$ is either empty or a co-contig chunk of\/ $\Phi$.
\end{corollary}

\begin{proof}
 Observe that if $\X$ is a co-contig chunk of $\Phi$, then $\overline{\N}(\X)$ is a range of $\Phi$ by \Lemma~\ref{lem:overline N}.  If $\overline{\N}(\X) = \emptyset$, then $\X = \{B\}$ for some universal semiblock $B$ of $\G(\Phi)$; otherwise, $\overline{\N}(\X)$ is a co-contig chunk of $\Phi$.  In the latter case, $\overline{\N}(\overline{\N}(\X)) \neq \emptyset$ is also a co-contig chunk of $\Phi$, by \Lemma~\ref{lem:overline N}.  Therefore, since $\X$ is a co-contig chunk of $\Phi$ before the main loop of Algorithm~\ref{alg:co-bipartition of CA}, we obtain that $\X$ and $\Y$ are both co-contig chunks of $\Phi$ after every step of the main loop of Algorithm~\ref{alg:co-bipartition of CA}.
\end{proof}

\begin{figure}[htb!]
 \centering
 \includegraphics{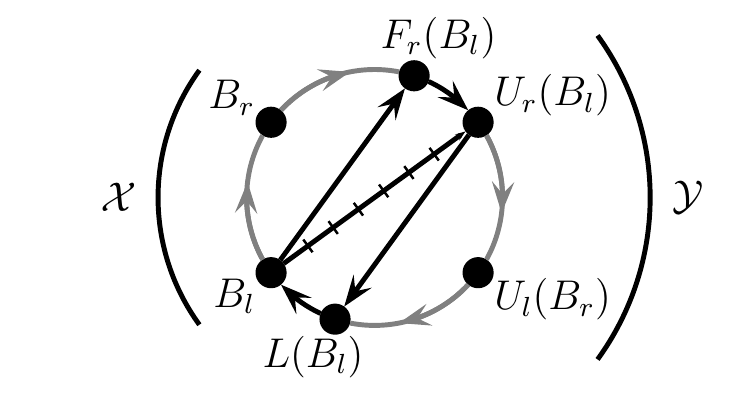}
 \caption{In a round representation $\Phi$, each co-component of $\G(\Phi)$ is represented by a co-contig $\Phi|(\X \cup \Y)$ that is described by a co-contig pair $\Bip{\X}{\Y}$.  A semiblock $B$ of $\G(\Phi)$ is a left co-end semiblock if and only if $U_r(U_r(B)) = B$ (\eg\ $B_l$ and $U_r(B_l)$) in the figure.}\label{fig:co-contig}
\end{figure}

The above corollary allows us to define co-contigs as the analogous of contigs.  Let $\Phi$ be a round representation of a co-bipartite round graph $\G$, and $\X$ and $\Y$ be two ranges of $\B(\Phi)$.  Say that $\Bip{\X}{\Y}$ is a \Definition{co-contig pair} of $\Phi$, and that $\X$ and $\Y$ are \Definition{co-contig ranges} of $\Phi$, when $\Bip{\X}{\Y}$ is a co-bipartition of some co-component of $\G$.  When $\Bip{\X}{\Y}$ is a co-contig pair, $\Phi|(\X \cup \Y)$ is a \Definition{co-contig} of $\Phi$ that is \Definition{described} by $\Bip{\X}{\Y}$ and that \Definition{represents} $\G[\X \cup \Y]$.  We also refer to $\Phi$ as a co-contig to indicate that $\G$ is co-connected.  The following corollary shows the similarity between contigs and co-contigs (see \Figure~\ref{fig:co-contig}).  Since we use this corollary almost as a definition of co-contigs, we will make no references to it.

\begin{corollary}[see also~\cite{Huang1992}]{co-contigs represent}
  Let\/ $\Phi$ be a round representation. If $\G(\Phi)$ is co-bipartite, then every co-component of $\G(\Phi)$ is represented by a co-contig of\/ $\Phi$.
\end{corollary}

\begin{proof}
  Let $B \in \B(\Phi)$, and $\X$ and $\Y$ be the families of semiblocks obtained by the execution of Algorithm~\ref{alg:co-bipartition of CA} with input $\G(\Phi)$ and $B$.  By \Lemma~\ref{lem:co-bip-alg}, $\Bip{\X}{\Y}$ is a co-bipartition of the co-component $\G$ of $\G(\Phi)$ that contains $B$.  On the other hand, by \Corollary~\ref{cor:X and Y co-bipartite}, both $\X$ and $\Y$ are co-contig ranges.  Consequently, $\Bip{\X}{\Y}$ is a co-contig pair describing the co-contig that represents $\G$.
\end{proof}

If $\X = [B_l, B_r]$ is a co-contig range of $\Phi$, then $B_l$ and $B_r$ are the \Definition{left and right co-end semiblocks} of $\X$, respectively.  A co-contig range $\X$ has no co-end semiblocks when $\X = \emptyset$.  By the corollary above, every $B \in \B(\Phi)$ belongs to a unique co-contig range of $\Phi$.  We say that $B$ is a \Definition{(left or right) co-end semiblock} of $\Phi$ when $B$ is a (left or right) co-end semiblock of co-contig range to which it belongs.

By definition, every universal semiblock of $\G$ is a left and right co-end semiblock of $\Phi$.  \Lemma~\ref{lem:overline N} can be used to determine if $B$ is a co-end semiblock of $\Phi$ when $B$ is not universal in $\G$.  Just observe that the co-contig $\Gamma$ containing $B$ is described by a co-contig pair $\Bip{[B_l, B_r]}{[W_l, W_r]}$.  By \Lemma~\ref{lem:overline N}, $[W_l, W_r] = (F_r(B_l), F_l(B_r))$, hence $W_l = U_r(B_l)$ and $W_r = U_l(B_r)$.  Analogously, $B_l = U_r(W_l)$ and $B_r = U_l(W_r)$, see \Figure~\ref{fig:co-contig}.  Consequently, $B$ is a left co-end semiblock if and only if $B = U_r(U_r(B))$, while $B$ is a right co-end block if and only if $B = U_l(U_l(B))$.  On the other hand, if $B = U_r(U_r(B))$, then $\G(\Phi)$ is necessarily co-bipartite, because $\Bip{[B, F_r(B)]}{[U_r(B), B)}$ is a co-bipartiton of $\G$.

\begin{observation}
  $B$ is a co-end semiblock if and only if $B = U_r(U_r(B))$ or $B = U_l(U_l(B))$.
\end{observation}

\Lemma~\ref{lem:overline N} and \Corollary~\ref{cor:X and Y co-bipartite} show how to simulate Algorithm~\ref{alg:co-bipartition of CA} when a contig $\Phi$ is given.  Because $O(1)$ time evaluation of $U_r$ and $U_l$ is required, and base contigs provide no means for efficiently evaluating these functions when $\Phi$ is linear, we divide the implementation in two, according to whether $\Phi$ is circular or linear.  When $\Phi$ is circular, Algorithm~\ref{alg:co-bipartition of CA} can be simulated as in Algorithm~\ref{alg:co-bipartition implementation}.  Note that Algorithm~\ref{alg:co-bipartition implementation} does not require $\G(\Phi)$ to be co-bipartite for accepting $\Phi$ as an input.  When $\G(\Phi)$ is not co-bipartite, then a message indicating so is obtained.  On the other hand, when $\G(\Phi)$ is co-bipartite, the algorithm outputs a co-contig pair $\Bip{\X}{\Y}$ such that $B \in \X$.  

\begin{algorithm}[!ht]
 \caption{Co-contig containing a semiblock $B$.}\label{alg:co-bipartition implementation}

 \textbf{Input:} a semiblock $B$ of a circular base contig $\Phi$.

 \textbf{Output:} the co-contig pair $\Bip{\X}{\Y}$ of $\Phi$ such that $B \in \X$ if $\G(\Phi)$ is co-bipartite.  If $\G(\Phi)$ is not co-bipartite, then an error message is obtained.
 
 \mbox{}

 \begin{AlgorithmSteps}
  \Step{Set $\X := [B, B]$, $\Y := \emptyset$. \label{alg:co-bipartition implementation:step:1}}
  \Step{If $U_r(B) = F_l(B)$ then output $\Bip{\X}{\emptyset}$ and halt.\label{alg:co-bipartition implementation:step:2}}
  \Step{Define the function $\overline{\bullet}$ that, given a range $[B_l, B_r]$, outputs $[U_r(B_l), U_l(B_r)]$.}
  \Step{Perform the following operations for at most $d_{\G(\Phi)}(B)$ iterations, while $\Y \neq \overline{\X}$.}
  \IncreaseIndent
   \Step{Set $\Y := \overline{\X}$.\label{alg:co-bipartition implementation:step:3}}
   \Step{Set $\X := \overline{\Y}$.\label{alg:co-bipartition implementation:step:4}}
  \DecreaseIndent
  \Step{If $\X = \overline{\Y}$, then output $\Bip{\X}{\Y}$; otherwise, output an error message.\label{alg:co-bipartition implementation:total steps}}
 \end{AlgorithmSteps}
\end{algorithm}

Discuss the correctness of Algorithm~\ref{alg:co-bipartition implementation}.  Step~\ref{alg:co-bipartition implementation:step:2} checks whether $B$ is universal in $\G(\Phi)$; if so, it outputs the co-contig pair $\Bip{[B, B]}{\emptyset}$.  Otherwise, let $\X = [B_l, B_r]$ at some point of the execution, and suppose $\G(\Phi)$ is co-bipartite.  By invariant, $B_l$ is not universal, thus $U_r(B_l) \neq F_l(B_l)$. Then, by \Lemma~\ref{lem:overline N}, $\overline{\N_{\G(\Phi)}}(\X) = \overline{\X} = [U_r(B_l), U_l(B_r)]$.  Furthermore, since at least one semiblock is inserted into $\X$ at each iteration of the main loop, and the co-component containing $B$ has at most $d_{\G(\Phi)}(B)$ semiblocks, Algorithm~\ref{alg:co-bipartition implementation} effectively simulates Algorithm~\ref{alg:co-bipartition of CA} when $\G(\Phi)$ is co-bipartite.  On the other hand, if $\X = \overline{\Y}$ at Step~\ref{alg:co-bipartition implementation:total steps}, then $\Bip{\X}{\Y}$ is a co-contig pair representing $\Phi|(\X \cup \Y)$.  Therefore, Algorithm~\ref{alg:co-bipartition implementation} halts with the error message only when $\G(\Phi)$ is not co-bipartite.  Summing up, Algorithm~\ref{alg:co-bipartition implementation} is correct.

For the implementation, co-contig chunks are represented by a pair of pointers, referencing the leftmost and rightmost semiblocks in the range.  Of course, the empty range is implemented with a pair of pointers referencing $NULL$.  Clearly, mappings $U_r$ and $U_l$ take $O(1)$ time because $\Phi$ is circular.  On the other hand, the main loop is executed for at most $|\X|$ iterations if $\G(\Phi)$ is co-bipartite, while is executed for $d_{\G(\Phi)}(B)$ iterations otherwise.  

The case in which $\Phi$ is linear can be solved using the following lemma.

\begin{lemma}\label{lem:cobipartite-linear}
  Let\/ $\Phi$ be a linear contig and $B \in \B(\Phi)$.  Then, $\G(\Phi)$ is co-bipartite if and only if $B_l = F_l(N_l(F_l(B)))$ and $B_r = F_r(N_r(F_r(B_l)))$ are the left and right end semiblocks of\/ $\Phi$, respectively.  Furthermore, if\/ $\G(\Phi)$ is co-bipartite, then\/ $\Bip{[B_l, F_l(B_r))}{(F_r(B_l), B_r]}$ and\/ $\Bip{[B, B]}{\emptyset}$, for $B \in [F_l(B_r), F_r(B_l)]$, are all the co-contig pairs of\/ $\G(\Phi)$. 
\end{lemma}

\begin{proof}
 Let $W_l$ and $W_r$ be the left and right end semiblock of $\Phi$, respectively.  Suppose first that $\G(\Phi)$ is co-bipartite.  If $N_l(F_l(W_r)) = F_l(W_r)$ or $N_r(F_r(W_l)) = W_r$, then $\G(\Phi)$ is a clique and the lemma holds.  Otherwise, $N_l(F_l(W_r)) \nto W_r$ and $W_l \nto N_r(F_r(W_l))$, thus $W_l \to N_l(F_l(W_r))$ and $N_r(F_r(W_l)) \to W_r$ because $\G(\Phi)$ is co-bipartite.  Hence, $B_l = W_l$, $B_r = W_r$, and the furthermore part holds.  For the converse, observe that both $[B_l, F_l(B_r))$ and $(F_l(B_r), B_r]$ are cliques of $\G(\Phi)$.  
\end{proof}

\begin{figure}[ht!]
  \centering\includegraphics{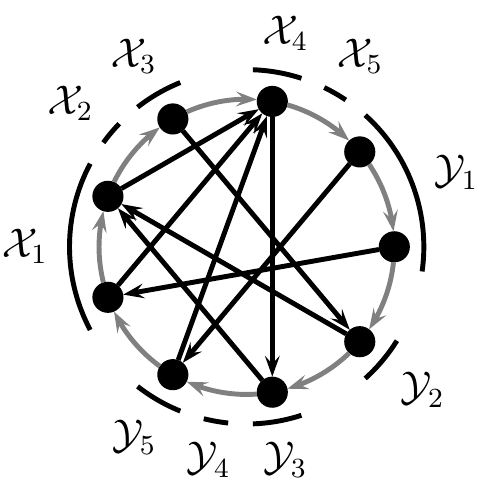}
  \caption{A natural ordering $\Bip{\X_1}{\Y_1}, \ldots, \Bip{\X_5}{\Y_5}$ of the co-contigs of a round representation $\Phi$. (Only the edges from semiblocks to their near and far neighbors are shown; the former are gray, the latter are black.)  Even though $\X_2$, $\X_5$, and $\Y_4$ are empty co-contigs ranges, they occupy a well determined location in the ordering, that depends on the locations of $\Y_2$, $\Y_5$, and $\X_4$, respectively.  Hence, $\Bip{\X_1}{\Y_1}, \ldots, \Bip{\X_5}{\Y_5}$ is a natural ordering only of $\Phi$.}\label{fig:co-contigs algorithm}
\end{figure}

In any round representation $\Phi$, the family of contigs admits an ordering $\Phi_1, \ldots, \Phi_s$ such that $\B(\Phi) = \B(\Phi_1) \Cat \ldots \Cat \B(\Phi_s)$.  The family of co-contigs of $\Phi$ satisfies an analogous condition.  When $\G(\Phi)$ is co-bipartite, an ordering $\bB = \Bip{\X_1}{\Y_1}, \ldots, \Bip{\X_s}{\Y_s}$ of the co-contig pairs of $\Phi$ is said to be \Definition{natural} if $\B(\Phi) = \X_1 \Cat \ldots \Cat \X_s \Cat \Y_1, \ldots \Cat \Y_s$ (see \Figure~\ref{fig:co-contigs algorithm}).  It is not hard to see that, among all the round representations of $\G(\Phi)$, $\bB$ is a natural ordering only of $\Phi$.  

To end this part, Algorithm~\ref{alg:co-bipartition implementation} and \Lemma~\ref{lem:cobipartite-linear} are combined into Algorithm~\ref{alg:co-components of PCA}, whose purpose is to determine if a round graph is co-bipartite.  The input of Algorithm~\ref{alg:co-components of PCA} is a round representation $\Phi$, and the output is either a message, if $\G(\Phi)$ is not co-bipartite, or a natural ordering of the co-contig pairs of $\Phi$, otherwise.  

Recall that the only disconnected co-bipartite graph is the graph formed by the union of two cliques.  So, the only co-bipartite round representation $\Phi$ that is not a contig has two contig ranges $\X$ and $\Y$, each of which is a co-contig range.  Steps \ref{alg:co-components of PCA:step 01}--\ref{alg:co-components of PCA:step 03} find the co-contig pair formed by these co-contig ranges.  The case in which $\Phi$ is linear and $\G(\Phi)$ is co-bipartite is solved in Step~\ref{alg:co-components of PCA:linear}.  If the algorithm reaches Step~\ref{alg:co-components of PCA:step 1}, it is because either $\Phi$ is circular or $\G(\Phi)$ is not co-bipartite.  So, Algorithm~\ref{alg:co-bipartition implementation} applied on $\G(\Phi)$ outputs a co-contig pair if and only if $\Phi$ is circular and $\G(\Phi)$ is co-bipartite.  Steps \ref{alg:co-components of PCA:circular begin}--\ref{alg:co-components of PCA:circular end} are then used to obtain a natural ordering $\bB = \Bip{\X_1}{\Y_1}, \ldots, \Bip{\X_s}{\Y_s}$ of its co-contig pairs.  For this, Step~\ref{alg:co-components of PCA:circular begin} choses a left co-end semiblock $B_l$, and defines $\Bip{\X_1}{\Y_1}$ to be the co-contig pair $\Bip{\X}{\Y}$ such that $B_l \in \X$.  Suppose that, after some iterations of Loop~\ref{alg:co-components of PCA: loop init}--\ref{alg:co-components of PCA:other co-contigs}, $\Bip{\X_1}{\Y_1}, \ldots, \Bip{\X_i}{\Y_i}$ has already been determined for $i \geq 1$.  Moreover, suppose $\X_i \neq \emptyset$, and let $B_r$ the right co-end block of $\X_i$.  Clearly, $W_l = R(B_r)$ is the left co-end semiblock of some co-contig range $\X_{i+k+1}$.  Note that $k = 0$ if and only if $(F_l(B_r), F_r(W_l)) = \emptyset$.  Otherwise, every semiblock in $(F_l(B_r), F_r(W_l))$ is universal in $\G(\Phi)$.  Thus, for every $1 \leq j \leq k$, it follows that $\X_{i+j} = \emptyset$ and $\Y_{i+j}$ is formed by the $j$-th semiblock to the right of $F_l(B_r)$.  Step~\ref{alg:co-components of PCA:universal co-contigs} is responsible of extending $\bB$ with $\Bip{\X_{i+j}}{\Y_{i+j}}$, for every $1 \leq j \leq k$.  Following, Step~\ref{alg:co-components of PCA:other co-contigs} adds $\Bip{\X_{i+k+1}}{\Y_{i+k+1}}$ as a co-contig pair as well.  Finally, Step~\ref{alg:co-components of PCA:universal co-contigs bis} inserts the co-contig pairs not found by the loop into $\bB$.  Summing up, Algorithm~\ref{alg:co-components of PCA} is correct.

\begin{algorithm}[!ht]
 \caption{Co-contigs of a round representation $\Phi$}\label{alg:co-components of PCA}

 \textbf{Input:} a base round representation $\Phi$.

 \textbf{Output:} if $\G(\Phi)$ is not co-bipartite, then a message.  Otherwise, a natural ordering of the co-contig pairs of $\Phi$.

 \mbox{}
 
 \begin{AlgorithmSteps}
  \Step{If $\Phi$ has at least three contigs, then halt with a message.\label{alg:co-components of PCA:step 01}}
  \Step{If $\Phi$ has two contig ranges $\X_1$ and $\X_2$, then:}
  \IncreaseIndent
   \Step{If, for $i \in \{1,2\}$, $F_l(B)$ and $F_r(F_l(B))$ are end semiblocks for some $B \in \X_i$,  then output $\Bip{\X_1}{\X_2}$ and halt.  Otherwise, output a message and halt.}\label{alg:co-components of PCA:step 03}
  \DecreaseIndent
  \Step{If $B_l := F_l(N_l(F_l(B)))$ and $B_r := F_r(N_r(F_r(B_l)))$ are end semiblocks, for $B \in \B(\Phi)$, then output $\Bip{[B_l, F_l(B_r))}{[F_r(B_l), B_r]}$, $\Bip{\X_1}{\emptyset}$, $\ldots$, $\Bip{\X_k}{\emptyset}$, where $\X_1 \Cat \ldots \Cat \X_k = [F_l(B_r), F_r(B_l)]$, and halt.}\label{alg:co-components of PCA:linear}
  \Step{Apply Algorithm~\ref{alg:co-bipartition implementation} to some semiblock of $\Phi$.  If Algorithm~\ref{alg:co-bipartition implementation} halts in error, then output a message and halt.  Otherwise, a co-contig pair $\Bip{[B_l, B_r]}{\Y}$ is obtained.\label{alg:co-components of PCA:step 1}}
  \Step{Set $\bB := \{\Bip{[B_l, B_r]}{\Y}\}$.}\label{alg:co-components of PCA:circular begin}
  \Step{While $B_r \neq F_r(B_l)$:}\label{alg:co-components of PCA: loop init}
  \IncreaseIndent
   \Step{Apply Algorithm~\ref{alg:co-bipartition implementation} to $W_l = R(B_r)$, to obtain a new co-contig pair $\Bip{[W_l, W_r]}{\Y}$.}\label{alg:co-components of PCA:iteration}
   \Step{Add $\Bip{\emptyset}{\Y_1}, \ldots, \Bip{\emptyset}{\Y_k}$ at the end of $\bB$, for $\Y_1 \Cat \ldots \Cat \Y_k = (F_l(B_r), F_r(W_l))$.}\label{alg:co-components of PCA:universal co-contigs}
   \Step{Add $\Bip{[W_l, W_r]}{\Y}$ at the end of $\bB$ and set $B_r := W_r$.}\label{alg:co-components of PCA:other co-contigs} 
  \DecreaseIndent
  \Step{Add $\Bip{\emptyset}{\Y_1}, \ldots, \Bip{\emptyset}{\Y_k}$ at the end of $\bB$, for $\Y_1 \Cat \ldots \Cat \Y_k = (F_l(B_r), B_l)$.}\label{alg:co-components of PCA:universal co-contigs bis}
  \Step{Output $\bB$.}\label{alg:co-components of PCA:circular end}
 \end{AlgorithmSteps}
\end{algorithm}

\begin{lemma}\label{lem:natural ordering}
  If\/ $\Phi$ is a round representation of a co-bipartite graph, then its family of co-contigs admits a natural ordering.
\end{lemma}

With respect to the time complexity of Algorithm~\ref{alg:co-components of PCA}, observe that every step of the algorithm takes constant time, except for Steps~\ref{alg:co-components of PCA:linear}, \ref{alg:co-components of PCA:step 1}, \ref{alg:co-components of PCA:iteration}, \ref{alg:co-components of PCA:universal co-contigs}, and~\ref{alg:co-components of PCA:universal co-contigs bis}.  If $\Phi$ is a $u$-universal contig, then Steps \ref{alg:co-components of PCA:linear}, \ref{alg:co-components of PCA:universal co-contigs}, and~\ref{alg:co-components of PCA:universal co-contigs bis} consume up to $O(u)$ time.  On the other hand, Step~\ref{alg:co-components of PCA:step 1} takes $O(\Delta(\G(\Phi)))$ time, while Step \ref{alg:co-components of PCA:iteration} takes $O(\#[W_l, W_r]|)$ time, because it is executed only when $\G(\Phi)$ is co-bipartite.  Finally, Loop \ref{alg:co-components of PCA: loop init}--\ref{alg:co-components of PCA:other co-contigs} is executed exactly once for each co-contig of $\Phi$.  

\begin{lemma}\label{lem:co-contigs}
  Determining whether a round graph $\G$ is co-bipartite takes $O(\Delta(\G))$ time, when a base round representation\/ $\Phi$ of $\G$ is given.  Furthermore, when $\G$ is co-bipartite, a natural ordering of co-contig pairs of\/ $\Phi$ is obtained in $O(\Delta(\G))$ time. 
\end{lemma}

\subsection{Co-components of incremental round graphs}
\label{subsec:co-components incremental}

In this part of the section we deal with the problem of finding all the co-components of a $u$-universal round graph $\G$ when a semiblock $B$ is to be inserted into $\G$.  The key idea is to prove that, for $\MH = \G \cup \{B\}$ to be round, $B$ has to be adjacent to a large number of semiblocks of $\G$.  Even more, $B$ does not have non-neighbors in more than two non-universal co-components of $\G$.  So, a simple modification of Algorithm~\ref{alg:co-components of PCA} yields an $O(d_\MH(B)+u)$ time algorithm that, given a round representation $\Phi$ of $\G$, outputs all the co-contig ranges of $\Phi$ or claims that $\MH$ is not round.

\begin{lemma}\label{lem:high-degree}
 Let $B$ be a semiblock of a round graph $\MH$ such that $\MH \setminus \{B\}$ is co-bipartite and not straight, $\Phi$ be a base contig representing $\MH \setminus \{B\}$, and\/ $\Gamma$ be a co-contig of\/ $\Phi$.  If the main loop of Algorithm~\ref{alg:co-bipartition implementation} takes $p$ iterations to stop when applied to a semiblock in $\B(\Gamma)$, then $p \leq d_{\G(\Gamma)}(B) + 2$.
\end{lemma}

\begin{proof}
  The lemma is clearly true for $p \leq 2$, so consider $p > 2$.  Let $B_1 = B$, and $B_i$ and $W_i$ ($2 \leq i \leq p$) be the leftmost semiblocks of ranges $\X$ and $\Y$ prior to the $i$-th iteration of the main loop of Algorithm~\ref{alg:co-bipartition implementation}, respectively. Recall that $W_i = U_r^\Phi(B_{i-1})$, while $B_{i} = U_r^\Phi(W_i)$, for $2 \leq i \leq p$. Thus, if $B_i = B_{i-1}$, then $W_{i+1} = W_i$ and $B_{i+1} = B_i$.  The rightmost semiblocks of $\X$ and $\Y$ follow a similar invariant.  So, we may assume w.l.o.g.\ that $B_i \neq B_{i-1}$ for $2 \leq i \leq p$, and thus $W_i \neq W_{i-1}$ for $3 \leq i \leq p$.  Now, since $B_{i-1}$ is not universal in $\G(\Phi)$ ($2 \leq i \leq p$), it follows that $B_{i} \in (W_{i}, B_{i-1})$.  Thus, since $\X$ is a co-contig chunk at each iteration of Algorithm~\ref{alg:co-bipartition implementation} by \Corollary~\ref{cor:X and Y co-bipartite}, it follows that $B_p \to B_1$.  Therefore, since $W_2 \nto B_1$, we obtain that $W_2, B_p, \ldots, B_1$ appear in this order in $\B(\Phi)$.  Similarly, $B_1, W_p, \ldots, W_2$ appear in this order in $\B(\Phi)$ (see \Figure~\ref{fig:high-degree}).  Summing up, since $W_i = U_r^\Phi(B_{i-1})$ and $B_{i} = U_r^\Phi(W_i)$ ($2 \leq i \leq p$), we obtain that $B_1, W_2, B_2, W_3, B_3 \ldots, W_{p}, B_{p}$ induce a path in $\overline{\G(\Gamma)}$.  

  \begin{figure}
    \centering\includegraphics{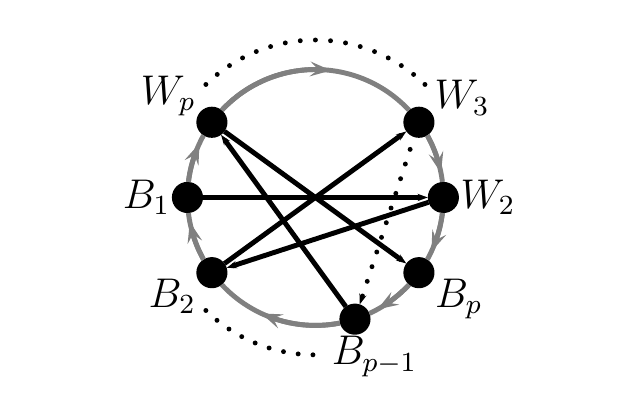}
    \caption{Ordering of the semiblocks $B_1, \ldots, B_p, W_2, \ldots, W_p$ inside $\Phi$.  The edges of $\overline{\G(\Gamma)}$ are shown.}\label{fig:high-degree}
  \end{figure}

  Rename the semiblocks of the above path to $B_1, \ldots, B_{2p-1}$. By definition, $B_1, B_2, B_4$ and $B_5$ induce a hole in $\MH$, thus $B$ is adjacent to at least one of these semiblocks by \Theorem~\ref{thm:forbiddens PCA}.  Let $a$ be the minimum such that $B_a$ is adjacent to $B$.  If $a \geq 4$ then $B_1, B_2, B_a$ and $B$ induce a $K_{1,3}$ in $\MH$, contradicting \Theorem~\ref{thm:forbiddens PCA}.  Consequently $a \leq 3$.  If $B$ is adjacent to $B_i$ for every even $i > a$ or for every odd $i > a$, then the result follows.  Hence,  suppose that $B$ has two non-neighbors $B_i$ and $B_j$, where $j-i > 0$ is odd and $i > a$.  Of all the possible combinations, take $i$ and $j$ so that $B$ is adjacent to $B_h$, for every $i < h < j$.  By construction, $B_i, \ldots, B_j, B$ is a hole of $\overline{\MH}$ with odd length, thus $B_a$ cannot be adjacent to all these semiblocks, by \Theorem~\ref{thm:forbiddens PCA}.  Consequently, $i = a+1$, and $h - j$ is even for every $j < h \leq p$ such that $B_h$ is not adjacent to $B$.  Therefore, $p \leq d_{\G(\Gamma)}(B) + 2$.
\end{proof}

\begin{lemma}\label{lem:at-most-three-coadj}
 If $B$ is a semiblock of a round graph $\MH$, then the non-universal semiblocks of $\MH \setminus \{B\}$ that are not adjacent to $B$ lie in at most two co-components of $\MH \setminus \{B\}$.
\end{lemma}

\begin{proof}
 On the contrary, suppose that there are three non-universal semiblocks $B_1, B_2, B_3$ that are not adjacent to $B$ and lie in different co-components of $\G = \MH \setminus \{B\}$.  Call $B_{i+3} \in V(\G)$ to a non-neighbor of $B_i$ for $i \in \{1,2,3\}$. 

 If $B$ is adjacent to $B_4, B_5$ and $B_6$, then $B_1, \ldots, B_6, B$ induce a subgraph isomorphic to $\overline{H_5}$ (see \Figure~\ref{fig:forbiddens PCA}) in $\MH$.  If $B$ is adjacent to $B_i$ and not to $B_j$, for $i,j \in \{4, 5, 6\}$, then $B_{j-3}, B_j, B$ and $B_i$ induce a $K_{1,3}$ in $\MH$.  Finally, if $B$ is not adjacent to $B_i$ and to $B_j$, for $4 \leq i < j \leq 6$, then $B_i, B_j, B_{i-3}, B_{j-3}, B$ induce a $C_4$ plus an isolated vertex. Whichever the case, $\MH$ is not a round graph by \Theorem~\ref{thm:forbiddens PCA}.
\end{proof}

These lemmas imply a lower bound for the degree of $B$ in $\MH$, as it follows from the next corollary.

\begin{corollary}\label{cor:co-components v-block complexity}
 Let $B$ be a semiblock of a round graph $\MH$, $\Phi$ be a base round representation of $\MH \setminus B$, and $s$ be the number of times that Algorithm~\ref{alg:co-bipartition implementation} is invoked when Algorithm~\ref{alg:co-components of PCA} is applied to\/ $\Phi$.  For $i = 1, \ldots, s$, denote by $p_i$ the number of iterations that the main loop of Algorithm~\ref{alg:co-bipartition implementation} requires for the $i$-th invocation.  If\/ $\Phi$ is $u$-universal, then \[s + \sum_{i=1}^s{p_i} \leq u + 2d_\MH(B) + 4.\]
\end{corollary}

\begin{proof}
 If $\Phi$ is not a circular contig or $\MH \setminus \{B\}$ is not co-bipartite, then $s = 0$ and the corollary is trivially true.  When $\Phi$ is a circular contig and $\MH \setminus \{B\}$ is co-bipartite, a new co-contig pair is found each time Algorithm~\ref{alg:co-components of PCA} invokes Algorithm~\ref{alg:co-bipartition implementation}. Thus, $\Phi$ contains at least $s$ co-contigs $\Gamma_1, \ldots, \Gamma_s$ that are found by invocations of Algorithm~\ref{alg:co-components of PCA} (and it contains other co-contigs not found by these invocations).  By relabeling the co-contigs if required, suppose $\Gamma_1, \ldots, \Gamma_u$ contain universal semiblocks.  Suppose also that $B$ is adjacent in $\MH$ to all the semiblocks of $\Gamma_{u+1}, \ldots \Gamma_{j}$, while $B$ is not adjacent to at least one semiblock in each one of $\Gamma_{j+1}, \ldots, \Gamma_s$.  

 By definition, $p_i = 0$ for every $1 \leq i \leq u$, while $p_i \leq d_{\G(\Gamma_i)}(B)$ for every $1 \leq i \leq j$.  On the other hand, \Lemma~\ref{lem:high-degree} implies that $p_i \leq d_\G(\Gamma_i)(B) + 2$ for every $j < i \leq s$, while \Lemma~\ref{lem:at-most-three-coadj} implies that $s - j \leq 2$.  Therefore \[\sum_{i=1}^s{(1+p_i)} \leq u + \sum_{i=u+1}^j{d_{\G(\Gamma_i)}(B)} + \sum_{i=j+1}^s{(2d_{\G(\Gamma_i)}(B) + 2)} \leq  u + 2d_\MH(B) + 4.\]
\end{proof}

Algorithm~\ref{alg:co-components v-block} takes a $u$-universal round representation $\Phi$ of $\MH \setminus \{B\}$ and $d_\MH(B)$, and outputs a natural ordering of the co-contig pairs of $\Phi$ or claims that $\MH$ is not a round graph.  The correctness of this algorithm follows from \Corollary~\ref{cor:co-components v-block complexity}, and its time complexity is $O(d_\MH(B)+u)$.  Note that the algorithm requires $\G(\Phi)$ to be co-bipartite and that it could output the co-contig pairs of $\Phi$ even when $\MH$ is not round.  

\begin{algorithm}
 \caption{Co-contigs of an incremental round representation $\Phi$}\label{alg:co-components v-block}

 \textbf{Input:} a base round representation $\Phi$ of a co-bipartite graph $\MH \setminus B$, and the number $d_{\MH}(B)$.

 \textbf{Output:} either a natural ordering of the co-contig pairs of $\Phi$, or a message indicating that $\MH$ is not round.

 \mbox{}

 \begin{AlgorithmSteps}
  \Step{Set $s := p := 0$.}
  \Step{Apply Algorithm~\ref{alg:co-components of PCA} while $s+p \leq 2d_\MH(B) + 4$.  For each invocation of Algorithm~\ref{alg:co-bipartition implementation} add $1$ to $s$ if the obtained co-contig has at least two semiblocks.  Similarly, for each iteration of the main loop of Algorithm~\ref{alg:co-bipartition implementation} add $1$ to $p$.}
  \Step{If $s+p > 2d_\MH(B) + 4$, then output an error message; otherwise, output the obtained co-contig pairs.}
 \end{AlgorithmSteps}
\end{algorithm}

\begin{lemma}\label{lem:incremental co-contigs}
  Let $\MH$ be a semiblock graph such that $\MH \setminus \{B\}$ is $u$-universal and co-bipartite, $B \in V(\MH)$, and\/ $\Phi$ be a base round representation of $\MH \setminus \{B\}$.  If\/ $\Phi$ and $d_\MH(B)$ are given as input, then a natural ordering of the co-contig pairs of\/ $\Phi$, or a message indicating that $\MH$ is not round, is obtained in $O(d_{\MH}(B)+u)$ time.
\end{lemma}

Algorithm~\ref{alg:co-components v-block} can be applied on $\Phi$ even when $\MH \setminus \{B\}$ is not co-bipartite.  In such case, Algorithm~\ref{alg:co-components v-block} halts in $O(d_\MH(B)+u)$ time claiming that $\MH$ is not round.  If $\MH$ is known to be round, such output is wrong, and it can be concluded that $\MH \setminus \{B\}$ is not co-bipartite.  Thus, the following lemma follows as well.

\begin{lemma}\label{lem:decremental co-contigs}
  Let $\MH$ be a round graph, $B \in V(\MH)$, and\/ $\Phi$ be a $u$-universal base round representation of $\MH \setminus \{B\}$.  If\/ $\Phi$ and $d_\MH(B)$ are given as input, then it takes $O(d_\MH(B)+u)$ time to determine whether $\MH \setminus \{B\}$ is co-bipartite.  Furthermore, when $\MH \setminus \{B\}$ is co-bipartite, a natural ordering of the co-contig pairs of\/ $\Phi$ is obtained.
\end{lemma}

\subsection{Split and join of co-contigs}\label{subsec:divide-and-conquer}

In this last part we present two algorithms that can be used to traverse all the round representations of co-bipartite round graphs.  These algorithms are based on the characterization of co-bipartite round graphs given in \Section~\ref{subsec:co-components PCA graphs}, and resemble the work by Huang~\cite{Huang1992}.

Let $\Phi$ be a round representation of a co-bipartite round graph, and $\Bip{\X_1}{\Y_1}, \ldots, \Bip{\X_s}{\Y_s}$ be a natural ordering of the co-contigs of $\Phi$.  For ranges $\X$ and $\Y$ of $\B(\Phi)$, say that $\Bip{\X}{\Y}$ is a \Definition{co-bipartition pair} of $\Phi$, and that $\X$ and $\Y$ are \Definition{co-bipartition ranges} of $\Phi$, when $\X = \X_i, \ldots, \X_j$ and $\Y = \Y_i, \ldots, \Y_j$, for $1 \leq i,j \leq s$.  If $\Bip{\X}{\Y}$ is a co-bipartition pair, then $\Phi|(\X \cup \Y)$ is the representation of $\Phi$ \Definition{described} by $\Bip{\X}{\Y}$. Note that, by definition, $\Phi|(\X \cup \Y)$ represents a subgraph $\G$ of $\G(\Phi)$ induced by several of its co-components.  In other words, $\G(\Phi) = \G + (\G(\Phi) \setminus V(\G))$.  Moreover, $\Bip{\X}{\Y}$ is a co-bipartition of $\G$.

Suppose $\Phi$ has a co-bipartition pair $\Bip{\X_1}{\X_3}$ not describing $\Phi$, and let $\Gamma = \Phi|(\X_1 \cup \X_3)$ and $\Lambda = \Phi \setminus (\X_1 \cup \X_3)$.  Observe that $\Lambda$ is also described by a co-bipartition pair inside $\Phi$.  To see why, suppose $\X_i \neq \emptyset$ ($i \in \{1,3\}$) and let $B_l$ and $B_r$ be its leftmost and rightmost semiblocks.  By definition, $\B(\Phi) = \X_i \Cat \X_{i+1} \Cat \X_{i+2} \Cat \X_{i+3}$, where $\X_{i+1} = (B_r, F_r(B_l)]$ and $\X_{i+3} = [F_l(B_r), B_l)$.  Hence, $\Lambda = \Phi | (\X_{i+1} \cup \X_{i+3})$ and $\Bip{\X_{i+1}}{\X_{i+3}}$ is a co-bipartition pair of $\Phi$ describing $\Lambda$.  The \Definition{split problem} consists of transforming $\Phi$ into $\Bip{\Gamma}{\Lambda}$, when $\Bip{\X_1}{\X_3}$ is given as input. 

Consider how do $\Gamma$ and $\Lambda$ look like inside $\Phi$.  For this, let $\Omega \in \{\Gamma, \Lambda\}$ be described by the co-bipartition range $\Bip{\X_i}{\X_{i+2}}$ ($1 \leq i \leq 4$), $B \in \X_i$, and $B_l^i$ and $B_r^i$ be the leftmost and rightmost semiblocks of $\X_i$.  Since $\Bip{\X_i}{\X_{i+2}}$ is a co-bipartition range, either $F_r^\Phi(B_r^i) = F_r^\Phi(B_l^i)$ or $F_r^\Phi(B_r^i) \in \X_{i+2}$.  Consequently,
\begin{equation}\label{eq:split gamma}
  F_r^\Omega(B) = 
   \begin{cases}
     B_r^i & \text{if } F_r^\Phi(B) = F_r^\Phi(B_l^i) \\
     F_r^\Phi(B) & \text{otherwise.} 
   \end{cases}
\end{equation}

Algorithm~\ref{alg:separation of G+H} solves the split problem for the case in which $\X_1$, $\X_2$, $\X_3$, and $\X_4$ are all nonempty.  The other cases are similar, and we omit them for the sake of simplicity.  

\begin{algorithm}[!ht]
 \caption{Split of a co-bipartition pair}\label{alg:separation of G+H}

 \textbf{Input:} a co-bipartition pair $\Bip{\X_1}{\X_3}$ of a base contig $\Phi$ such that $\X_1 = [B_l^1, B_r^1]$, $\X_3 = [B_l^3, B_r^3]$, $(B_r^1, B_l^3)$, and $(B_r^3, B_l^1)$ are all nonempty.

 \textbf{Output:} $\Phi$ is transformed into $\Phi|(\X_1 \cup \X_3)$ and $\Phi \setminus (\X_1 \cup \X_3)$.

 \mbox{}

 \begin{AlgorithmSteps}
 \Step{Let $B_l^{i+1} := R(B_r^i)$ and $B_r^{i-1} := L(B_l^i)$, for $i \in \{1,3\}$.}
 \Step{For $i \in \{1,2,3,4\}$, set $F_r(B) := B_r^{i+3}$ for every $B \in [F_l(B_r^{i+2}), B_l^{i+1})$.}\label{alg:separation of G+H: preemptive}
 \Step{For $i \in \{1,2,3,4\}$, simultaneously set $S_r(B_r^i) := S_r(B_r^{i+1})$.\label{alg:separation of G+H:step1}}
 \Step{For $i \in \{1,2,3,4\}$, set $F_l(B) := B_l^{i+3}$ for every $B \in (B_r^{i+3}, F_r(B_l^{i+2})]$.}\label{alg:separation of G+H: preemptive 2}
 \Step{For $i \in \{1,2,3,4\}$, simultaneously set $S_l(B_l^{i}) := S_l(B_l^{i-1})$.\label{alg:separation of G+H:step2}}
 \Step{Update the near pointers so that, for $i \in \{1,2,3,4\}$, $R(B_r^i) = B_l^{i+2}$.}
 \end{AlgorithmSteps}
\end{algorithm}

To discuss the correctness of Algorithm~\ref{alg:separation of G+H}, let $\Omega \in \{\Gamma, \Lambda\}$ be the round representation described by $\Bip{\X_i}{\X_{i+2}}$, $1 \leq i \leq 4$, and consider the effects of Steps \ref{alg:separation of G+H: preemptive}~and~\ref{alg:separation of G+H:step1} on $B \in \X_i$.  To begin, observe that $F_r(B_l^i) = B_r^{i+1}$ because $\X_{i+1} \neq \emptyset$.  If $F_r^\Phi(B) \not\in \{B_r^{i+1}, B_r^{i+2}\}$, then $F_r^\Omega(B) = F_r^\Phi(B)$ by (\ref{eq:split gamma}), and $F_r(B)$ is not changed by Steps \ref{alg:separation of G+H: preemptive}~and~\ref{alg:separation of G+H:step1}.  If $F_r^\Phi(B) = B_r^{i+1}$, then Step~\ref{alg:separation of G+H: preemptive} makes no changes to $F_r(B)$, thus Step~\ref{alg:separation of G+H:step1} sets $F_r(B) = B_r^i$, which is correct by (\ref{eq:split gamma}).  Finally, if $F_r^\Phi(B) = B_r^{i+2}$, then Step~\ref{alg:separation of G+H: preemptive} preemptively sets $F_r(B) = B_r^{i+3}$ so that, after Step~\ref{alg:separation of G+H:step1}, $F_r(B)$ references $B_r^{i+2}$ which is also correct by (\ref{eq:split gamma}).  Thus, the update of the right far pointers is correct. Similar arguments applied on $\Phi^{-1}$ are enough to conclude that Steps \ref{alg:separation of G+H: preemptive 2}--\ref{alg:separation of G+H:step2} correctly update all the left far pointers.  Therefore, Algorithm~\ref{alg:separation of G+H} is correct. 

With respect to the time complexity of Algorithm~\ref{alg:separation of G+H}, observe that the semiblocks in $[F_l(B_r^{i+2}), B_l^{i+1}) \cup (B_r^{i+3}, F_r(B_l^{i+2})]$ are all universal in $\G(\Phi)$.   That is, the semiblocks updated by Steps \ref{alg:separation of G+H: preemptive}~and~\ref{alg:separation of G+H: preemptive 2} are all universal in $\G(\Phi)$.  On the other hand, $R(B) = N_r(B)$ and $L(B) = N_l(B)$, for any $B \in \B(\Phi)$, because $\Phi$ has to be circular to be admitted as input of Algorithm~\ref{alg:separation of G+H}.  Consequently, Algorithm~\ref{alg:separation of G+H} requires $O(u)$ time when it is applied to $u$-universal contigs.  

\begin{lemma}\label{lem:separation of G+H}
  Let\/ $\Phi$ be a\/ $u$-universal base round representation that has a co-bipartition pair $\Bip{\X}{\Y}$ not describing $\Phi$.  If\/ $\Bip{\X}{\Y}$ is given as input, then\/ $\Phi$ can be transformed into\/ $\Phi|(\X\cup\Y)$ and\/ $\Phi\setminus(\X\cup\Y)$ in $O(u)$ time.
\end{lemma}

The \Definition{join problem} is the inverse of the split problem.  Roughly speaking, the goal of the join problem is to transform $\Bip{\Gamma}{\Lambda}$ back into $\Phi$.  However, contrary to the split problem, $\Gamma$ and $\Lambda$ can be joined in many ways that yield different representations of $\G(\Phi)$. 

Let $\Bip{\X_1}{\X_3}$ be a co-bipartition pair describing $\Gamma$, and $\Bip{\X_2}{\X_4}$ be a co-bipartition pair describing $\Lambda$.  For $1 \leq i \leq 4$, define $W_r^i$ to be the rightmost semiblock of $\X_{i} \Cat \X_{i+1}$.  Note that $W_r^i$ is well defined only if $\X_i \Cat \X_{i+1} = \emptyset$.  The \Definition{$\Bip{\X_1}{\X_2}$-join of $\Bip{\Gamma}{\Lambda}$} is the round representation $\Phi$ with $\B(\Phi) = \X_1 \Cat \X_2 \Cat \X_3 \Cat \X_4$ such that, for $B \in \X_i$ ($1 \leq i \leq 4$), the semiblock $F_r(B)$ is defined as follows. Suppose $\Bip{\X_i}{\X_{i+2}}$ describes the representation $\Omega \in \{\Gamma, \Lambda\}$, and let $B_l^i$ and $B_r^i$ be the leftmost and rightmost semiblocks of $\X_i$.  Then,
\[
  F_r^\Phi(B) = 
   \begin{cases}
     W_r^i          & \text{if } F_r^\Omega(B) = B_r^i \\
     F_r^\Omega(B)  & \text{otherwise.} \\
   \end{cases}
\]

The join problem consists of computing $\Phi$ when $\Bip{\X_1}{\X_3}$ and $\Bip{\X_2}{\X_4}$ are given as input.  It is not hard to see that $\Phi$ is indeed a round representation.  Furthermore, $\Bip{\X_1}{\X_3}$ and $\Bip{\X_2}{\X_4}$ are co-bipartition pairs of $\Phi$ representing $\Gamma$ and $\Lambda$, respectively.  In other words, $\Gamma$ can be split from $\Phi$ so as to generate back the pair $\Bip{\Gamma}{\Lambda}$.  From an algorithmic point of view, $\Phi$ is computed by reversing the effects that Algorithm~\ref{alg:separation of G+H} has when splitting $\Gamma$ from $\Phi$.  Hence, the join problem can be solved in $O(u)$ time when $\Gamma$ and $\Lambda$ are $u$-universal.

\begin{lemma}\label{lem:join of G+H}
  Let\/ $\Gamma$ and\/ $\Lambda$ be $u$-universal base round representations described by co-bipartition pairs $\Bip{\X}{\overline{\X}}$ and $\Bip{\Y}{\overline{\Y}}$, respectively.  If $\Bip{\X}{\overline{\X}}$ and $\Bip{\Y}{\overline{\Y}}$ are given as input, then it takes $O(u)$ time to transform\/ $\Gamma$ and\/ $\Lambda$ into the $\Bip{\X}{\Y}$-join of $\Bip{\Gamma}{\Lambda}$.
\end{lemma}

As previously mentioned, the $\Bip{\X_1}{\X_2}$-join of $\Bip{\Gamma}{\Lambda}$ is always a round representation $\Phi$ of $\G(\Gamma)+\G(\Lambda)$.  The join operation can be used to obtain different representation of $\G(\Phi)$.  Indeed, the $\Bip{\X_1}{\X_4}$-join of $\Bip{\Gamma}{\Lambda}$ is also a representation of $\G(\Phi)$ that needs not be equal to $\Phi$.  Similarly, the $\Bip{\X_1^{-1}}{\X_2}$-join of $\Bip{\Gamma^{-1}}{\Lambda}$ also represents $\G(\Phi)$.  Since every round representation admits a natural ordering by \Lemma~\ref{lem:natural ordering}, it turns out all the round representation of $\G(\Phi)$ can be obtained by joining the co-contig pairs of $\Phi$ in different ways.  Let $\bB = \Bip{\X_1}{\X_{s+1}}, \ldots, \Bip{\X_s}{\X_{2s}}$ be a natural ordering of the co-contigs pairs of $\Phi$, and denote $\X_i^1 = \X_i$ for $1 \leq i \leq 2s$.  Say that the ordering $\bW = \Bip{\Y_1}{\Y_{s+1}}, \ldots, \Bip{\Y_s}{\Y_{2s}}$ is a \Definition{natural permutation} of $\bB$ when there is a permutation $w$ of $\{1, \ldots, 2s\}$ and a mapping $y$ from $\{1, \ldots, 2s\}$ to $\{-1,1\}$ such that:
\begin{enumerate}[(i)]
  \item $w(s+i) = s+w(i)$ and $y(s+i) = s+y(i)$ for every $1 \leq i \leq s$, and
  \item $\Y_i = \X_{w(i)}^{y(i)}$ for every $1 \leq i \leq 2s$.
\end{enumerate}
The following characterization follows from \Lemma~\ref{lem:natural ordering} and the fact that the join operation always yields a round representation.

\begin{theorem}[see also \cite{Huang1992}]\label{thm:characterization round representations}
  Let\/ $\Phi$ be a round representations of a co-bipartite graphs, and\/ $\bB$ be a natural ordering of the co-contig pairs of\/ $\Phi$.  Then, $\Gamma$ is a round representation of $\G(\Phi)$ if and only if some natural permutation of\/ $\bB$ is also a natural ordering of\/ $\Gamma$.
\end{theorem}

The triplet $\langle \bB, w, y \rangle$ is referred to as the \Definition{$\bB$-encoding of\/ $\bW$}.  The following lemma shows how to traverse the round representations of $\G(\Phi)$ from $\Phi$, taking advantage of the $\bB$-encodings.

\begin{lemma}\label{lem:transformation of representations}
  Let\/ $\Phi$ and\/ $\Gamma$ be $u$-universal base round representations of a co-bipartite graph, and\/ $\bB$ and\/ $\bW$ be natural orderings of the co-contig pairs of\/ $\Phi$ and\/ $\Gamma$, respectively.  If a\/ $\bB$-encoding of\/ $\bW$ is given, then\/ $\Bip{\Phi}{\Phi^{-1}}$ can be transformed into\/ $\Bip{\Gamma}{\Gamma^{-1}}$ in $O(u|\bB|)$ time.  
\end{lemma}

\begin{proof}
  Let $\langle \bB, w, y \rangle$ be the $\bB$-encoding of $\bW$, $\bB = \Bip{\X_1}{\X_{s+1}}, \ldots, \Bip{\X_s}{\X_{2s}}$, and $\X_i^1 = \X_i$ ($1 \leq i \leq 2s$).  The algorithm is composed by two main steps.  The first step is to iteratively apply \Lemma~\ref{lem:separation of G+H} on $\Phi$ and its co-contig pairs so as to obtain the co-contig $\Phi_{i}^1 = \Phi|(\X_i \cup \X_{s+i})$ for $i = 1, \ldots, s$.  Similarly, each co-contig $\Phi_i^{-1}$ is obtained by applying \Lemma~\ref{lem:separation of G+H} on $\Phi^{-1}$.  The second step is to to obtain $\Gamma$ by the iterative application of \Lemma~\ref{lem:join of G+H}, as follows.  Let $\Y_i = \X^{y(i)}_{w(i)}$ and $\Lambda_i = \Phi^{y(i)}_{w(i)}$ for $i = 1, \ldots, 2s$.  By definition, $\bW = \Bip{\Y_1}{\Y_{s+1}}, \ldots, \Bip{\Y_s}{\Y_{2s}}$, while $\Lambda_i$ is the co-contig of $\Gamma$ described by $\Bip{\Y_i}{\Y_{s+i}}$.  Thus, initially $\Gamma_1 = \Lambda_1$.   The $i$-th time \Lemma~\ref{lem:join of G+H} is applied, the $\Bip{\Y_1 \Cat \ldots \Cat \Y_i}{\Y_{i+1}}$-join of $\Bip{\Gamma_i}{\Lambda_{i+1}}$, called $\Gamma_{i+1}$, is computed.  At the end, $\Gamma = \Gamma_s$.  Analogously, \Lemma~\ref{lem:join of G+H} is applied to join the remaining co-contigs into $\Gamma^{-1}$.  Just observe that $\Bip{\Y_s^{-1}}{\Y_{2s}^{-1}}, \ldots, \Bip{\Y_1^{-1}}{\Y_{s+1}^{-1}}$ is a natural ordering of $\Gamma^{-1}$.

  By \Lemmas \ref{lem:separation of G+H}~and~\ref{lem:join of G+H}, $O(us)$ time is required by this algorithm if $w$ and $y$ are stored in such a way that, for $1 \leq i \leq 2s$, $w(i)$ and $y(i)$ take $O(1)$ time.
\end{proof}

\Lemma~\ref{lem:transformation of representations} can be used to traverse all the round representations of any co-bipartite round graph.  In this article, though, we use it only when a vertex is to be inserted into a co-connected graph.  In such case, only $O(1)$ round representations need to be traversed.  The lemma below shows the traversal algorithm; we emphasize that some round representations could be traversed several times by this algorithm.

\begin{lemma}\label{lem:traversal of representations}
  Let $\MH$ be a co-connected semiblock graph such that $\MH \setminus \{B\}$ is co-bipartite, $B \in V(\MH)$, and\/ $\Phi$ be a base round representation of $\MH \setminus \{B\}$.  If\/ $\Phi, \Phi^{-1}$ and $d_\MH(B)$ are given as input, then the family of round representations $\Gamma, \Gamma^{-1}$ of $\G(\Phi)$ can be traversed, or a message indicating that $\MH$ is not round can be obtained, in $O(d_{\MH}(B))$ time.
\end{lemma}

\begin{proof}
  The algorithm is composed by a preprocessing phase and a main loop.  The preprocessing phase begins applying  \Lemma~\ref{lem:incremental co-contigs}, with input $\Phi$ and $d_\MH(B)$, so as to find a natural ordering $\bB$ of the co-contigs of $\Phi$.  Recall that a message indicating that $\MH$ is not round can be obtained.  In such case, this message is forwarded and the algorithm is halted.  On the other hand, if $\bB$ is obtained, then the preprocessing phase continues with an evaluation of $|\bB|$.  By \Lemma~\ref{lem:at-most-three-coadj}, if $|\bB| > 3$, then $\MH$ is not round.  Thus, the algorithm is halted with a message indicating that $\MH$ is not round when $|\bB| > 3$.  When $|\bB| \leq 3$, the preprocessing phase finishes and the main loop begins.  The main loop traverses all the $\bB$-encodings while it generates the round representations of $\G(\Phi)$.  Suppose $\langle \bB, w, y \rangle$ is the next $\bB$-encoding to be traversed, and let $\bW$ be the natural permutation encoded by $\langle \bB, w, y \rangle$.  Initially, \Lemma~\ref{lem:transformation of representations} is applied with input $\langle \bB, w, y \rangle$, $\Phi$, and $\Phi^{-1}$.  As a result, $\Bip{\Phi}{\Phi^{-1}}$ is transformed into round representations $\Bip{\Gamma}{\Gamma^{-1}}$ of $\G(\Phi)$, that are provided to the invoking procedure for its traversal.  When the invoking procedure asks for a new pair of representations, $\Bip{\Gamma}{\Gamma^{-1}}$ is transformed back to $\Bip{\Phi}{\Phi^{-1}}$, and then a new $\bB$-encoding is processed.  For the transformation of $\Bip{\Gamma}{\Gamma^{-1}}$ into $\Bip{\Phi}{\Phi^{-1}}$, notice that $\bB$ is the natural permutation of $\bW$ encoded by $\langle \bW, w^{-1}, y \circ w^{-1}\rangle$.  Thus, \Lemma~\ref{lem:transformation of representations} is applied with input $\langle \bW, w^{-1}, y \circ w^{-1}\rangle$.

  By \Lemmas \ref{lem:incremental co-contigs}~and~\ref{lem:transformation of representations}, taking into account that the main loop is executed only when $|\bB| = O(1)$, the above algorithm requires $O(d_{\MH}(B))$ time, as desired.
\end{proof}

\section{Incremental recognition of proper circular-arc graphs}
\label{sec:incremental}

In this section we describe the algorithms that make up the incremental recognition algorithm of PCA graphs, namely the insertion of a new vertex (\Section~\ref{subsec:vertex insertion}) or a new edge (\Section~\ref{subsec:edge insertion}).  In this section, each contig $\Phi$ is implemented as an augmented base contig in which:
\begin{itemize}
  \item every block $B$ is associated with an \Definition{end pointer} $E^{\Phi}(B)$ such that $E^{\Phi}(B) = NULL$ if $B$ is not an end block, while $E^{\Phi}(B)$ references the other end block of the contig containing $B$ otherwise,
  \item there is a \Definition{co-bipartite pointer} $CB^\Phi$ referencing a left co-end block of $\Phi$.  If $\G(\Phi)$ has a universal block $B$, then $CB^\Phi = B$, while if $\G(\Phi)$ is not co-bipartite, then $CB^\Phi = NULL$.
\end{itemize}
As usual, we omit the superscript when no confusions arise.  The end pointers are used and maintained by the HSS algorithm while $\Phi$ is linear.  When $\Phi$ is circular, the end pointers are not required (indeed, they are all null).  However, they must be maintained in case an edge insertion yields a linear contig.  On the other hand, the co-bipartite pointer is unknown by the HSS algorithm, and it is required for the insertion of edges.  After each modification done by the HSS algorithm, the co-bipartite pointer needs to be updated.  We write that $\Phi$ is an \Definition{incremental contig} to emphasize that $\Phi$ is a base contig augmented with end and co-bipartite pointers.  Similarly, an \Definition{incremental round representation} is a base round representation $\Phi$ whose contigs are incremental contigs.  

As mentioned in \Section~\ref{sec:dyn-pca data structure}, two incremental round block representations $\Phi, \Phi^{-1}$, both satisfying the straightness property, are stored to represent an incremental graph $G$.  Recall that, by the straightness property, either $\Phi$ is straight or $G$ is not a PIG graph.  In the former case, the HSS algorithms can be applied on $\Phi$ and $\Phi^{-1}$.

\subsection{The impact of a new vertex}
\label{subsec:vertex insertion}

To begin this section, we show how to insert a new vertex into a PCA graph.  Given a vertex $v$ of a graph $H$ and two round block representations $\Phi$ and $\Phi^{-1}$ of $H \setminus \{v\}$, we ought to update $\Phi$ and $\Phi^{-1}$ into round block representations $\Psi$ and $\Psi^{-1}$ of $H$.  In this part we mostly deal with the case in which $\Psi$ is a circular contig, though $\Phi$ can be a linear contig, because the other case is solved, with exception of the co-bipartite pointer, by the HSS algorithm.  

Let $B$ be a semiblock of $H \setminus \{v\}$.  Say that $v$ is \Definition{adjacent} to $B$ when $B \cap N(v) \neq \emptyset$, while $v$ is \Definition{co-adjacent} to $B$ when $B \setminus N(v) \neq \emptyset$.  In other words, $v$ is adjacent to $B$ if $v$ has some neighbor in $B$, while it is co-adjacent if it has some non-neighbor in $B$.  When $v$ is adjacent to all the vertices in $B$, then $v$ is \Definition{fully adjacent} to $B$.  Similarly, when $v$ is adjacent to none of the vertices in $B$, then $v$ is \Definition{not adjacent} to $B$.  Observe that $v$ is fully adjacent to $B$ if and only if $v$ is not co-adjacent to $B$.  

For any contig $\Gamma$ representing $H \setminus \{v\}$, say that $v$ is \Definition{insertable into\/ $\Gamma$} when $v$ is adjacent to two semiblocks $B_a \neq B_b$ of $\B(\Gamma)$ such that:
\begin{enumerate}[(i)]
  \item $v$ is fully adjacent to all the semiblocks in $(B_a, B_b)$,
  \item $v$ is not adjacent to the semiblocks in $(B_b, B_a)$, and 
  \item $\Bip{B_a \cap N(v)}{B_b \cap N(v)}$ is refinable in $\Gamma$.
\end{enumerate}
When $v$ is insertable into $\Gamma$, the $\{v\}$-refinement of $\Bip{B_a \cap N(v)}{B_b \cap N(v)}$ in $\Gamma$ is referred to as an \Definition{insertion of $v$ into\/ $\Gamma$}.  By definition, $\Psi$ is  an insertion of $v$ into $\Gamma$ only if $\Psi$ is circular.  Furthermore, $\Psi$ is the unique insertion of $v$ into $\Gamma$ unless $v$ is a universal vertex of $H$.  The insertion of a vertex is, in some sense, the inverse of the compressed removal.

\begin{observation}
  $\Psi$ is an insertion of $v$ into\/ $\Gamma$ if and only if\/ $\Gamma$ is the compressed removal of\/ $\{v\}$ from\/ $\Psi$.
\end{observation}

A proof of the above observation is implicit in the the following lemma, that highlights the importance of insertable contigs.

\begin{lemma}\label{lem:insertable contig}
  Let $H$ be a\/ $0$-universal graph that is not PIG, and $v \in V(H)$.  Then, $H$ is a PCA graph if and only if $H \setminus \{v\}$ admits a block contig\/ $\Gamma$ into which $v$ is insertable.  Furthermore, if $v$ is insertable into\/ $\Gamma$, then the insertion of $v$ into\/ $\Gamma$ is a block contig representing $H$. 
\end{lemma}

\begin{proof}
  Suppose $H$ is a PCA graph and let $B$ be the block of $H$ that contains $v$.  Since $H$ is not a PIG graph, it is represented by some circular block contig $\Psi$.  Let $\Gamma$ be the compressed removal of $v$ from $\Psi$, and $B_a$ and $B_b$ be the semiblocks of $\Gamma$ containing $F_l^\Psi(B)$ and $F_l^\Psi(B)$, respectively.  Observe that $\Gamma$ is $1$-universal, thus, by \Corollary~\ref{cor:indistinguishable}, $\Gamma$ is a block contig representing $H \setminus \{v\}$.  On the other hand, by definition, $v$ is adjacent to $B_a$ and $B_b$ and: (i) $v$ is fully adjacent to all the blocks in $(B_a, B_b)$, (ii) $v$ is adjacent to no block in $(B_b, B_a)$, and (iii) $\Psi$ is the $\{v\}$-refinement of $\Bip{B_a \cap N(v)}{B_b \cap N(v)}$ in $\Gamma$.  That is, $v$ is insertable into $\Gamma$.

  For the converse, observe that the insertion $\Psi$ of $v$ into $\Gamma$ is $0$-universal and compressed.  Therefore, by \Corollary~\ref{cor:indistinguishable}, $\Psi$ is a block contig.  Moreover, by definition, if $B\in \Psi$ is the block containing $v$, then $N[B]$ equals the range $[B_a \cap N(v), B_b \cap N(v)]$ of $\Psi$.  Consequently, $\Psi$ represents $H$.
\end{proof}

Algorithm~\ref{alg:vertex insertion co-connected} can be used to test if $v$ is insertable into $\Gamma$.  Furthermore, if $v$ is insertable, then $\Gamma$ gets transformed in the insertion of $v$ into $\Gamma$.  Its correctness follows by definition. 

\begin{algorithm}[!htb]
 \caption{Insertion of an insertable vertex into a contig.}\label{alg:vertex insertion co-connected}

 \textbf{Input:} $N_H(v)$ and an incremental contig $\Gamma$ representing a graph $H \setminus \{v\}$ such that $H$ is not PIG.

 \textbf{Output:} if $v$ is insertable into $\Gamma$, then $\Gamma$ is updated into the incremental insertion of $v$ into $\Gamma$; otherwise, the algorithm halts in error.

 \mbox{}

 \begin{AlgorithmSteps}
  \Step{Let $\N$ and $\F$ be the families containing the semiblocks adjacent and fully adjacent to $v$, respectively.}\label{alg:vertex insertion co-connected:N and F}
  \Step{If $\N$ is not a range of $\B(\Gamma)$, then halt in error.}\label{alg:vertex insertion co-connected:range}
  \Step{Let $B_a$ and $B_b$ be the leftmost and rightmost semiblocks in $\N$.}
  \Step{If $(B_a, B_b) \not\subseteq \F$, then halt in error.}\label{alg:vertex insertion co-connected:B_a B_b in F}
  \Step{Let $B_l = B_a \cap N_H(v)$ and $B_r = B_b \cap N_H(v)$.}
  \Step{Transform $\Gamma$ into the $\{v\}$-refinement of $\Bip{B_l}{B_r}$ if possible, and halt in error otherwise.}\label{alg:vertex insertion co-connected:refinement}
  \Step{Set $E(B) = NULL$ for every $B \in \B(\Gamma)$.}\label{alg:vertex insertion co-connected:end pointers}
  \Step{If $\G(\Gamma)$ is co-bipartite, then let $CB^\Gamma$ reference a left co-end block of $\Gamma$.}\label{alg:vertex insertion co-connected:co-bipartite pointer}
 \end{AlgorithmSteps}
\end{algorithm}

Algorithm~\ref{alg:vertex insertion co-connected} is implemented in a way rather similar to the HSS algorithm.  To compute $\N$ and $\F$, each vertex $w \in N_H(v)$ is traversed so as to find the semiblock $B$ containing $w$.  If $w$ is the first traversed vertex of $B$, then $B$ is inserted into $\N$, while if all the vertices of $B \setminus \{w\}$ were already traversed, then $B$ is inserted into $\F$.  Thus, Step~\ref{alg:vertex insertion co-connected:N and F} takes $O(d_H(v))$ time.  For Step~\ref{alg:vertex insertion co-connected:range}, select a semiblock in $\N$ and traverse $\B(\Phi)$ to the left until the last semiblock $B_a$ in $\N$ is found.  Next, traverse $\B(\Phi)$ to the right until the last semiblock $B_b$ in $\N$ is found.  The range $(B_a, B_b)$ must contain all the semiblocks in $\N$ to satisfy the condition of Step~\ref{alg:vertex insertion co-connected:range}.  For Step~\ref{alg:vertex insertion co-connected:B_a B_b in F}, we ought to traverse $(B_a, B_b)$ so as to find if there is some semiblock outside $\F$.  Similarly, for Step~\ref{alg:vertex insertion co-connected:refinement}, $[B_a, B_b]$ is traversed to determine that it contains at most one universal semiblock.  If not, then, by definition, $\Bip{B_l}{B_r}$ is not refinable in $\Gamma$, and the algorithm halts. Otherwise, $\Gamma$ is tested to be refinable in $O(d_H(v))$ time by invoking \Lemma~\ref{lem:refinable complexity}.  The update of the end pointers in Step~\ref{alg:vertex insertion co-connected:end pointers} is done by examining only $O(1)$ semiblocks of $\Psi$.  Indeed, by \Lemma~\ref{lem:semiblock insertion}, $N_l(B)$ and $N_r(B)$ are the only possible end semiblocks of\/ $\Psi \setminus \{B\}$.  Hence, the only possible end pointers of $\Phi$ not referencing $NULL$ at this point correspond to the semiblocks in \{$N_l(\{v\})$, $N_r(\{v\})$, $N_l(N_l(\{v\}))$,$N_r(N_r(\{v\}))$\} (the end pointers of the last two could be non-null when either $N_l(v)$ and $N_l(N_l(\{v\}))$ or $N_r(v)$ and $N_r(N_r(\{v\}))$ were separated).  Finally, for Step~\ref{alg:vertex insertion co-connected:co-bipartite pointer}, just apply Algorithm~\ref{alg:co-bipartition implementation} to the semiblock containing $v$.  Summing up, Algorithm~\ref{alg:vertex insertion co-connected} takes $O(d_H(v))$ time.  

\begin{lemma}\label{lem:insertion of insertable}
  Let $H$ be a non-PIG graph, $v \in V(H)$, and\/ $\Gamma$ be an incremental contig representing $H \setminus \{v\}$.  If\/ $\Gamma$ is given as input, then it takes $O(d_H(v))$ to determine if $v$ is insertable into\/ $\Gamma$. Furthermore, if $v$ is insertable into\/ $\Gamma$, then the incremental insertion of $v$ into\/ $\Gamma$ is also obtained in $O(d_H(v))$ time.
\end{lemma}

\Lemmas \ref{lem:insertable contig}~and~\ref{lem:insertion of insertable} reduce the problem of inserting $v$, when $H$ is $0$-universal, to the problem of finding a block contig representing $H \setminus \{v\}$ into which $v$ is insertable.   The following lemma discusses how can such block contig be found when $H$ is co-connected.

\begin{lemma}\label{lem:add-vertex-co-connected}
  Let $H$ be a co-connected non-PIG graph, $v \in V(H)$, and\/ $\Phi$ be an incremental block contig of $H \setminus \{v\}$.  If\/ $\Phi, \Phi^{-1}$ and $N_H(v)$ are given as input, then it takes $O(d_H(v))$ to determine whether $H$ is a PCA graph.  Furthermore, if $H$ is a PCA graph, then two incremental block contigs $\Psi, \Psi^{-1}$ representing $H$ can be obtained in $O(d_H(v))$ time. 
\end{lemma}

\begin{proof}
  The algorithm works by traversing all the block contigs $\Gamma, \Gamma^{-1}$ that represent $H \setminus \{v\}$.  For each $\Gamma$, $v$ is queried to be insertable into $\Gamma$, using \Lemma~\ref{lem:insertion of insertable} with input $\Gamma$ and $N_H(v)$.  By definition, $v$ is insertable into $\Gamma$ if and only if $v$ is insertable into $\Gamma^{-1}$.  Thus, by \Lemma~\ref{lem:insertable contig}, $H$ is PCA if and only if $v$ is insertable into one such $\Gamma$.  Furthermore, the insertions of $v$ into $\Gamma$ and $\Gamma^{-1}$ are one the reverse of the other.  Consequently, by \Lemma~\ref{lem:insertable contig}, the furthermore part is fulfilled by taking $\Psi$ and $\Psi^{-1}$ as the insertions of $v$ in $\Gamma$ and $\Gamma^{-1}$, respectively.  

  Two cases are considered for the traversal of the block contigs  $\Gamma, \Gamma^{-1}$ that represent $H \setminus \{v\}$.  First, if $H \setminus \{v\}$ is not co-bipartite, then $\Phi$ and $\Phi^{-1}$ are the unique block contigs representing $H \setminus \{v\}$.  Thus, $\Phi, \Phi^{-1}$ are the only traversed block contigs.  On the other hand, if $H \setminus \{v\}$ is co-bipartite, then \Lemma~\ref{lem:traversal of representations} is applied on $\Phi, \Phi^{-1}$ and $d_H(v)$.  If \Lemma~\ref{lem:traversal of representations} outputs a message indicating that $H$ is not PCA, then no contig is traversed and the algorithm halts indicating that $H$ is not PCA.

  Recall that $H \setminus \{v\}$ is co-bipartite if and only if $CB^\Phi \neq NULL$; thus it takes $O(1)$ time to determine if $H \setminus \{v\}$ is co-bipartite.  By \Lemma~\ref{lem:traversal of representations}, $O(d_H(v))$ time is required to traverse all the block contigs representing $H \setminus \{v\}$.  Only $O(1)$ block contigs representing $H \setminus \{v\}$ are traversed, and querying whether $v$ is insertable into each block contig takes $O(d_H(v))$ time, by \Lemma~\ref{lem:insertion of insertable}.  Therefore, the described algorithm takes $O(d_H(v))$ time.
\end{proof}

The remaining case is when $H$ is not co-connected, which can be solved by observing the following corollary of \Theorem~\ref{thm:forbiddens PCA}.

\begin{corollary}\label{cor:add-vertex-not-co-connected}
  Let $H$ be a graph that is not co-connected, and $H_v$ be the subgraph of $H$ induced by the co-component containing a given vertex $v$.  Then, $H$ is a PCA graph if and only if $H \setminus V(H_v)$ and $H_v$ are co-bipartite PCA graphs.
\end{corollary}

Algorithm~\ref{alg:vertex insertion} can be used to insert $v$ into the block contigs $\Phi, \Phi^{-1}$ representing $G$.  The algorithm works as follows.  First, Steps \ref{alg:vertex insertion:universal separation begin}--\ref{alg:vertex insertion:split 2} split $\Phi$ into two round block representations $\Phi_v$ and $\Lambda$ so that $G(\Phi_v) \cup \{v\}$ is the subgraph of $H$ induced  by the co-component of $H$ containing $v$.  Similarly $\Phi^{-1}$ is split into $\Phi^{-1}_v$ and $\Lambda^{-1}$.  For the sake of simplicity, the algorithm allows round representations and graphs to be empty.  So, for instance, $\Phi_v$ is empty when $v$ is universal in $H$, while $\Lambda$ is empty when $H$ is co-connected.  Second, Step~\ref{alg:vertex insertion:psi_v} test whether $G(\Phi_v) \cup \{v\}$ is a PCA graph.  If negative, then $H$ is not a PCA graph, while if affirmative, then there are two possibilities.  If $\Lambda = \emptyset$, then $H = G(\Phi_v) \cup \{v\}$ is already known to be a PCA graph.  Otherwise, by \Corollary~\ref{cor:add-vertex-not-co-connected}, $H$ is PCA if and only if $G(\Phi_v)$ is co-bipartite. Step~\ref{alg:vertex insertion:co-bipartite} checks if $G(\Phi_v)$ is co-bipartite when $\Lambda \neq \emptyset$.  Thus, Algorithm~\ref{alg:vertex insertion} correctly determines that $H$ is PCA.  On the other hand, if $H$ is PCA, then $\Phi$ and $\Phi^{-1}$ are correctly updated into $\Psi$ and $\Psi^{-1}$.

\begin{algorithm}[htb!]
 \caption{Insertion of a vertex $v$ into a block contig.}\label{alg:vertex insertion}

 \textbf{Input:} incremental block contigs $\Phi, \Phi^{-1}$ that satisfy the straightness property and represent a graph $H \setminus \{v\}$, and the set $N_H(v)$.

 \textbf{Output:} if $H$ is a PCA graph, then $\Phi, \Phi^{-1}$ are updated into incremental block contigs $\Psi, \Psi^{-1}$ that satisfy the straightness property and represent $H$; otherwise, the algorithm halts in error.

 \mbox{}

 \begin{AlgorithmSteps}
  \Step{If $v$ is adjacent to the universal block $B$ of $\G(\Phi)$, then:}\label{alg:vertex insertion:universal separation begin}
  \IncreaseIndent
    \Step{Separate $B$ into $\Bip{B \cap N(v)}{B \setminus N(v)}$ in $\Phi$.}\label{alg:vertex insertion:universal separation 1}
    \Step{Separate $B$ into $\Bip{B \setminus N(v)}{B \cap N(v)}$ in $\Phi^{-1}$.}\label{alg:vertex insertion:universal separation end}
  \DecreaseIndent
  \Step{Split $\Phi$ into $\Phi_v$ and $\Lambda = \Phi\setminus \B(\Phi_v)$ so that $G(\Phi_v) \cup \{v\}$ is co-connected and $v$ is universal in $G(\Lambda) \cup \{v\}$.}\label{alg:vertex insertion:split 1}
  \Step{Split $\Phi^{-1}$ into $\Phi_v^{-1}$ and $\Lambda^{-1}$.}\label{alg:vertex insertion:split 2}
  \Step{Determine if $G(\Phi_v) \cup \{v\}$ is a PCA graph.  If false, then halt in error; otherwise, transform $\Phi_v$ and $\Phi_v^{-1}$ into incremental block contigs $\Psi$ and $\Psi^{-1}$ representing $G(\Phi_v) \cup \{v\}$ that satisfy the straightness property.}\label{alg:vertex insertion:psi_v}
  \Step{If $\B(\Lambda) \neq \emptyset$, then:}
  \IncreaseIndent
    \Step{If $G(\Psi)$ is not co-bipartite, then halt in error.}\label{alg:vertex insertion:co-bipartite}
    \Step{Join $\Lambda$ into $\Psi$ and $\Lambda^{-1}$ into $\Psi^{-1}$ keeping the straightness property.}\label{alg:vertex insertion:join}
  \DecreaseIndent
  \Step{Output $\Psi$ and $\Psi^{-1}$.}\label{alg:vertex insertion:output}
 \end{AlgorithmSteps}
\end{algorithm}

Discuss the implementation and the time complexity of Algorithm~\ref{alg:vertex insertion}.  Accessing the co-bipartite pointer, the condition of Step~\ref{alg:vertex insertion:universal separation begin} takes $O(1)$ time.  For Steps \ref{alg:vertex insertion:universal separation 1}~and~\ref{alg:vertex insertion:universal separation end}, \Lemma~\ref{lem:semiblock separation} is applied on $B$ and $N(v)$, for both $\Phi$ and $\Phi^{-1}$, thus $O(d_H(v))$ time is consumed.  Before executing Step~\ref{alg:vertex insertion:split 1}, \Lemma~\ref{lem:incremental co-contigs} is applied on $\Phi$ and $d(v)$.  If a message indicating that $H$ is not PCA is obtained, then Algorithm~\ref{alg:vertex insertion} halts in error; otherwise, a natural ordering of the co-contig pairs of $\Phi$ is obtained.  Similarly, the co-contig pairs of $\Phi^{-1}$ are obtained.  Then, Step~\ref{alg:vertex insertion:split 1} is executed as follows.  First, the blocks fully-adjacent to $v$ are traversed so as to compute the family $\bB = \{\Bip{\X_1}{\Y_1}, \ldots, \Bip{\X_r}{\Y_r}\}$ of co-contig pairs containing blocks co-adjacent to $v$.  Such traversal requires $O(d_H(v))$ time if it is executed as in Algorithm~\ref{alg:vertex insertion co-connected}.  Second, \Lemma~\ref{lem:separation of G+H} is applied to each of the co-contig pairs of $\bB$, to split $\Phi$ into $\Lambda$ and $\Phi_i = \Phi|(\X_i \cup \Y_i)$, for $1 \leq i \leq r$.  Finally, as in the proof of \Lemma~\ref{lem:transformation of representations}, \Lemma~\ref{lem:join of G+H} is applied $r-1$ times so as to join $\Phi_1, \ldots, \Phi_r$ into $\Phi_v$.  By \Lemmas~\ref{lem:incremental co-contigs}, \ref{lem:separation of G+H}~and~\ref{lem:join of G+H}, Step~\ref{alg:vertex insertion:split 1} requires $O(d_H(v))$ time.  A similar procedure is applied for Step~\ref{alg:vertex insertion:split 2}.  Let $N_v(v)$ be the set of neighbors of $v$ in $G(\Phi_v)$.  For Step~\ref{alg:vertex insertion:psi_v}, first it is queried whether $G(\Phi_v) \cup \{v\}$ is a PIG graph or not.  This query takes $O(d_H(v))$ time, because $G(\Phi_v) \cup \{v\}$ is a PIG graph if and only if either $\Phi_v = \emptyset$ or $\Phi_v$ is straight and the HSS algorithm applied on $\Phi_v$, $\Phi_v^{-1}$ and $N_v(v)$ is successful.  Then, Step~\ref{alg:vertex insertion:psi_v} takes one of three paths according to the result of the query.  In the first case, $G(\Phi_v) \cup \{v\}$ is a PIG graph and $\Psi$ and $\Psi^{-1}$ are obtained as a byproduct of the execution of the HSS algorithm (the case $\Phi_v = \emptyset$ is trivial).  Clearly, $\Psi$ and $\Psi^{-1}$ satisfy the straightness property, and their co-bipartite pointers can be updated in $O(1)$ time with \Lemma~\ref{lem:cobipartite-linear}.  The second case applies when $G(\Phi_v) \cup \{v\}$ is not PIG and $\Phi_v$ is not a contig.  In this case, $G(\Phi_v) \cup \{v\}$ is not a PCA graph, so neither is $H$, and the algorithm is halted.  The third case is when $G(\Phi_v) \cup \{v\}$ is not PIG and $\Phi_v$ is a contig.  In this case, \Lemma~\ref{lem:add-vertex-co-connected} is applied on $\Phi_v$, $\Phi_v^{-1}$, and $N_v(v)$.  Since $G(\Phi_v) \cup \{v\}$ is not PIG, the obtained contigs satisfy the straightness property.  Whichever paht is taken by Algorithm~\ref{alg:vertex insertion}, Step~\ref{alg:vertex insertion:psi_v} takes $O(d_H(v))$ time.  Step~\ref{alg:vertex insertion:co-bipartite} takes $O(1)$ time with the co-bipartite pointer of $\Psi$.  Finally, Step~\ref{alg:vertex insertion:join} takes $O(1)$ time by executing \Lemma~\ref{lem:join of G+H} on co-bipartition pairs of $\Psi$ and $\Lambda$, and on the corresponding co-bipartition pairs of $\Psi^{-1}$ and $\Lambda^{-1}$.  With respect to the straightness property, note that $H$ is a PIG graph only if both $\Psi$ and $\Gamma$ are co-contigs.  So, it takes $O(1)$ time to decide how their co-contig pairs should by joined so as to satisfy the straightness property.  Summing up, Algorithm~\ref{alg:vertex insertion} takes $O(d_H(v))$ time.                                                                                                                                                                                                                                                                                                                                                                                                                                                                                                                                                                                                                                                                                                                                                                                                                                                                                                                                                                                                                                                                                                                                                       

\begin{theorem}\label{thm:vertex insertion}
  The problem of deciding whether an incremental graph is a PCA graph takes $O(1)$ time per inserted edge, when only the insertion of vertices is allowed.
\end{theorem}

\Theorem~\ref{thm:vertex insertion} solves the following problems implicitly posed in~\cite{DengHellHuangSJC1996}.  First, can proper circular-arc graphs be recognized in linear time by an incremental algorithm?  Second, such an incremental algorithm, follows the same ideas as the DHH algorithm?

\subsection{The impact of a new edge}
\label{subsec:edge insertion}

In this part we complete the incremental PCA recognition algorithm by showing how to insert an edge in constant time, whenever possible.  That is, given two vertices $v, w$ of a graph $G$ that admits a round block representation $\Phi$, the problem is to transform $\Phi, \Phi^{-1}$ into round block representations $\Psi, \Psi^{-1}$ of $H = G \cup \{vw\}$.  This problem has already been addressed by the HSS algorithm for the case in which $G$ is disconnected.  Let $B$ and $W$ be the blocks of $\Phi$ that contain $v$ and $w$, respectively.  For the remaining cases, we prove that $H$ is PCA if only if either (i) $B$ and $W$ are connectable, or (ii) $B$ and $W$ are \Definition{almost-connectable}, or (iii) one of $v,w$ is universal in $H$ and $G$ is co-bipartite.  

\begin{figure}[htb!]
 \centering
  \begin{tabular}{c@{\hspace{1cm}}c@{\hspace{1cm}}c}
    \includegraphics{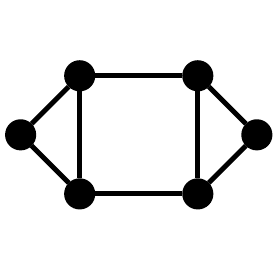} & \includegraphics{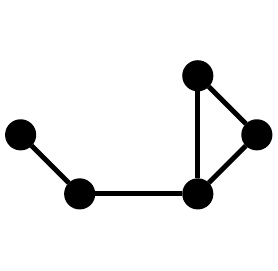} & \includegraphics{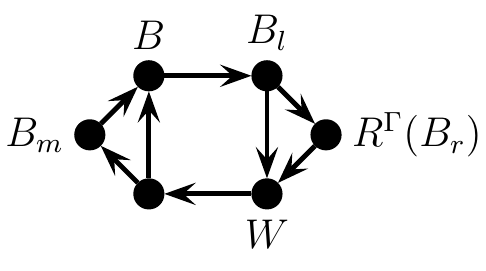} \\
    (a) & (b) & (c)
  \end{tabular}

 \caption{(a) a co-domino graph, (b) a co-P graph, and (c) $\Tophi$ for a round representation $\Phi$ with $\G(\Phi)$ isomorphic to a co-domino graph.}\label{fig:co-domino}
\end{figure}

In \Section~\ref{sec:basic manipulation} we defined $\Bip{B}{W}$ to be connectable when their vertices can be connected without affecting the order of $\B(\Phi) \setminus \{B,W\}$.  We define $\Bip{B}{W}$ to be almost-connectable when their vertices can be connected after slightly modifying the order of $\B(\Phi) \setminus \{B, W\}$.  Co-domino and co-P graphs are required for this definition; see \Figure~\ref{fig:co-domino}.  

Let $\Phi$ be a block contig representing a graph $\G$ that has two non-adjacent blocks $B, W$.  Define  $\G_B$ to be the co-component of $\G \setminus \{W\}$ that contains $B$, $\G_{BW} = \G[V(\G_v) \cup W]$, and $\G_W = \G_{BW} \setminus \{B\}$.  Say that $\Bip{B}{W}$ is \Definition{almost-connectable} when $|B| = |W| = 1$, $\G_B$ and $\G_W$  are isomorphic to co-P graphs, and $\G_{BW}$ is isomorphic to co-domino graph.  We claim that $\G(\Phi) \cup \{vw\}$ is a PCA graph when $\Bip{B}{W}$ is almost-connectable.  To see why, we show how to build a \Definition{connection of $\Bip{B}{W}$ in\/ $\Phi$}.

By definition, $B$ is not adjacent to exactly two blocks of $\Phi$, namely $U_l(B)$ and $U_r(B)$.  One of these semiblocks is $W$.  By taking the reverse of $\Phi$, if required, suppose $W = U_l(B)$.  Now, since $\G_B$ and $\G_{W}$ are isomorphic to co-P graphs and $\G_{BW}$ is isomorphic to a co-domino graph, it follows that $U_r(B)$ and $F_r(B)$ are indistinguishable in $\Phi \setminus \{B\}$.  Hence, changing the order between $U_r(B)$ and $F_r(B)$ in $\Phi \setminus \{B\}$, we obtain a contig $\Gamma$ that represents $\G(\Phi) \setminus \{B\}$.  Let $B_l = F_r^\Phi(B)$, $B_r = L^\Phi(B_l)$, and $B_m = U_l^\Phi(B_l)$.  Since all the non-neighbors of $B_l$ in $\G(\Phi)$ belong to $\G_W$, it follows that $B_m$ is a block of $\G_W$.  Furthermore, $U_r^\Phi(B) = R^\Gamma(B_r)$, thus $B_m = U_l^\Gamma(R^\Gamma(B_r))$ because $B_l$ and $U_r^\Phi(B)$ are indistinguishable in $\Gamma$ (see \Figure~\ref{fig:co-domino}~(c)).  Hence, $R^\Gamma(B_m) = F_l^\Gamma(B_l)$ and we obtain, by \Lemma~\ref{lem:receptive representation}, that $\Bip{B_l}{B_r}$ is receptive in $\Gamma$.  Moreover, if $\Psi$ is the $B$-reception of $\Bip{B_l}{B_r}$ in $\Gamma$, then $\Psi$ is a block contig representing $\G(\Phi) \cup \{vw\}$.  We refer to $\Psi$ as the \Definition{connection of $\Bip{B}{W}$ in\/ $\Phi$}.  The connection of $\Bip{B}{W}$ is defined analogously when $W = U_r^\Phi(B)$.  

By \Lemmas \ref{lem:semiblock removal}~and~\ref{lem:receptive representation}, $\Psi$ can be computed in $O(1)$ time if $B$ and $W$ are given.  By \Theorem~\ref{thm:forbiddens PIG}, $\Phi$ is not linear, while $\Psi$ is linear only if all the non-neighbors of $W$ belong to $\G_W$.  Thus, $\Psi$ is linear only if it has exactly six blocks.  In such case, the end pointers of $\Psi$ can be updated in $O(1)$ time.  On the other hand, $\G(\Phi)$ is co-bipartite by \Theorem~\ref{thm:forbiddens PCA} and the fact that $\G_B$ is a co-component of $\G \setminus W$.  Hence, $\G(\Psi)$ is co-bipartite and the co-end block referenced by $CB^\Phi$ is also a co-end block of $\Psi$.  That is, the co-bipartite pointer needs not be updated, so it takes $O(1)$ time to build the incremental connection of $\Bip{B}{W}$.

\begin{lemma}\label{lem:almost-connection}
  Let\/ $\Phi$ be a incremental block contig, $B, W \in \B(\Phi)$ be non-adjacent in $\G(\Phi)$, and $v \in B$ and $w \in W$.  If $\Bip{B}{W}$ is almost-connectable in\/ $\Phi$, then the connection of $\Bip{B}{W}$ in\/ $\Phi$ is a block contig representing $\G(\Phi) \cup \{vw\}$.  Furthermore, if $B$ and $W$ are given, then the incremental connection of $\Bip{B}{W}$ can be computed in $O(1)$ time.
\end{lemma}

We are now ready to characterize when $H = G \cup \{vw\}$ is a PCA graph.  Because different approaches are used, the proof is divided in several cases. We begin considering the case in which $H$ is a PIG graph.

\begin{lemma}\label{lem:edge insertion 1}
  Let\/ $\Phi$ be a block contig, $B, W \in \B(\Phi)$ be non-adjacent in $\G(\Phi)$, and $v \in B$ and $w \in W$.  If $G(\Phi) \cup \{vw\}$ is a PIG graph, then either
  \begin{enumerate}[(i)]
    \item both $\Bip{B}{W}$ and $\Bip{W}{B}$ are almost-connectable, or
    \item one of $\Bip{B}{W}$ and $\Bip{W}{B}$ is connectable.
  \end{enumerate}
\end{lemma}

\begin{proof}
  If $G(\Phi)$ is PIG, then (ii) follows~\cite{HellShamirSharanSJC2001}.   Suppose $G(\Phi)$ is not PIG for the rest of the proof.  Thus, by \Theorem~\ref{thm:forbiddens PIG}, some family $\B \subseteq \B(\Phi)$ induces a $C_{|\B|}$ ($|\B| \geq 4$) or an $S_3$ in $\G(\Phi)$.  Let $H = G(\Phi) \cup \{vw\}$ and, for each $B' \in \B(\Phi)$, denote by $b(B')$ any vertex of $B$, different from $v$ and $w$ if possible.    By \Theorem~\ref{thm:forbiddens PIG}, $\{b(B) \mid B \in \B\}$ induces neither a $C_{|\B|}$ nor an $S_3$ in $H$.  Hence, $B, W \in \B$, $B = \{v\}$, and $W = \{w\}$.

  Suppose, to obtain a contradiction, that $\B$ induces an $S_3$ in $\G(\Phi)$.  That is, $\B$ $=$ $\{B_1,$ $\ldots$, $B_6\}$, $B_1 \to B_2$, $B_2 \to B_3$, $B_3 \to B_1$, and $B_i$ is not adjacent to $B_{i+3}$ in $\G(\Phi)$, for $1 \leq i \leq 3$.  Then, w.l.o.g, $B_4 = \{w\}$ and $B_i = \{v\}$ for $i \in \{1,5\}$.   If $i = 5$, then  $v, w, b(B_1), b(B_2)$ induce a $C_4$ in $H$; otherwise $v, w, b(B_5), b(B_6)$ induce a $K_{1,3}$ in $H$.  Both contradict \Theorem~\ref{thm:forbiddens PIG}, thus $\B$ does not induce an $S_3$ in $\G(\Phi)$.  Hence, $\B$ induces a $C_{|\B|}$ in $\G(\Phi)$.

  Let $\B = \{B_1, \ldots B_{|\B|}\}$ where $B_i \to B_{i+1}$ for $1 \leq i \leq |\B|$, and suppose, w.l.o.g., that $B_1 = \{v\}$ and $B_j = \{w\}$ for some $3 \leq j < k$.  By definition, both $\{v, b(B_{i+1}), \ldots, b(B_{j-1}), w\}$ and $\{w, b(B_{j+1}), \ldots, b(B_{i-1}), v\}$ induce cycles in $H$.  Therefore, by \Theorem~\ref{thm:forbiddens PIG}, $j = 3$ and $|\B| = 4$.  We now prove that if neither $\Bip{B_1}{B_3}$ nor $\Bip{B_3}{B_1}$ are connectable in $\Phi$, then (i) follows.   By taking the reverse of $\Phi$ if required, we are left with only two cases.
  \begin{description}
    \item [Case 1:] $U_r(B_1) \neq B_3$ and $U_l(B_1) \neq B_3$.  If $U_r(B_1) \to B_4$, then $b(U_r(B_1)), b(B_4), v, b(B_2)$ induce a $C_4$ in $H$, while if $B_2 \to U_l(B_1)$, then $b(B_2), b(U_l(B_1)), b(B_4), v$ induce a $C_4$ in $H$.  By \Theorem~\ref{thm:forbiddens PIG}, neither case can happen.  Now, if $U_r(B_1) \to U_l(B_1)$, then $b(U_r(B_1)), b(U_l(B_1)), b(B_4), v, b(B_2)$ induce a $C_5$ in $H$, while if $U_r(B_1) \nto U_l(B_1)$, then $v, w, b(U_r(B_1)), b(U_l(B_1))$ induce a $K_{1,3}$ in $H$.  Again, both possibilities contradict \Theorem~\ref{thm:forbiddens PIG}, thus Case~1 is impossible.

    \item [Case 2:] $U_r(B_1) \neq B_3$ and $U_r(B_3) \neq B_1$.  As in Case~1, by \Theorem~\ref{thm:forbiddens PIG}, $U_r(B_1) \nto B_4$ and $U_r(B_3) \nto B_2$.  Hence, $\W = \B \cup \{U_r(B_1), U_r(B_3)\}$ induces a co-domino in $\G(\Phi)$.  Suppose there is a block $Z$ not adjacent to some block in $\W$, and consider the following possibilities for the position of $Z$ in $\B(\Phi)$.
    \begin{description}
      \item [Case 2.1:] $Z \in (B_1, F_r(B_4)]$.  In this case, $U_r(B_3) \to Z$.  By \Theorem~\ref{thm:forbiddens PIG}, $b(Z)$, $b(B_2)$, $b(B_3)$, $b(B_4)$ do not induce a $C_4$ in $H$, thus $Z \to B_3$ and, in particular, $Z \to U_r(B_1)$.  But then, $Z$ is adjacent to all the blocks of $\W$, a contradiction.
      \item [Case 2.2:] $Z \in (F_r(B_4), F_r(U_r(B_3))]$.  In this case, $Z \nto B_4$.  If $Z \to B_3$, then $b(Z)$, $w$, $b(B_4)$, $b(U_r(B_3))$ induce a $C_4$ in $H$, while if $Z \nto B_3$, then $b(Z)$, $b(B_2)$, $w$, $b(B_4)$, $b(U_r(B_3))$ induce a $C_5$ in $H$.  Hence, by \Theorem~\ref{thm:forbiddens PIG}, this case cannot happen.
      \item [Case 2.3:] $Z \in (F_r(U_r(B_3)), B_2)$.  In this case, $U_r(B_3) \nto Z$.  By \Theorem~\ref{thm:forbiddens PIG}, $v$, $b(Z)$, $w$, $b(U_r(B_3))$ do not induce a $K_{1,3}$ in $H$, thus $Z \to B_3$.  So, since $Z$ and $B_2$ are not indistinguishable in $\Phi$, we obtain that either $F_l(Z) \neq F_l(B_2)$ or $F_r(Z) \neq F_r(B_2)$. The latter is impossible because Case~2.1 is obtained by replacing $Z$ with $F_r(B_2)$, $B_1$ with $B_3$, and $B_4$ with $B_2$.  For the former case, observe that $B_3$ and $F_l(Z)$ are not adjacent because $B_1 \nto B_3$ and $U_r(B_3) \nto Z$.  Then, $b(F_l(Z)), b(Z), w, b(B_4)$ induce a $C_4$ in $H$, again contradicting \Theorem~\ref{thm:forbiddens PIG}.  Hence, Case~2.3 is also impossible.
      \item [Case 2.4:] $Z \in (B_2, F_r(B_1)]$.  Notice that $Z \nto B_4$ because $U_r(B_1) \nto B_4$.  Therefore, by replacing $B_2$ with $Z$, we are a case analogous to Case~2.3.
      \item [Case 2.5:] $Z \in (F_r(B_1), B_3)$.  In this case, $B_1 \nto Z$.  By \Theorem~\ref{thm:forbiddens PIG}, $v$, $b(B_2)$, $b(Z)$, $b(B_4)$ do not induce a $C_4$ in $H$, thus $Z \nto B_4$.  Then, since $Z$ and $U_r(B_1)$ are not indistinguishable, we obtain that either $F_l(U_r(B_1)) \neq F_l(Z)$ or $F_r(U_r(B_1)) \neq F_r(Z)$.  The former is impossible since, replacing $Z$ with $F_l(U_r(B_1))$ and observing that $F_l(U_r(B_1)) \in (B_1, F_r(B_1))$, one of the Cases 2.1--2.4 would hold.  The latter is also impossible because Case~2.3 is obtained by replacing $Z$ with $F_r(Z)$,  $B_1$ with $B_3$, and $B_2$ with $B_4$.  
    \end{description}
    All the remaining cases for the position of $Z$ inside $\B(\Phi)$ are analogous to one of the cases 2.1--2.5; just replace $B_1$ with $B_3$ and $B_2$ with $B_4$.  Therefore, such $Z$ does not exist, which implies that $\W$ induces a co-component of $\G(\Phi)$ and (i) follows.
  \end{description}
\end{proof}

Next we deal with the case in which $G \setminus \{v\}$ and $G \setminus \{w\}$ are both co-connected.  The following lemma, that analyses the positions of $v$ and $w$ after $vw$ is inserted, is required.

\begin{lemma}\label{lem:edge removal 1}
  Let\/ $\Psi$ be a circular block contig, $B, W \in \B(\Psi)$ be such that $B \to W$, and $v \in B$ and $w \in W$.  If $G(\Psi) \setminus \{vw\}$ is a PCA graph and both $G(\Psi)$ and $G(\Psi) \setminus \{v\}$ are co-connected, then $F_r^\Psi(B) = W$.
\end{lemma}

\begin{proof}
  Let $H = G(\Psi)$.  Suppose $G = H \setminus \{vw\}$ is a PCA graph and yet $F_r^\Psi(B) \neq W$.  Let $\Phi$ be the compressed removal of $v$ from $\Psi$.  Recall that $\Psi$ is the insertion of $v$ into $\Phi$, thus $\Phi$ has blocks $B_a \neq B_b$ such that (i) $B_m \subseteq N_H(v)$ for every $B_m \in (B_a, B_b)$, (ii) $B_m \cap N_H(v) = \emptyset$ for every $B_m \in (B_b, B_a)$, and (iii) $\Psi$ is the $\{v\}$-refinement of $\Bip{B_a \cap N_H(v)}{B_b \cap N_H(v)}$ in $\Phi$.  Therefore, $W \neq B_a$ because $B \Topsi W$, while $W \neq B_b$ because $W \neq F_r^\Psi(B)$.   Thus, since $N_H(v) \cap W \neq \emptyset$, it follows that $W$ is a block of $\Phi$ that belongs to $(B_a, B_b)$. On the other hand, $\G(\Phi)$ is $1$-universal and connected, because $\Psi$ is co-connected and circular.  Hence, by \Lemma~\ref{cor:indistinguishable}, $\Phi$ is a block contig representing $G \setminus \{v\}$.  Then, since $G \setminus \{v\}$ is co-connected, it follows, by \Theorem~\ref{thm:unique-PCA-models}, that $\Phi$ and $\Phi^{-1}$ are the unique block contigs representing $G \setminus \{v\}$.  Consequently, by \Lemma~\ref{lem:insertable contig}, taking into account that $G$ is a PCA graph, there is an insertion of $v$ into $\Phi$ that represents $G$.  That is, $\Phi$ has blocks $B_c \neq B_d$ such that (i) $B_m \subseteq N_G(v)$ for every $B_m \in (B_c, B_d)$ and (ii) $B_m \cap N_G(v) = \emptyset$ for every $B_m \in (B_c, B_d)$.   Since $W \not\in \{B_a, B_b\}$, it follows that $B_a \cap N_G(v) \neq \emptyset$ and $B_b \cap N_G(v) \neq \emptyset$, thus $[B_a, B_b] \subseteq [B_c, B_d]$.  But then, $W \subseteq N_G(v)$, a contradiction.
\end{proof}

The following lemma deals with the case in which both $G \setminus \{v\}$ and $G \setminus \{w\}$ are co-connected.

\begin{lemma}\label{lem:edge insertion 2}
  Let\/ $\Phi$ be a block contig, $B, W \in \B(\Phi)$ be non-adjacent in $\G(\Phi)$, and $v \in B$ and $w \in W$.  If $G(\Phi) \cup \{vw\}$ is a non-interval PCA graph and both $G(\Phi) \setminus \{v\}$ and $G(\Phi) \setminus \{w\}$ are co-connected, then either\/ $\Bip{B}{W}$ or\/ $\Bip{W}{B}$ is connectable.
\end{lemma}

\begin{proof}
  Suppose $G(\Phi) \cup \{vw\}$ admits a circular block contig $\Psi$.  Since both $G(\Phi) \setminus \{v\}$ and $G(\Phi) \setminus \{w\}$ are co-connected and $G(\Phi) \cup \{vw\}$ is not a PIG graph, it follows that $G(\Phi) \cup \{vw\}$ is co-connected.  Let $B^\Psi$ and $W^\Psi$ be the blocks of $\Psi$ that contain $v$ and $w$, respectively, and suppose, w.l.o.g., that $B^\Psi \Topsi W^\Psi$.  Applying  \Lemma~\ref{lem:edge removal 1} on both $\Psi$ and $\Psi^{-1}$, we obtain that $\Bip{B^\Psi}{W^\Psi}$ is disconnectable.  Thus, the disconnection $\Gamma$ of $\Bip{\{v\}}{\{w\}}$ in $\Psi$ is a block contig representing $\G(\Phi)$ in which $\Bip{B}{W}$ is connectable.  Consequently, since $\Phi \in \{\Gamma, \Gamma^{-1}\}$ by \Theorem~\ref{thm:unique-PCA-models}, the result follows.
\end{proof}

For the third case, suppose neither $G \setminus \{v\}$ nor $G \setminus \{w\}$ are co-connected.

\begin{lemma}\label{lem:edge insertion 3}
  Let\/ $\Phi$ be a block contig, $B, W \in \B(\Phi)$ be non-adjacent in $\G(\Phi)$, and $v \in B$ and $w \in W$.  If $G \cup \{vw\}$ is a non-interval PCA graph, $G(\Phi)$ is co-connected and none of $G \setminus \{v\}$ and $G \setminus \{w\}$ is co-connected, then either\/ $\Bip{B}{W}$ or\/ $\Bip{W}{B}$ is connectable.
\end{lemma}

\begin{proof}
  Observe that, since $\G_B = \G(\Phi) \setminus \{B\}$ is not co-connected, then there is some block $\overline{B} \not\in N(B)$ that is not in the same co-component of $\G_B$ as $W$.  Similarly, there is some block $\overline{W} \not\in N(W)$ that is not in the same co-component of $\G_W = \G(\Phi) \setminus \{W\}$ as $B$.  Clearly, $B, \overline{W}, \overline{B}, W$ induce a $P_4$ in $\G(\Phi)$.  Without loss of generality, suppose $B \Tophi \overline{W}$, thus $\overline{W} \Tophi \overline{B}$ and $\overline{B} \Tophi W$.  

  Since $B$ is not universal in $\G(\Phi)$, it follows that $B \Ntophi U_l(B)$. If $U_l(B) \neq W$, then $U_l(B) \in (W, B)$, implying that $U_l(B) \not\in N(\overline{W})$.  But then, $\overline{W}$, $U_l(B)$, and $B$ belong to the same co-component of $\G_W$, a contradiction.  Therefore, $U_l(B) = W$.  Applying the same arguments on $\Phi^{-1}$, we obtain that $\Bip{W}{B}$ is connectable.
\end{proof}

It remains to consider the case in which $G \setminus \{v\}$ is co-connected and $G \setminus \{w\}$ is not co-connected.  For the sake of simplicity, we divide this case according to whether $H$ is co-connected (\ie, $v$ is not universal) or not (\ie, $v$ is universal).  We considering first the case $H$ co-connected.

\begin{lemma}\label{lem:edge insertion 4}
  Let\/ $\Phi$ be a block contig, $B, W \in \B(\Phi)$ be non-adjacent in $\G(\Phi)$, and $v \in B$ and $w \in W$.  If $G(\Phi) \cup \{vw\}$ is a non-interval PCA graph and both $G(\Phi) \setminus \{v\}$ and $G(\Phi) \cup \{vw\}$ are co-connected, then either:
  \begin{enumerate}[(i)]
    \item $\Bip{B}{W}$ is almost-connectable, or
    \item one of $\Bip{B}{W}$ and $\Bip{W}{B}$ is connectable.
  \end{enumerate}
\end{lemma}

\begin{proof}
  If $G(\Phi) \setminus \{w\}$ is co-connected, then (ii) follows from \Lemma~\ref{lem:edge insertion 2}.  Consider, then, the case in which $G(\Phi) \setminus \{w\}$ is not co-connected.  Notice that $W = \{w\}$ because $G(\Phi) \cup \{vw\}$ is co-connected.  By \Lemma~\ref{lem:co-connected->co-bipartite}, $G(\Phi) \setminus \{w\}$ is co-bipartite, thus $B \in \X$ for some co-contig pair $\Bip{\X}{\Y}$ of $\Phi \setminus \{W\}$. Since $G(\Phi) \cup \{vw\}$ is co-connected and $W = \{w\}$, it follows that $\Y \neq \emptyset$.  Let $W_l$ be the left co-end block of $\Y$ and $\Gamma = \Phi|(\X \cup \Y \cup \{W\})$.  Clearly, $G(\Gamma) \cup \{vw\}$ is an induced subgraph of $G(\Phi) \cup \{vw\}$, while, by construction, $G(\Gamma)$, $G(\Gamma) \setminus \{v\}$, and $G(\Gamma) \setminus \{w\}$ are all co-connected.  Then, by \Lemmas \ref{lem:edge insertion 1}~and~\ref{lem:edge insertion 2}, either (a)~$\Bip{B}{W}$ is almost-connectable in $\Gamma$, or (b)~one between $\Bip{B}{W}$ and $\Bip{W}{B}$ is connectable in $\Gamma$.  In case (a), $\Bip{B}{W}$ is almost-connectable in $\Phi$ because $\G(\Gamma \setminus \{W\})$ is the co-component of $\G(\Phi) \setminus \{W\}$ that contains $B$.  Suppose, then, that (b) holds.  Moreover, by taking the reverse of $\Phi$ if required, suppose $\Bip{B}{W}$ is connectable in $\Gamma$, \ie, $U_r^\Gamma(B) = W$ and $U_l^\Gamma(W) = B$.  So, $W \in (B, W_l)$ in both $\Gamma$ and $\Phi$.   If $B$ is the right co-end block of $\X$, then either $W = R^\Phi(B)$ or $R^\Phi(B) \not\in \B(\Gamma)$ and $R^\Phi(B) \Tophi W_l$.  Whichever the case, $\Bip{B}{W}$ is connectable in $\Phi$.  Otherwise, if $B$ is not the right co-end block of $\X$, then $R^\Phi(B) = R^\Gamma(B)$ because $\X$ is a range of $\B(\Phi)$.  Therefore, since $R^\Gamma(B) \Tophi W$, it follows that $\Bip{B}{W}$ is connectable in $\Phi$.
\end{proof}

Finally, we consider the case in which $G \setminus \{v\}$ is co-connected but $H$ is not.

\begin{lemma}\label{lem:edge insertion 5}
  Let $G$ be a PCA graph and $v,w \in V(G)$ be non-adjacent.  If $G \cup \{vw\}$ is a non-interval PCA graph that is not co-connected and $G \setminus \{v\}$ is co-connected, then $v$ is universal in $G \cup \{vw\}$ and $G$ is co-bipartite.
\end{lemma}

\begin{proof}
  Suppose $G \cup \{vw\}$ is a PCA graph.  Since $G \cup \{vw\}$ is not co-connected, it follows that $G \setminus \{v\}$ is co-bipartite by \Lemma~\ref{lem:co-connected->co-bipartite}.  Also, $v$ is universal in $G \cup \{vw\}$ and so $G$ is co-bipartite, because $G \setminus \{v\}$ is co-connected.
\end{proof}

\Lemmas \ref{lem:edge insertion 1}--\ref{lem:edge insertion 5} are summed up as follows.  

\begin{lemma}\label{lem:edge insertion}
  Let\/ $\Phi$ be a block contig, $B, W \in \B(\Phi)$ be non-adjacent in $\G(\Phi)$, and $v \in B$ and $w \in W$. Then, $G(\Phi) \cup \{vw\}$ is a PCA graph if and only if either:
  \begin{enumerate}[(i)]
    \item one of $\Bip{B}{W}$ or $\Bip{W}{B}$ is almost-connectable, \label{lem:edge insertion:co-domino}
    \item one of $\Bip{B}{W}$ or $\Bip{W}{B}$ is connectable, or\label{lem:edge insertion:connectable}
    \item one of\/ $\{v,w\}$ is universal in $G(\Phi) \cup \{vw\}$ and $G(\Phi)$ is co-bipartite. \label{lem:edge insertion:universal}
  \end{enumerate}
\end{lemma}

\begin{proof}
  Suppose $H = G(\Phi) \cup \{vw\}$ is a PCA graph, and let $\Gamma$ be a round block representation of the co-component of $G(\Phi)$ that contains both $B$ and $W$.  If $\G(\Gamma)$ is disconnected, then $\Gamma \neq \Phi$, thus $\G(\Phi)$ is not co-connected.  Consequently, by \Lemma~\ref{lem:co-connected->co-bipartite}, $|\B(\Gamma)| = 2$ and (\ref{lem:edge insertion:connectable}) follows.  On the other hand, if $\G(\Gamma)$ is connected, then one of (\ref{lem:edge insertion:co-domino})--(\ref{lem:edge insertion:universal}) holds for $\Gamma$ by \Lemmas~\ref{lem:edge insertion 1}--\ref{lem:edge insertion 5}.  By definition, if (\ref{lem:edge insertion:co-domino}) holds for $\Gamma$, then it also holds for $\Phi$.  On the other hand, if $\Gamma$ satisfies (\ref{lem:edge insertion:universal}), then the universal vertex of $G(\Gamma) \cup \{vw\}$ is also universal in $G(\Phi) \cup \{vw\}$ while, by \Lemma~\ref{lem:co-connected->co-bipartite}, $G(\Phi)$ is co-bipartite.  That is, $\Phi$ satisfies (\ref{lem:edge insertion:universal}).  Finally, if $\Gamma$ satisfies (\ref{lem:edge insertion:connectable}), then either $\Gamma = \Phi$ and (\ref{lem:edge insertion:connectable}) follows, or $\Phi$ is not co-connected.  In the latter case, by \Lemma~\ref{lem:co-connected->co-bipartite}, $\Gamma$ is described by a co-contig pair $\Bip{\X}{\Y}$.  Since $\Bip{\X}{\Y}$ is also a co-contig pair of $\Phi$, (\ref{lem:edge insertion:connectable}) holds for $\Phi$.

  The converse follows from \Theorem~\ref{thm:forbiddens PCA} and \Lemmas \ref{lem:connection}~and~\ref{lem:almost-connection}.
\end{proof}

Algorithm~\ref{alg:edge insertion} transforms the block contig $\Phi$ representing $G$ into a compressed contig $\Psi$ representing $H = G \cup \{vw\}$, whenever possible.  Its correctness follows from \Lemma~\ref{lem:edge insertion}.  In particular, Steps \ref{alg:edge insertion:v universal}--\ref{alg:edge insertion:v universal end} check statement (\ref{lem:edge insertion:universal}) and transform $\Phi$ into $\Psi$ when (\ref{lem:edge insertion:universal}) holds.  On the other hand, Step~\ref{alg:edge insertion:connectable} checks statements (\ref{lem:edge insertion:co-domino})~and~(\ref{lem:edge insertion:connectable}) and transforms $\Phi$ into $\Psi$.  The correctness of this step follows by \Lemmas~\ref{lem:connection}~and~\ref{lem:almost-connection}.  

\begin{algorithm}[htb!]
  \caption{Insertion of an edge $vw$ into a block contig.}\label{alg:edge insertion}

  \textbf{Input:} an incremental block contig $\Phi$, and two non-adjacent vertices $v,w$ of $G(\Phi)$.

  \textbf{Output:} if $G(\Phi) \cup \{vw\}$ is a PCA graph, then $\Phi$ is updated into a compressed contig $\Psi$ of $G(\Phi) \cup \{vw\}$; otherwise, the algorithm halts in error.

  \mbox{}

  \begin{AlgorithmSteps}
    \Step{If $d(v) < d(w)$, then swap $v$ and $w$.}
    \Step{Let $B$ and $W$ be the blocks containing $v$ and $w$, respectively.}
    \Step{If $v$ is universal in $G(\Phi) \cup \{vw\}$ and $G(\Phi)$ is co-bipartite (\ie\ $CB \neq NULL$):}\label{alg:edge insertion:v universal}
    \IncreaseIndent
      \Step{Transform $\Phi$ into the compressed removal of $v$.}\label{alg:edge insertion:v removal}
      \Step{Build the $\{v\}$-reception of $\{CB, L(CB)\}$ in $\Phi$ and halt.}\label{alg:edge insertion:v universal end}
    \DecreaseIndent
    \Step{If $\Bip{B}{W}$ or $\Bip{W}{B}$ is either connectable or almost-connectable, then transform $\Phi$ into the corresponding connection of $\{\{v\}, \{w\}\}$ or $\{\{w\}, \{v\}\}$ in $\Phi$ and halt.}\label{alg:edge insertion:connectable}
    \Step{Halt in error.}
  \end{AlgorithmSteps}
\end{algorithm}

Consider the time complexity of Algorithm~\ref{alg:edge insertion}.   If Step~\ref{alg:edge insertion:v removal} is executed, then $W = U_l(B) = U_r(B) = \{w\}$, because $v$ is universal in $G \cup \{vw\}$.  Consequently, Steps \ref{alg:edge insertion:v removal}~and~\ref{alg:edge insertion:v universal end} take $O(1)$ time, by \Lemmas \ref{lem:compressed removal}~and~\ref{lem:semiblock insertion}, respectively. To check if $\Bip{B}{W}$ is almost-connectable for Step~\ref{alg:edge insertion:connectable}, apply a BFS-traversal on $\overline{\G(\Phi)}$ starting from $B$ and without surpassing $W$.  If more than $6$ blocks are found, then $\Bip{B}{W}$ is not almost-connectable, thus $O(1)$ time is required for this check.  Finally, the connection of $\Phi$ in Step~\ref{alg:edge insertion:connectable} takes $O(1)$ time, by \Lemmas \ref{lem:connection complexity}~and~\ref{lem:almost-connection}.  Therefore, Algorithm~\ref{alg:edge insertion} requires $O(1)$ time.

Clearly, the only possible new universal vertices of $H$ are $v$ and $w$.  Recall $CB^\Phi$ references the universal block of $\Phi$ prior the execution of Algorithm~\ref{alg:edge insertion}, if any. Then, evaluating if $v$ and $w$ are universal in $H$,  moving $v$ and $w$ into an universal block, and updating $CB$ so as to point to this block is doable in $O(1)$ time.  That is, the output $\Psi$ of Algorithm~\ref{alg:edge insertion} can be transformed into a block contig representing $H$ in $O(1)$ time.   Observe that $\Psi$ satisfies the straightness invariants.  If $\Phi$ is linear and $\Psi$ is circular, the end pointers of the blocks of $\Phi$ should be nullified in $O(1)$ time.  As it happens after the insertion of a vertex, the only possible end blocks of $\Phi$ in this case are $B$ and $W$.

\begin{lemma}\label{lem:nolouse3}
  Let\/ $\Phi$ be a linear block contig, $B, W \in \B(\Phi)$ be non-adjacent in $\G(\Phi)$, and $v \in B$ and $w \in W$.  If $G(\Phi) \cup \{vw\}$ is a non-interval PCA graph, then $B$ and\/ $W$ are the end blocks of\/ $\Phi$.
\end{lemma}

\begin{proof}
  Since $G(\Phi) \cup \{vw\}$ is a non-interval PCA graph, then, by \Theorem~\ref{thm:forbiddens PIG}, it contains an induced $S_3$ or $C_4$ that, as in \Lemma~\ref{lem:edge insertion 1}, contains both $v$ and $w$.  Then, neither $v$ nor $w$ is universal in $G(\Phi) \cup \{vw\}$.  Therefore, by \Lemma~\ref{lem:edge insertion}, one of $\Bip{B}{W}$ or $\Bip{W}{B}$ is connectable in $\Phi$.  Then, since the respective connection of $B$ and $W$ in $\Psi$ is not an interval model, it follows that $B$ and $W$ are the end blocks of $\Phi$.
\end{proof}

The co-end pointer of the data structure should also be updated for $\Psi$.  As discussed before, $CB$ has been already updated when $v$ or $w$ is universal, while it needs not be updated when $G(\Phi)$ is co-bipartite.  When $\Psi$ is linear and co-bipartite, the co-bipartite pointer is updated in $O(1)$ time using \Lemma~\ref{lem:cobipartite-linear}.  In the remaining case, the co-end block can be updated to reference either $B$ or $W$ as follows.

\begin{lemma}\label{lem:nolouse4}
  Let $G$ be a PCA graph that is not co-bipartite, $v,w \in V(G)$ be non-adjacent, and $B$ and\/ $W$ be the blocks of $G \cup \{vw\}$ that contain $v$ and $w$, respectively.  If $G \cup \{vw\}$ is a\/ $0$-universal co-bipartite graph that admits a circular block contig\/ $\Psi$, then one of $B$ and\/ $W$ is a left co-end block of\/ $\Psi$.
\end{lemma}

\begin{proof}
  Since $G$ is not co-bipartite, it is co-connected by \Lemma~\ref{lem:co-connected->co-bipartite}.  Furthermore, $G \cup \{vw\}$ is also co-connected, because it is co-bipartite.  Then, $\Psi$ has exactly two left co-end blocks, say $B_l$ and $U_r^\Psi(B_l)$.  Moreover, $\Psi$ and $\Psi^{-1}$ are the unique block contigs representing $G \cup \{vw\}$, by \Theorem~\ref{thm:unique-PCA-models}.  On the other hand, by \Lemma~\ref{lem:edge insertion}, the blocks containing $v$ and $w$ in $G$ are connectable in any block contig $\Phi$ representing $G$.  Hence, $\Psi$ is the connection of either $\Bip{\{v\}}{\{w\}}$ or $\Bip{\{w\}}{\{v\}}$ in $\Phi$.  Assume the former, thus $\Bip{B}{W}$ is disconnectable in $\Psi$ and $\Phi$ is the disconnection of $\Bip{\{v\}}{\{w\}}$ in $\Psi$.  If $B \not\in \{B_l, U_r^\Psi(B_l)\}$, then $B_l$ and $U_r^\Psi(B_l)$ are included in blocks of $B_l^\Phi$ and $U_r^\Phi(B_l^\Phi)$ such that $B_l^\Phi = U_r^\Phi(U_r^\Phi(B_l^\Phi))$.  Therefore, $G(\Phi)$ is co-bipartite, a contradiction.
\end{proof}

To solve the incremental recognition problem, $\Phi^{-1}$ also has to be transformed into $\Psi^{-1}$.  The algorithm discussed above transforms $\Phi$ into $\Psi$.  To transform $\Phi^{-1}$ into $\Psi^{-1}$, this algorithm cannot be blindly applied because a round representation $\Psi^* \neq \Psi^{-1}$ representing $\G(\Psi)$ could be obtained.  Instead, some careful is required.  For instance, if both $\Bip{B}{W}$ and $\Bip{W}{B}$ are connectable or in $\Phi$, and $\Psi$ is computed as the connection of $\Bip{B}{W}$ in Algorithm~\ref{alg:edge insertion}, then $\Psi^{-1}$ has to be taken as the connection of $\Bip{W}{B}$ in $\Phi^{-1}$.  Analogously, the $\{v\}$-reception of $\Bip{L^\Phi(CB^\Phi)}{CB^\Phi}$ has to be build at Step~\ref{alg:edge insertion:v universal end} while producing $\Psi^{-1}$.  It is not hard to see that all these decisions for transforming $\Phi^{-1}$ into $\Psi^{-1}$ can be applied in $O(1)$ time.  

Combining Algorithm~\ref{alg:edge insertion} with the HSS algorithm for the case in which $\Phi$ is not a contig, the main theorem of this section is obtained.

\begin{theorem}\label{thm:edge insertion}
  The problem of deciding whether an incremental graph is a PCA graph takes $O(1)$ time per inserted edge.
\end{theorem}

\section{Decremental recognition of proper circular-arc graphs}
\label{sec:decremental}

This section is devoted to the methods that compose the decremental recognition algorithm of PCA graphs, \ie, the removal of a vertex (\Section~\ref{subsec:vertex removal}) or an edge (\Section~\ref{subsec:edge removal}).  For this section, each contig $\Psi$ is implemented with an augmented base contig in which:
\begin{itemize}
  \item each semiblock $B$ is associated with a \Definition{co-end pointer} $CE^{\Psi}(B)$ such that $CE^{\Psi}(B) = NULL$ if $B$ is not a co-end semiblock, while $CE^{\Psi}(B)$ references the other co-end semiblock of the co-contig range containing $B$ otherwise.
\end{itemize}
As usual, we omit the superscript when no confusions arise.  As it happens with the co-bipartite pointer of the incremental algorithm, the co-end pointers of $\Psi$ are ignored by the HSS algorithm, and should be restored each time the HSS algorithms are applied on $\Psi$.   

We write that $\Psi$ is a \Definition{decremental contig} to emphasize that $\Psi$ is a base contig augmented with co-end pointers.  Similarly, a \Definition{decremental round representation} is a base round representation whose contigs are decremental contigs.  For the algorithms in this section, two decremental round block representations $\Psi, \Psi^{-1}$, both satisfying the straightness property, are stored to represent a decremental graph $H$.

\subsection{The impact of a removed vertex}
\label{subsec:vertex removal}

In this part we describe an algorithm that transforms a round block representation $\Psi$ of a graph $H$ into a round block representation $\Phi$ of $H \setminus \{v\}$, for any given $v \in V(H)$.  The algorithm is divided in three major phases.  The first phase computes a round block representation $\Phi$ of $H \setminus \{v\}$, without caring about its co-end pointers or the straightness property.  Then, the second and third phases restore the straightness property and update the co-end pointers of $\Phi$, respectively.  Each of these phase is described below.
\begin{description}
  \item [First phase.]  Let $B$ be the block of $\Psi$ that contains $v$.  This phase is composed by four steps.  First, the universal block $U$ of $\Psi$, if any, is located by traversing $[F_l(B), F_r(B)]$.  Let $U = \emptyset$ if such an universal block does not exist.  Second, the universal block of $G(\Psi) \setminus (U \cup \{v\})$, if any, is located as follows.  Let $W = F_r(R(B))$.  If $B = \{w\}$ and $F_l(L(B)) = W$, then $W$ is universal in $G(\Psi) \setminus \{v\}$; otherwise, $U$ contains all the universal vertices of $G(\Psi) \setminus \{v\}$.  Let $W = \emptyset$ in the latter case.   Third, \Lemma~\ref{lem:compressed removal} is applied on $\{v\}$ to obtain the compressed removal $\Phi$ of $v$ from $\Psi$.  By construction, $\Phi$ is $2$-universal and either $U \cup W = \emptyset$ or $U \cup W$ is the universal block of $G(\Phi)$.  If $W \neq \emptyset$, then the forth step merges $U$ and $W$ by moving the vertices from $U$ into $W$ and then removing $U$.  After this step, $\Phi$ is $1$-universal and compressed, thus $\Phi$ is a round block representation of $H \setminus \{v\}$, by \Corollary~\ref{cor:indistinguishable}.  By \Lemma~\ref{lem:compressed removal}, the third step takes $O(d_H(v))$ time, while the remaining steps take $O(d_H(v))$ time with an standard implementation.

  \item[Second phase.]  The second phase restores the straightness invariant of $\Phi$.  If $\Phi$ is straight or $H \setminus \{v\}$ is not PIG, then $\Phi$ already satisfies the straightness invariant.  So, the second phase is applied only when $\Phi$ is circular, and its goal is to determine whether $H \setminus \{v\}$ is a PIG graph.  If so, then $\Phi$ should be transformed into a linear block contig representing $H \setminus \{v\}$.  By \Theorems \ref{thm:unique-PCA-models}~and~\ref{thm:forbiddens PIG}, $H \setminus \{v\}$ is a PIG graph only if $\G(\Phi)$ contains a universal block $U$.  Furthermore, by Theorem~4.10 of~\cite{LinSoulignacSzwarcfiter2011}, $H \setminus \{v\}$ is a PIG graph if and only if $F_r(L(U)) = U = F_l(R(U))$.  Moreover, by \Lemma~\ref{lem:receptive representation}, $\Bip{R(U)}{L(U)}$ is receptive in $\Phi \setminus U$ and the $U$-reception of $\Bip{R(U)}{L(U)}$ in $\Phi \setminus \{U\}$ is a linear block contig representing $H \setminus \{v\}$~\cite{LinSoulignacSzwarcfiter2011}.  The universal block $U$, if existing, was already computed in the first phase.  Then, this phase checks whether $F_r(L(U)) = U = F_l(R(U))$ and, if so, it builds the $U$-reception of $\Bip{R(U)}{L(U)}$ in $\Phi \setminus \{U\}$.  By \Lemmas \ref{lem:semiblock removal}~and~\ref{lem:semiblock insertion}, this phase takes $O(1)$ time. 

  \item[Third phase.]  The last phase computes the co-end pointers of $\Phi$.  Note that, by definition, all the co-end blocks of $\Psi$ are also co-end blocks of $\Phi$.  Then, for this phase, first \Lemma~\ref{lem:decremental co-contigs} is applied on $\Phi$ and $d_H(v)$.  If $\Phi$ is not co-bipartite, then all the co-end pointers of $\Phi$ correctly reference $NULL$.  On the other hand, if $\Phi$ is co-bipartite, then a natural ordering of its co-contigs is obtained.  Such natural ordering is traversed to update the co-end pointers that reference an incorrect location.  The application of \Lemma~\ref{lem:decremental co-contigs} and the traversal of its output take $O(d_H(v))$ time.
\end{description}

The described algorithm applied on $\Psi^{-1}$ yields the round representation $\Phi^{-1}$.  Hence, the main theorem of this section follows.  

\begin{theorem}\label{thm:vertex removal}
  The problem of deciding whether a decremental graph is a PCA graph takes $O(1)$ time per removed edge, when only the removal of vertices is allowed.
\end{theorem}

\subsection{The impact of a removed edge}
\label{subsec:edge removal}

In this part we complete the recognition algorithm for decremental PCA graphs, by showing how to process the removal of an edge.  This time the input is formed by two block contigs $\Psi, \Psi^{-1}$, representing a graph $H$, and an edge $vw \in E(H)$, and the goal is to compute two round block representations $\Phi, \Phi^{-1}$ of $H \setminus \{vw\}$, whenever possible.  

In essence, the removal of $vw$ follows the inverse path that would be taken to insert $vw$ into a representation of $H \setminus \{vw\}$.  That is, if $B$ and $W$ are respectively the blocks of $H$ that contain $v$ and $w$, then either (i) $B$ and $W$ are \Definition{almost-disconnectable}, (ii) or $B$ is the universal block of $H$, or (iii) $B$ and $W$ are disconnectable in some block contig representing $H$.  It is important to remark that (iii) needs not be satisfied by $\Psi$ when $H$ does not satisfy (i) and (ii).  However, as we shall see, in such case $B$ and $W$ are co-end blocks of different co-contigs of $\Psi$.  Thus, $\Psi, \Psi^{-1}$ can be transformed into block contigs $\Gamma, \Gamma^{-1}$ representing $H$ such that $\Bip{B}{W}$ is disconnectable in $\Gamma$.  The lemma below discusses this transformation.

\begin{lemma}\label{lem:disconnectable contig}
  Let\/ $\Psi$ be a decremental contig, and $B$ and $W$ be co-end blocks of different co-contigs of $\Psi$.  If\/ $\Psi, \Psi^{-1}$ are given as input, then it takes $O(1)$ time to transform\/ $\Psi, \Psi^{-1}$ into decremental block contigs\/ $\Gamma, \Gamma^{-1}$ such that\/ $\Bip{B}{W}$ is disconnectable in\/ $\Gamma$.
\end{lemma}

\begin{proof}
  Let $\Bip{\X}{\overline{\X}}$ be the co-contig pair of $\Psi$ such that $B \in \X$.  The transformation algorithm has five steps.  The first step is to obtain the other co-end block of $\X$ by accessing the co-end pointer of $B$.  Second, the co-end blocks of $\overline{\X}$ are obtained via the mappings $U_r$ and $U_l$.  For the third step, $\Psi$ is split into $\Lambda = \Psi|(\X \cup \overline{X})$ and $\Omega = \Psi \setminus (\X \cup \overline{\X})$, while $\Psi^{-1}$ is split into $\Lambda^{-1}$ and $\Omega^{-1}$.  The forth step is to exchange $\Lambda$ with $\Lambda^{-1}$ and $\X$ with $\X^{-1}$, if required, so as to make $B$ the left co-end block of $\X$.  Similarly, $\Omega$ and $\Omega^{-1}$ are exchanged so as to make $W$ a right co-end block.  The final step is to compute $\Gamma$ as the $\Bip{\X}{[F_l(W), W]}$-join of $\Bip{\Lambda}{\Omega}$ and $\Gamma^{-1}$ as the $\Bip{[W, F_r(W)]}{\X^{-1}}$-join of $\Bip{\Omega^{-1}}{\Lambda^{-1}}$.   By construction, $\Gamma$ and $\Gamma^{-1}$ are mutually reverse round representations of $G(\Psi)$, and $\Bip{B}{W}$ is disconnectable in $\Gamma$.  For the time complexity, observe that the first and forth step require $O(1)$ time, the second step takes $O(1)$ time even when $\Psi$ is linear by \Lemma~\ref{lem:cobipartite-linear}, and the third and final steps take $O(1)$ time by \Lemmas \ref{lem:separation of G+H}~and~\ref{lem:join of G+H}, respectively.
\end{proof}

Almost-disconnectable blocks are defined as the inverse of almost-connectable blocks.  Let $\Phi$ be a block contig, $\Bip{B}{W}$ be almost-connectable in $\Phi$, and $\Psi$ be the connection of $\Bip{B}{W}$ in $\Phi$.  Note that, by definition, $B, W$ are blocks of $\Psi$.  We say that $\Bip{B}{W}$ is \Definition{almost-disconnectable} in $\Psi$ and that $\Phi$ is the \Definition{disconnection} of $\Bip{B}{W}$ in $\Phi$.  Suppose $W = U_l^\Phi(B)$, and let $\Gamma$ be the contig obtained from $\Psi \setminus \{B\}$ by swapping $F_l^\Psi(B)$ and $U_l^\Psi(B)$.  Since $\Psi$ is the connection of $\Phi$, it follows that $\Gamma = \Phi \setminus \{B\}$.  Furthermore, $F_l^\Psi(B) = F_r^\Phi(B)$, while $R^\Psi(W) = F_l^\Phi(B)$. That is, $\Phi$ is the $B$-refinement of $\Bip{R^\Psi(W)}{F_l^\Psi(B)}$ in $\Gamma$.  An analogous condition holds when $W = U_r^\Phi(B)$.

By \Lemmas \ref{lem:semiblock removal}~and~\ref{lem:receptive representation}, $\Phi$ can be computed in $O(1)$ time if $B$ and $W$ are given.  Recall that both $G(\Phi)$ and $G(\Psi)$ are co-bipartite.  Furthermore, $G(\Phi)$ has the same co-components as $G(\Psi)$ with the exception that an edge is missing from one of the co-components.  Therefore, $\Phi$ and $\Psi$ share the same co-contig pairs.  In other words, the co-end pointers of $\Phi$ need not be updated after disconnecting $B$ and $W$. 

\begin{lemma}\label{lem:almost-disconnection}
  Let\/ $\Psi$ be a decremental block contig, $B, W \in \B(\Psi)$ be adjacent in $\G(\Psi)$, and $v \in B$ and $w \in W$.  If $\Bip{B}{W}$ is almost-disconnectable in\/ $\Phi$, then the disconnection of $\Bip{B}{W}$ in\/ $\Psi$ is a block contig representing $\G(\Psi) \setminus \{vw\}$.  Furthermore, if $B$ and $W$ are given, then the decremental disconnection of $\Bip{B}{W}$ can be computed in $O(1)$ time.
\end{lemma}
 
The following lemma characterizes those block contigs that admit the removal of an edge.

\begin{lemma}\label{lem:edge removal}
  Let\/ $\Psi$ be a block contig, $B, W \in \B(\Psi)$ be adjacent in $\G(\Psi)$, and $v \in B$ and $w \in W$. Suppose $d(v) \geq d(w)$.  Then, $G(\Psi) \setminus \{vw\}$ is a PCA graph if and only if either:
  \begin{enumerate}[(i)]
    \item one of $\Bip{B}{W}$ and $\Bip{W}{B}$ is almost-disconnectable, \label{lem:edge removal:co-A}
    \item one of $\Bip{B}{W}$ and $\Bip{W}{B}$ is disconnectable, \label{lem:edge removal:disconnectable}
    \item $v$ is universal in $G(\Psi)$, $W \setminus \{v\} \neq \{w\}$ and either\/ $\Bip{W \setminus \{v,w\}}{L(W)}$ or\/ $\Bip{R(W)}{W \setminus \{v,w\}}$ is refinable in the compressed removal of\/ $\{v\}$ from\/ $\Psi$, or \label{lem:edge removal:refinable}
    \item $v$ is universal in $G(\Psi)$, $W \setminus \{v\} = \{w\}$, and\/ $\Bip{R(W)}{L(W)}$ is refinable in the compressed removal of\/ $\{v\}$ from\/ $\Psi$, or \label{lem:edge removal:refinable bis}
    \item $B$ and $W$ are co-end blocks of different co-contigs of\/ $\Psi$.\label{lem:edge removal:co-ends}
  \end{enumerate}
\end{lemma}

\begin{proof}
  Suppose $G(\Psi) \setminus \{vw\}$ admits a round block representation $\Phi$ and let $B^\Phi$ and $W^\Phi$ be the blocks of $\Phi$ containing $v$ and $w$, respectively.  Then, by \Lemma~\ref{lem:edge insertion}, either
  \begin{enumerate}[(a)]
    \item $G(\Phi)$ is disconnected,
    \item one of $\Bip{B^\Phi}{W^\Phi}$ or $\Bip{W^\Phi}{B^\Phi}$ is almost-connectable in $\Phi$,
    \item one of $\Bip{B^\Phi}{W^\Phi}$ or $\Bip{W^\Phi}{B^\Phi}$ is connectable in $\Phi$, or
    \item $v$ is universal in $G(\Psi)$ and $G(\Phi)$ is co-bipartite, or
  \end{enumerate}
  If (a) holds, then $G(\Psi)$ is a PIG graph and (\ref{lem:edge removal:disconnectable}) follows~\cite{HellShamirSharanSJC2001}.  The remaining cases are considered below.
  \begin{description}
    \item [(b) holds.] By relabeling $v$, $w$, $B$ and $W$ if required, suppose $\Bip{B^\Phi}{W^\Phi}$ is almost-connectable in $\Phi$.  Note that $B^\Phi$ and $W^\Phi$ are blocks of $G(\Psi)$, thus $B^\Phi = B$ and $W^\Phi = W$.  Let $\B$ be the family of blocks that induce the co-component of $G(\Psi) \setminus W$ containing $B$.  Observe that the subgraph of $\G(\Psi)$ induced by $\B \cup \{W\}$ is isomorphic to a co-A graph; see \Figure~\ref{fig:co-A}~(a).  By \Theorem~\ref{thm:unique-PCA-models}, such induced round graph admits two round representation, one the reverse of the other; see \eg\ \Figure~\ref{fig:co-A}~(b).  Let $\Gamma$ be the contig obtained from $\Psi \setminus \{B\}$ by swapping $F_l^\Psi(B)$ and $U_l^\Psi(B)$.  It is not hard to see that $\Bip{R^\Psi(W)}{F_l^\Psi(B)}$ is receptive in $\Gamma$, and that its reception is a disconnection of $\Bip{B}{W}$ in $\Psi$.  That is, $\Bip{B}{W}$ is almost-disconnectable in $\Psi$.  

    \begin{figure}[htb!]
      \centering
      \begin{tabular}{c@{\hspace{1cm}}c}
        \includegraphics{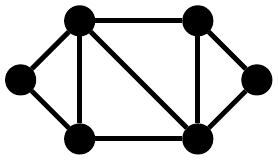} & \includegraphics{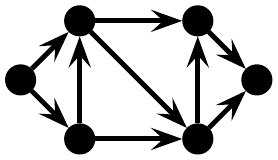} \\
        (a) & (b) 
      \end{tabular}
      \caption{(a) a co-A graph, (b) $\Topsi$ for a round representation $\Psi$ with $\G(\Psi)$ isomorphic to a co-A graph.}\label{fig:co-A}
    \end{figure}
  
    \item [(c) holds.]  By reversing $\Phi$ if required, suppose $\Bip{B^\Phi}{W^\Phi}$ is connectable in $\Phi$.  If $B$ is universal in $\Psi$, then $B$ is a co-end block of $\Psi$.  Also is $W$, because $\Bip{B}{W}$ is disconnectable in the connection of $\Bip{\{v\}}{\{w\}}$ in $\Phi$.  That is (\ref{lem:edge removal:co-ends}) follows.  For the rest of this case, suppose $B$ is not universal in $\Psi$, so neither is $W$.  Let $\B$ be the family of blocks that induce the co-component of $\G(\Phi)$ containing $B^\Phi$, $\Gamma= \Phi|\B$, and $\Lambda$ be the connection of $\Bip{\{v\}}{\{w\}}$ in $\Gamma$.  Since $B$ and $W$ are not universal in $\G(\Psi)$, it follows that $B$ and $W$ are blocks of $\Lambda$ and $\Bip{B}{W}$ is disconnectable in $\Lambda$.  Furthermore, $G(\Lambda) = G(\Gamma) \cup \{vw\}$ is a subgraph of $G(\Psi) = G(\Phi) \cup \{vw\}$ induced by the vertices of either one or two co-components of $G(\Psi)$.  Consider the following alternatives for $G(\Lambda)$.
    \begin{description}
      \item [Case 1:] $G(\Lambda)$ is co-connected.  By \Theorem~\ref{thm:unique-PCA-models}, either $\Lambda = \Omega$ or $\Lambda$ is a co-contig of $\Omega$, for $\Omega \in \{\Psi, \Psi^{-1}\}$.  In the former case, $\Bip{B}{W}$ is disconnectable in $\Omega$ and (\ref{lem:edge removal:disconnectable}) follows.  In the latter case, $\Lambda$ is described by a co-contig pair $\Bip{\X}{\Y}$.  Moreover, since $\B(\Lambda) \neq \B(\Omega)$, it follows that $\B(\Gamma) \neq \B(\Phi)$, thus $\Gamma$ is a co-contig of $\Phi$ as well.  Consequently, $B^\Phi$ and $W^\Phi$ belong to different co-contig ranges of $\Gamma$, which implies that exactly one of $B, W$ belongs to $\X$, say $B \in \X$ and $W \in \Y$.  Hence, since $\Bip{B}{W}$ are disconnectable in $\Lambda$, it follows that $F_r^\Lambda(B) \in \Y$ and $F_l^\Lambda(W) \in \X$.  Thus, $F_r^\Psi(B) = F_r^\Lambda(B)$ and $F_l^\Psi(W) = F_l^\Lambda(W)$ and (\ref{lem:edge removal:disconnectable}) follows.  
      \item [Case 2:] $G(\Lambda)$ is not co-connected.  In this case, each co-contig of $\Lambda$ is a co-contig of $\Psi$.  Since $G(\Gamma)$ is co-connected, it follows that $B$ and $W$ belong different co-contigs of $\Lambda$, thus $B$ and $W$ belong to different co-contigs of $\Psi$.  On the other hand, since $\Bip{B}{W}$ is disconnectable in $\Lambda$ and $U_r^\Lambda(B)$ belongs to the same co-contig of $\Lambda$ as $B$, it follows that $R^\Lambda(W) = U_r^\Lambda(B)$.  That is, $W$ is a co-end block of $\Lambda$, which implies that $W$ is a co-end block of $\Psi$.  Analogously, $B$ is also a co-end block of $\Psi$, thus (\ref{lem:edge removal:co-ends}) follows.
    \end{description}

    \item [(d) holds.] If $B = W$, then $w$ is universal in $G(\Psi)$, and $L^\Psi(W)$ and $R^\Psi(W)$ are co-end blocks.  The, either (\ref{lem:edge removal:refinable}) or (\ref{lem:edge removal:refinable bis}) follows according to whether $W = \{v,w\}$ or not.  Suppose $B \neq W$ for the rest of this case.  Let $\Gamma$ be the co-contig of $\Phi$ containing $W^\Phi$, and $\Lambda$ be the compressed removal of $v$ from $\Gamma$.  By definition, $G(\Lambda)$ is the co-component of $G(\Phi) \setminus \{v\}$ containing $w$.  Since $G(\Phi) \setminus \{v\} = G(\Psi) \setminus \{v\}$ and $v$ is universal in $G(\Psi)$, it follows that $G(\Lambda)$ is a co-component of $G(\Psi)$.  So, by \Theorem~\ref{thm:unique-PCA-models}, $\Lambda$ is a co-contig of $\Omega$, for $\Omega \in \{\Psi, \Psi^{-1}\}$.  On the other hand, since $\Lambda$ is the compressed removal of $v$ from $\Gamma$, it follows that $v$ is insertable into $\Lambda$ so as to obtain $\Gamma$.  Then, by \Lemma~\ref{lem:insertable contig}, we obtain that either (1) $W = \{w\}$ and $\Bip{L^\Lambda(W)}{R^\Lambda(W)}$ is refinable in $\Lambda$, or (2) $W \neq \{w\}$ and one between $\Bip{W \setminus \{w\}}{L^\Lambda(W)}$ and $\Bip{R^\Lambda(W)}{W \setminus \{w\}}$ is refinable in $\Lambda$.  If $W$ is not a co-end block of $\Lambda$, then $L^\Lambda(W) = L^\Omega(W)$ and $R^\Lambda(W) = R^\Omega(W)$, thus (\ref{lem:edge removal:refinable}) or (\ref{lem:edge removal:refinable bis}) follows. On the other hand, if $W$ is a co-end block of $\Lambda$, then $W$ is a co-end block of $\Omega$ by definition, thus (\ref{lem:edge removal:co-ends}) follows. 

  \end{description}

  The converse follows from \Lemmas~\ref{lem:disconnection}, \ref{lem:insertable contig}, \ref{lem:disconnectable contig}~and~\ref{lem:almost-disconnection}.  
\end{proof}

Four phases are applied to transform $\Psi, \Psi^{-1}$ into the decremental round block representations $\Phi, \Phi^{-1}$ representing $H \setminus \{vw\}$.  Let $B$ and $W$ be the blocks of $H$ containing $v$ and $w$, respectively.  In the first phase, \Lemma~\ref{lem:disconnectable contig} is applied on $B$ and $W$ when $B$ and $W$ are co-end blocks of different co-contigs.  After this phase we can assume that $H \setminus \{vw\}$ is PCA if and only if $\Psi$, $B$ and $W$ satisfy one of conditions (\ref{lem:edge removal:co-A})--(\ref{lem:edge removal:refinable bis}) of \Lemma~\ref{lem:edge removal}.   In the second phase, Algorithm~\ref{alg:edge removal} is applied twice.  First it is applied on $\Psi$ and $vw$ so as to obtain $\Phi$, and then it is applied on $\Psi^{-1}$ and $vw$ so as to obtain $\Psi^{-1}$.  Observe that, as it happens with the edge insertion problem, the application of Algorithm~\ref{alg:edge removal} on $\Psi^{-1}$ must mimic the application of Algorithm~\ref{alg:edge removal} on $\Psi$.  Finally, the third and forth phases restore the straightness and co-end pointers, respectively.

The correctness of the above algorithm follows from \Lemma~\ref{lem:edge removal}.  In particular, Step~\ref{alg:edge removal:disconnectable} of Algorithm~\ref{alg:edge removal} checks statements  (\ref{lem:edge removal:co-A}) and (\ref{lem:edge removal:disconnectable}) on $\Psi$.  Following, statements (\ref{lem:edge removal:refinable}) and (\ref{lem:edge removal:refinable bis}) are checked by Steps \ref{alg:edge removal:v universal A}--\ref{alg:edge removal:v universal D}.   If any of (\ref{lem:edge removal:co-A})--(\ref{lem:edge removal:refinable bis}) holds, then $\Psi$ is accordingly transformed into a compressed round representation $\Phi$ of $\G(\Psi) \setminus \{vw\}$.  Note that, by definition, $\Phi$ is $1$-universal, thus, by \Corollary~\ref{cor:indistinguishable}, $\Phi$ is a round block representation of $\G(\Psi) \setminus \{vw\}$.  

\begin{algorithm}[htb!]
  \caption{Removal of an edge $vw$ from a block contig.}\label{alg:edge removal}

  \textbf{Input:} a decremental block contig $\Psi$, and an edge $vw$ of $G(\Psi)$.

  \textbf{Output:} if $\Psi, v$ and $w$ satisfy one of conditions (\ref{lem:edge removal:co-A})--(\ref{lem:edge removal:refinable bis}) of \Lemma~\ref{lem:edge removal}, then $\Psi$ is updated into a round block representation $\Phi$ of $G(\Psi) \setminus \{vw\}$; otherwise, the algorithm halts in error.

  \mbox{}

  \begin{AlgorithmSteps}
    \Step{If $d(v) < d(w)$, then swap $v$ and $w$.}
    \Step{Let $B$ and $W$ be the blocks containing $v$ and $w$, respectively.}
    \Step{If $\Bip{B}{W}$ or $\Bip{W}{B}$ is either disconnectable or almost-disconnectable, then transform $\Psi$ into the corresponding disconnection of $\Bip{\{v\}}{\{w\}}$ or $\Bip{\{w\}}{\{v\}}$ in $\Psi$ and halt.}\label{alg:edge removal:disconnectable}
    \Step{If $v$ is not universal in $G(\Psi)$, then halt in error.}\label{alg:edge removal:v universal A}
    \Step{Transform $\Psi$ into the compressed removal of $v$ and let $W' = W \setminus \{v,w\}$.}\label{alg:edge removal:v universal B}
    \Step{If $W' \neq \emptyset$, and either $\Bip{W'}{L(W)}$ or $\Bip{R(W)}{W'}$ is refinable, then transform $\Psi$ into the corresponding $\{v\}$-refinement of $\Bip{W'}{L(W)}$ or $\Bip{R(W)}{W'}$ in $\Psi$ and halt.}\label{alg:edge removal:v universal C}
    \Step{If $W' = \emptyset$ and $\Bip{R(W)}{L(W)}$ is refinable, then transform $\Psi$ into the $\{v\}$-refinement of $\Bip{R(W)}{L(W)}$ and halt.}\label{alg:edge removal:v universal D}
    \Step{Halt in error.}\label{alg:edge removal:v universal end}
    \DecreaseIndent
  \end{AlgorithmSteps}
\end{algorithm}

Discuss the time complexity of the four phases of the algorithm.  To determine if $B$ and $W$ are co-end blocks of different co-contigs, so as to apply \Lemma~\ref{lem:disconnectable contig} for the first phase, proceed as follows.  First, test whether $B$ and $W$ are co-end blocks of $\Psi$.  If so, then compute the family $\B$ formed by the co-end blocks of the co-contig that contains $B$.  As in \Lemma~\ref{lem:disconnectable contig}, $\B$ is computed in $O(1)$ time by means of the co-end pointers.  If $W \not\in \B$, then \Lemma~\ref{lem:disconnectable contig} is applied on $B$ and $W$.  By \Lemma~\ref{lem:disconnectable contig} the first phase takes $O(1)$ time.  

The implementation of Step~\ref{alg:edge removal:disconnectable} of Algorithm~\ref{alg:edge removal} is  the same as the implementation of Step~\ref{alg:edge insertion:connectable} of Algorithm~\ref{alg:edge insertion}.  Just observe that $\Bip{B}{W}$ is almost-disconnectable if the co-component of $\G(\Psi) \setminus \{W\}$ containing $B$ is isomorphic to a co-P graph.  Thus, $O(1)$ time is required for Step~\ref{alg:edge removal:disconnectable} of Algorithm~\ref{alg:edge removal}.  The remaining steps of Algorithm~\ref{alg:edge insertion} take $O(1)$ time by \Lemmas \ref{lem:compressed removal}, \ref{lem:refinable complexity}, \ref{lem:disconnectable complexity}~and~\ref{lem:almost-disconnection}.  Consequently, the second phase also takes $O(1)$ time.  

The third phase is equivalent to the second phase of the vertex removal algorithm of \Section~\ref{subsec:vertex removal}.  Thus, $O(1)$ time is required for the third phase.  

For the forth phase, consider how do the co-contig pairs of $\Phi$ look like.  For this, let $B^\Phi$ and $W^\Phi$ be the blocks of $\Phi$ that contain $v$ and $w$, respectively, $\B$ be the family of blocks that induce the co-component of $\Phi$ containing $B^\Phi$, and $\Gamma = \Phi | \B$.  Clearly, if $G(\Psi)$ is not co-bipartite, then $G(\Phi)$ is neither co-bipartite.  Thus, the co-end pointers of $\Phi$ need not be updated when $G(\Psi)$ is not co-bipartite.  Suppose, then, that $G(\Psi)$ is co-bipartite.  If $\Lambda$ is a co-contig of $\Psi$ that contains neither $B$ nor $W$, then $\Lambda$ is also a co-contig of $\Phi$.  Therefore, the only co-end pointers that should be updated correspond to blocks of $\Gamma$. Let $\Bip{\X}{\overline{\X}}$ and $\Bip{\Y}{\overline{\Y}}$ be the co-contig pairs of $\Psi$ that respectively contain $B$ and $W$, prior to the execution of Algorithm~\ref{alg:edge removal}.  Denote by $B_l$ and $B_r$ the left and right co-end blocks of $\X$, and by $W_l, W_r$ the left and right co-end blocks of $\Y$.  The structure of $\Gamma$ depends on which of the conditions of \Lemma~\ref{lem:edge removal} holds.
\begin{description}
  \item [(\ref{lem:edge removal:co-A}) holds.]  This case need not be considered because, as argued before, $\Phi$ and $\Psi$ has the same co-contig pairs.
  \item [(\ref{lem:edge removal:disconnectable}) holds.]  Suppose, w.l.o.g., that $\Bip{B}{W}$ is disconnectable in $\Psi$.  Note that, by definition, $B = B_l$ if and only if $W = W_r$.  If $B \neq B_l$, then, by definition, $\Gamma$ is co-bipartite and $\Bip{[B_l, B_r \setminus \{v\}]}{[W_l \setminus \{w\}, W_r]}$ is a co-contig pair describing $\Gamma$.  On the other hand, if $B = B_l$, then there are two possibilities according to whether $\X = \Y$ or $\X \neq \Y$.  If $\X = \Y$, then there is a path of even length that joins $v$ and $w$ in $\overline{H}$, thus $H \setminus \{vw\}$ is not co-bipartite, \ie, $\Gamma = \Phi$ has no co-end blocks.  Finally, if $\X\neq\Y$, then $\Bip{\overline{\Y} \Cat \X}{\Y \Cat \overline{\X}}$ is a co-contig pair describing $\Gamma$.
  \item [(\ref{lem:edge removal:refinable}) holds.]  If $B = W$, then $B \setminus \{v,w\}$ is the universal block of $\Phi$, if any, while $\Gamma$ is described by the co-contig pair $\Bip{[\{v\}, \{v\}]}{[\{w\},\{w\}]}$.  If $B \neq W$ and, w.l.o.g., $\Bip{W \setminus \{w\}}{L^\Psi(W)}$ is refinable, then $B = B_l = B_r$ and $\overline{\X} = \emptyset$, while, by \Lemmas \ref{lem:receptive representation}~and~\ref{lem:compressed insertion}, $W_l = W$ and $L^\Psi(W)$ is the right co-end block of $\overline{\Y}$.  Therefore, $\Bip{[\{w\}, W_r]}{[B, L^\Psi(W)]}$ is a co-contig pair describing $\Gamma$.  
  \item [(\ref{lem:edge removal:refinable bis}) holds.]  This case is analogous to the previous case.
\end{description}
By the above discussion, only $O(1)$ co-end pointers of $\Gamma$ need to be updated.  Furthermore, the blocks corresponding to such pointers can be obtained in $O(1)$ time.  Indeed, either no co-end pointer needs to be updated or both $B$ and $W$ are co-end blocks of $\Psi$.  In the latter case, the co-end blocks whose co-end pointers require an update are obtainable with $CE^\Psi(B)$, $CE^\Psi(W)$, and the mappings $U_l$ and $U_r$.  Therefore, the whole algorithm takes $O(1)$ time.

\begin{theorem}\label{thm:edge removal}
  The problem of deciding whether a decremental graph is a PCA graph takes $O(1)$ time per removed edge.
\end{theorem}

\section{Fully dynamic recognition of proper circular-arc graphs}
\label{sec:dyn-pca connectivity}

The fully dynamic algorithm for the recognition of PCA graphs is obtained by combining the incremental and decremental algorithms described in \Sections \ref{sec:incremental}~and~\ref{sec:decremental}.  There is, however, a major incompatibility between the data structures used by these algorithms: incremental contigs are equipped with end pointers, while decremental contigs are equipped with co-end pointers.  It is not clear how the end and co-end pointers can coexist in an efficient fully dynamic algorithm.  In their fully dynamic algorithm for the recognition of PIG graph, Hell \etal\ discard the end pointers.  Instead, each base straight representation is equipped with a fully dynamic algorithm that solves the recognition and connectivity problems on union of paths graphs.  Our fully dynamic algorithm for the recognition of PCA graphs follows the same approach.  That is, end and co-end pointers are discarded, and each base round representation is augmented with fully dynamic algorithms that solve the recognition and connectivity problems on $2$-degree graphs.  \Section~\ref{subsec:2-degree} describes the fully dynamic algorithm for the recognition and connectivity of $2$-degree graphs, while \Section~\ref{subsec:fully dynamic} discusses the fully dynamic recognition of PCA graphs.

\subsection{Fully dynamic connectivity of 2-degree graphs}
\label{subsec:2-degree}

This part describes a simple fully dynamic algorithm that solves the recognition and connectivity problems for $2$-degree graphs (refer to~\cite{HellShamirSharanSJC2001} for a similar algorithm that works on union of paths graphs).  The input of the algorithm is a sequence of operations that involve: inserting or removing an isolated vertex, inserting or removing an edge, querying if two vertices belong to the same component, and obtaining the set of vertices of degree $1$ that belong to the same component as a given vertex.

Let $G$ be a $2$-degree graph.  By definition, each component $H$ of $G$ is either an induced path or an induced cycle.  For the implementation of $H$, two balanced (\eg\ red-black) trees $T$ and $T^{-1}$, and a boolean value $p$ are stored.  Tree $T$ has the same vertices as $H$, and these vertices are stored in such a way that the inorder traversal $v_1, \ldots, v_i$ of $T$ is a path of $H$.  Similarly, $T^{-1}$ has the same vertices as $H$, but its inorder traversal is $v_i, \ldots, v_{1}$.  On the other hand, $p$ is true if and only if $H$ is a path, \ie, $p$ is true when $v_1$ and $v_i$ are not adjacent in $H$.   Finally, $G$ is stored as a set containing one of the above triples for each of its components.  With this implementation, the insertion and removal of vertices are executed in $O(1)$ time, while the remaining operations take $O(\log n)$ time~\cite{CormenLeisersonRivestStein2009,Tarjan1983}.

\subsection{Fully dynamic contigs}
\label{subsec:fully dynamic}

In this part we complete the fully dynamic algorithm for the recognition of PCA graphs, by making the incremental and decremental algorithms compatible.  

Let $\Phi$ be a round block representation.  The \Definition{contigs graph} of $\Phi$ is the graph $C(\Phi)$ that has one vertex $v(B)$ for each $B \in \B(\Phi)$ such that $v(B)$ and $v(W)$ are adjacent in $C(\Phi)$ if and only if $B$ and $W$ belong to the same contig of $\Phi$ and are consecutive in $\B(\Phi)$.  The \Definition{co-contigs graph} $\overline{C}(\Phi)$ of $\Phi$ is defined in a similar manner.  There is a vertex $w(B)$ in $\overline{C}(\Phi)$ for each $B \in \B(\Phi)$, while the edges of $\overline{C}(B)$ depend on whether $\G(\Phi)$ is co-bipartite or not.  In the former case, $B$ and $W$ are adjacent in $\overline{C}(B)$ if and only if $B$ and $W$ belong to the same co-contig range and are consecutive in $\B(\Phi)$.  In the latter case, $B$ and $W$ are adjacent in $\overline{C}(B)$ if and only if $B$ and $W$ are consecutive in $\G(\Phi)$.  By definition, $C(\Phi)$ and $\overline{C}(\Phi)$ are $2$-degree graphs.

For the fully dynamic recognition, each dynamic PCA graph is implemented with two base round block representations $\Phi$, $\Phi^{-1}$, a \Definition{universal pointer} $U^\Phi$, and the graphs $C(\Phi)$ and $\overline{C}(\Phi)$ implemented as in \Section~\ref{subsec:2-degree}.  The universal pointer references the universal block of $\G(\Phi)$ if $\Phi$ is not $0$-universal, and references $NULL$ otherwise.  Also, pointers to $v(B)$ and $w(B)$ are stored together with $B$ in both $\Phi$ and $\Phi^{-1}$, while pointers to $B$ are stored with $v(B)$ and $w(B)$ in $C(\Phi)$ and $\overline{C}(\Phi)$, respectively.  Note that, by definition, $C(\Phi)$ is isomorphic to $C(\Phi^{-1})$, and $\overline{C}(\Phi)$ is isomorphic to $\overline{C}(\Phi^{-1})$.  Thus, $C(\Phi)$ and $\overline{C}(\Phi)$ can be regarded as the contigs and co-contigs graphs of $\Phi^{-1}$ as well.  

The fully dynamic algorithm works while a series of vertex insert, vertex remove, edge insert, and edge remove operations are executed.  For each such operation, the corresponding algorithms described in \Sections \ref{sec:incremental}~and~\ref{sec:decremental} are executed on $\Phi$ and $\Phi^{-1}$.  Suppose the following operation to be executed is a vertex insertion.  The vertex insertion algorithm described in \Section~\ref{sec:incremental} makes use of end pointers, which are present in neither $\Phi$ nor $\Phi^{-1}$.  Instead, if an end pointer of $B\in \B(\Phi)$ needs to be accessed, then a function $E^\Phi$ is executed with input $B$.  The output of $E^\Phi$ is $NULL$ if $B$ is not an end block of $\Gamma$, while it is the other end block of its contig otherwise.  That is, $E^\Phi$ emulates the behavior of the missing end pointer of $B$.  Note that $B$ is an end block of $\Phi$ if and only if $B$ has degree at most $1$ in $C(\Phi)$.  Thus, $E^\Phi$ requires $O(1)$ queries to $C(\Phi)$.  In a similar manner we can implement functions $CB^\Phi$ and $CE^\Phi$ that emulate the corresponding co-bipartite and co-end pointers that are missing in $\Phi$.  Just observe that $B$ is a co-end block of $\Phi$ if it has degree at most $1$ in $\overline{C}(\Phi)$.  

Besides accessing end pointers and co-end pointers, the incremental and decremental algorithms also transform $\Phi$ by inserting and removing semiblocks, and by changing the order of its semiblocks.  It is not hard to see how to maintain the universal pointer of $\Phi$.  On the other hand, each time a near pointer of $\Phi$ is modified, the contigs and co-contigs graphs may have to be updated as well.  However, each update of a near pointer of $\Phi$ involves only $O(1)$ insertion and removal of edges and vertices in $C(\Phi)$ and $\overline{C}(\Phi)$. 

Only $O(1)$ access to the end and co-end pointers and $O(1)$ modifications of the near pointers are applied by the incremental and decremental algorithms described in \Sections \ref{sec:incremental}~and~\ref{sec:decremental}. Therefore, by the discussions above, the main theorem of this article follows. 

\begin{theorem}\label{thm:fully dynamic}
  The problem of deciding if a fully dynamic graph $G$ is a PCA graph takes $O(\log n)$ time per inserted or removed edge.  Furthermore, the insertion and removal of a vertex $v$ take $O(d_G(v) + \log n)$ time each, while the problems of querying if two vertices of $G$ belong to the same component or if two vertices belong to the same co-component require $O(\log n)$ time each.  
\end{theorem}

\section{Further remarks}
\label{sec:conclusions}

In this article we presented an algorithm for the recognition of fully dynamic PCA graphs.  The algorithm is a generalization of the HSS algorithm because it has the same time bounds and it can answer in $O(1)$ time whether the dynamic graph is in fact a PIG graph.  The bottleneck for the recognition of both PIG and PCA graphs is an algorithm that solves the connectivity problem on fully dynamic $2$-degree graphs.  Any improvement on the connectivity algorithm gets immediately translated to an improvement on the recognition algorithm.  Hell \etal~\cite{HellShamirSharanSJC2001}  proved that at least $O(\log n/(\log\log n + \log b))$ amortized time per edge operation is required to solve the fully dynamic recognition problem of PIG graphs, when the cell probe model of computation with word-size $b$ is used.  Moreover, the connectivity problem on fully dynamic PIG graphs has the same lower time bound.  We conclude, therefore, that the recognition algorithm presented in this paper in near-optimal.

The recognition algorithm of this article can be generalized so as to recognize another interesting family of PCA graphs.  A \Definition{locally straight representation} is a round representation $\Phi$ such that $\Phi|[F_l(B), F_r(B)]$ is straight, for every $B \in \B(\Phi)$.  A \Definition{locally straight graph} is a round graph that admits a locally straight representation.  Similarly, a graph is a \Definition{proper Helly circular-arc (PHCA)} graph if it is isomorphic to $G(\Phi)$ for some locally straight representation $\Phi$.  PHCA and locally straight graphs were introduced and motivated by Lin \etal\ in~\cite{LinSoulignacSzwarcfiter2011}, where a simple characterization in terms of round representations is given.  A round representation $\Phi$ is locally straight if and only if $F_r(F_r(B)) \nto B$, for every $B \in \B(\Phi)$.  Also, a theorem analogous to \Theorems \ref{thm:unique-PIG-models}~and~\ref{thm:unique-PCA-models} holds for locally straight graphs.  That is, every locally straight graph admits at most two locally straight representations, one the reverse of the other.  These results can be used to extend the recognition algorithm of PCA graphs so that it can answer if the dynamic graph is PHCA in $O(1)$ time.  The details appear in~\cite{Soulignac2010}.

In this article we did not discuss the certification problem associated with the recognition of PCA graphs.  The goal of a certified algorithm is to provide some piece of evidence showing that the output of the algorithm is correct.  Such an evidence is called a certificate.  There are two kinds of certificates in a recognition problem, namely positive and negative certificates.  The former are given when the output is YES, \ie\ when the input graph belongs to the class, whereas the latter are given when the output is NO, \ie\ when the input graph does not belong to the class.  For instance, a certified algorithm for the recognition of PCA graphs could output round representations as positive certificates and forbidden induced subgraphs as negative certificates.  Several design issues related to the certification problem are discussed in~\cite{McConnellMehlhornNaeherSchweitzerCSR2011}.  The DHH algorithm always outputs a positive certificate.  Kaplan and Nussbaum~\cite{KaplanNussbaumDAM2009} developed an $O(n+m)$ time algorithm that finds a negative certificate.  Thus, the certification problem for static graphs is somehow solved.  However, the algorithm by Kaplan and Nussbaum is not able to produce a forbidden induced subgraph of the input graph.  We believe that our algorithm can be extended so as to provide such certification for static graphs in $O(n+m)$ time.  Moreover, we believe that it can even be extended so as to provide such certificates for incremental graphs in $O(1)$ time per inserted edge.  To begin a research in this direction, it could be useful to consider those places where the incremental recognition algorithm outputs NO.

\section*{Acknowledgments}

The author is grateful to Min Chih Lin for pointing out that the co-components algorithm can be solved in $O(\Delta)$ time, and to Jayme Szwarcfiter for asking whether the HSS algorithms can be generalized so as to recognize PHCA or PCA graphs.  These were key observations for beginning the research that gave life to this article.


\end{document}